\documentclass[twocolumn,tighten,twocolappendix]{aastex631} 

\usepackage{apjfonts} 
\usepackage{physics}
\usepackage{multirow}
\usepackage{graphicx}
\usepackage{amsmath}
\usepackage{amssymb}
\usepackage{color}
\usepackage{mathtools}
\usepackage{ulem} 
\usepackage[all]{hypcap}

\graphicspath{{./}{figures/}}

\newcommand{\rgc}{R_{\rm gc}}

\newcommand{\phigal}{\phi_g}
\newcommand{\phiclus}{\phi_c}
\newcommand{\phieff}{\phi_{\rm eff}}
\newcommand{\Ejacobi}{E_{\rm J}}
\newcommand{\Ecrit}{E_{\rm crit}}
\newcommand{\Etilde}{\tilde{E}}
\newcommand{\Myr}{{\rm Myr}}
\newcommand{\Gyr}{{\rm Gyr}}
\newcommand{\pc}{{\rm pc}}
\newcommand{\kpc}{{\rm kpc}}
\newcommand{\Msun}{M_\odot}

\newcommand{\kms}{{\rm km\,s}^{-1}}
\newcommand{\vect}[1]{\boldsymbol{\mathbf{#1}}}

\begin{document}

\title{Stellar Escape from Globular Clusters. II. Clusters May Eat Their Own Tails}

\author[0000-0002-9660-9085]{Newlin C. Weatherford}
\affil{Department of Physics \& Astronomy, Northwestern University, Evanston, IL 60208, USA}
\affil{Center for Interdisciplinary Exploration \& Research in Astrophysics (CIERA), Northwestern University, Evanston, IL 60208, USA}

\author[0000-0002-7132-418X]{Frederic A. Rasio}
\affil{Department of Physics \& Astronomy, Northwestern University, Evanston, IL 60208, USA}
\affil{Center for Interdisciplinary Exploration \& Research in Astrophysics (CIERA), Northwestern University, Evanston, IL 60208, USA}

\author[0000-0002-3680-2684]{Sourav Chatterjee}
\affil{Tata Institute of Fundamental Research, Homi Bhabha Road, Mumbai 400005, India}

\author[0000-0002-7330-027X]{Giacomo Fragione}
\affil{Department of Physics \& Astronomy, Northwestern University, Evanston, IL 60208, USA}
\affil{Center for Interdisciplinary Exploration \& Research in Astrophysics (CIERA), Northwestern University, Evanston, IL 60208, USA}

\author[0000-0003-4412-2176]{Fulya K{\i}ro\u{g}lu}
\affil{Department of Physics \& Astronomy, Northwestern University, Evanston, IL 60208, USA}
\affil{Center for Interdisciplinary Exploration \& Research in Astrophysics (CIERA), Northwestern University, Evanston, IL 60208, USA}

\author[0000-0002-4086-3180]{Kyle Kremer}
\affil{TAPIR, California Institute of Technology, Pasadena, CA 91125, USA}

\begin{abstract}
We apply for the first time orbit-averaged Monte Carlo star cluster simulations to study tidal tail and stellar stream formation from globular clusters, assuming a circular orbit in an unevolving, spherical Galactic potential. Treating energetically unbound bodies (potential escapers; PEs) as collisionless enables this fast but spherically symmetric method to capture asymmetric extratidal phenomena with exquisite detail. Reproducing stream features such as epicyclic overdensities, we show how `returning tidal tails' can form after the stream fully circumnavigates the Galaxy, enhancing the stream’s velocity dispersion by several ${\rm km\,s}^{-1}$ in our ideal case. While a truly clumpy, asymmetric, and evolving Galactic potential would greatly diffuse such tails, they warrant scrutiny as potentially excellent constraints on the Galaxy's history and substructure. Re-examining the escape timescale $\Delta t$ of PEs, we find new behavior related to chaotic scattering in the three-body problem; the $\Delta t$ distribution features sharp plateaus corresponding to distinct locally smooth patches of the chaotic saddle separating the phase space basins of escape. We study for the first time $\Delta t$ in an evolving cluster, finding that $\Delta t\sim(E_{\rm J}^{-0.1},E_{\rm J}^{-0.4})$ for PEs with (low, high) Jacobi energy $E_{\rm J}$, flatter than for a static cluster ($E_{\rm J}^{-2}$). Accounting for cluster mass loss and internal evolution lowers the median $\Delta t$ from ${\sim}10\,{\rm Gyr}$ to ${\lesssim}100\,{\rm Myr}$. We finally outline potential improvements to escape in the Monte Carlo method intended to enable the first large grids of tidal tail/stellar stream models from full globular cluster simulations and detailed comparison to stream observations.
\end{abstract}

\section{Introduction} \label{S:intro}
Recent astronomical surveys, especially the exquisite kinematics from \textit{Gaia} \citep{GaiaDR1}, have revealed extensive substructure in the Milky Way (MW), including nearly 100 stellar streams---for a recent review and catalog, see \cite{Helmi2020} and \cite{Mateu2023}, respectively. Many of these strands of stars sharing similar orbits are debris from either dwarf galaxies recently accreted and tidally disrupted by the MW or dissolved/dissolving globular clusters \citep[GCs; e.g.,][]{Bonaca2021}. Aiding this linkage, tidal tails directly emanate from ${\approx}15$ MWGCs \citep[e.g.,][]{Piatti2020}, some extending into full streams.

As kinematically cold structures sensitive to subtle perturbations on long timescales, stellar streams are outstanding probes of the MW's mass and gravitational potential, dark matter (DM) halo, internal substructure, and assembly history \citep[e.g.,][]{Koposov2010, Bonaca2014, Bonaca2018, Kupper2015, Bovy2016}. Stream morphology and kinematics can provide especially tantalizing hints to the nature of particulate DM \citep[e.g.,][]{Banik2021}; for example, cold DM should produce cuspier (more centrally dense) DM subhalos than alternatives such as warm DM \citep{Peebles1982}, which suppresses primordial density fluctuations on small scales, or ultra-light scalar `fuzzy' DM \citep{Hui2017}, which forms less small-scale substructure. So the frequency and morphology of gaps, and kinematic heating, induced in streams by passing DM subhalos depends on DM's identity \citep[e.g.,][]{Ibata2002, Johnston2002, Carlberg2009}.

Yet, capitalizing on the prospect of DM inference with stellar streams has proven challenging since many phenomena simultaneously and degenerately affect stream morphology and kinematics. Stream gaps, overdensities, and fanning naturally arise via the epicyclic trajectories of slow GC escapers even in static MW potentials without any clumpy substructure \citep[e.g.,][]{CapuzzoDolcetta2005,Kupper2008}, and also from perturbations by many \textit{baryonic} substructures, including giant molecular clouds \citep{Amorisco2016}, the MW spiral arms or disk \citep{Banik2019,Carlberg2023}, infalling MW satellites \citep{Garavito-Camargo2019}, and other GCs \citep{Doke2022}. Further complicating matters, MW substructure such as a rotating bar \citep{Hattori2016,Pearson2017} or even modified gravity \citep{Thomas2018,Kroupa2022} can induce stream asymmetry and observational artifacts can cause false gaps/overdensities \citep{Ibata2020}. Internal GC dynamics is impactful, too; in particular, dynamical heating from central black hole (BH) binaries can dramatically accelerate the evaporation rate at the end of a GC's life \citep[e.g.,][]{BanerjeeKroupa2011,Whitehead2013,Contenta2015,Chatterjee2017,Giersz2019,Weatherford2021}, thereby increasing the tidal tail density \citep{Gieles2021,Gieles2023,Roberts2024}. Disentangling the individual contributions of such diverse phenomena in observations (e.g., to measure the properties of DM subhalos specifically) will require comprehensive modeling of stream formation and disruption.

The importance of stellar streams to Galactic archeology strongly motivates further theoretical study of escape from GCs. This work is the second of a series on this topic that began with a thorough exploration of escape mechanisms \citep[][hereafter W23]{Weatherford2023a}. W23 emphasized more strongly high-speed ejection, relevant to production of runaway or hypervelocity stars in the MW halo. Focusing more on low-speed escape, we now apply for the first time orbit-averaged GC models---using the \cite{Henon1971a,Henon1971b} Monte Carlo method---to study stellar stream formation.

At first glance, orbit-averaged GC models are not an obvious choice for this application. The impacts of the GC potential and collisional dynamics steeply diminish beyond the tidal boundary, so it is common when simulating streams to neglect the GC's internal dynamics in favor of collisionless `streakline' or `particle spray' techniques \citep[e.g.,][]{Varghese2011,Kupper2012,Lane2012,Bonaca2014,Gibbons2014,Fardal2015,Shipp2018,Grondin2023}. Stars are simply sprayed at a prescribed rate and velocity distribution from a point representing the GC (or the Lagrange points on its tidal boundary) orbiting in an external Galactic potential. Key benefits of fully modeling the GC are then to accurately determine the stellar escape rates and velocities, and internal stream details such as stellar masses, luminosities, and binary properties. Direct summation $N$-body codes \citep[e.g., \texttt{nbody6}; ][]{Aarseth1999} are the ``gold standard" for such purposes due to their exquisite accuracy and accommodation of GC asymmetry (essential for tidal physics) and stages of the GC's life spent out of virial equilibrium or with small $N$ (essential for late-stage dissolution). Yet direct summation is computationally intensive; despite steady computing advances the method still has not been used to model a GC with a density and initial $N$ typical of MWGCs over a full Hubble time \citep[e.g.,][]{Wang2016,ArcaSedda2023}. So direct summation is ill-suited to generating large model grids of extratidal structures from dense GCs, necessary to optimally match specific MW streams or explore the many aforementioned factors that can significantly influence stream properties.

The Monte Carlo method is a much faster alternative to direct summation, capable of  simulating MWGCs of typical mass and density using $10^3$--$10^4$-times fewer CPU--hours. Of the two main algorithmic lineages \citep[excellently summarized by][]{Vasiliev2015}, the orbit-averaged version descended from \cite{Henon1971a,Henon1971b} is the only one still in regular use---specifically as the codes \texttt{MOCCA} and our \texttt{CMC} (respectively overviewed by \citealt{Giersz2013} and \citealt{CMCRelease}). Instead of direct orbit integration, H\'{e}non's method applies a statistical treatment of stellar dynamics by modeling the cumulative effect of many distant two-body encounters as a single effective scattering between neighboring bodies, radially sorted into pairs. This occurs each timestep on the relaxation timescale $t_r$ rather than the dynamical timescale, leveraging that distant two-body encounters dominate evolution in each body's energy $E$ and angular momentum $J$. Only especially strong encounters are handled via a small-$N$ direct integrator, greatly hastening the computation. At the end of the timestep, each body's position and velocity ($r,v_r,v_t)$ are randomly sampled from the time-averaged orbit consistent with the body's new $E$ and $J$.

H\'{e}non's method compares well to both direct summation and observations of typical GCs in most regimes \citep[e.g.,][]{Giersz2013,Rodriguez2016,CMCCatalog}, but imposes three key assumptions: (1) the radial sorting and orbit averaging impose spherical symmetry (but not velocity isotropy), (2) the statistical treatment imposes a large $N$ and virial equilibrium, preventing study of the final stages of GC dissolution, and (3) the orbit averaging neglects global evolution occurring on timescales $\ll t_r$, such as sharply evolving tides (tidal shocks). All three assumptions are relevant to escape, but most critically the assumed spherical symmetry means that unmodified the method cannot resolve tidal tails.

We test a simple solution to this challenge by removing bodies from \texttt{CMC} once they obtain sufficient energy to escape (usually in the GC's core) and evolving their trajectories from there in the full asymmetric tidal field. We explore how well this approach---implemented as a post-processing step applicable to any \texttt{CMC} simulation---reproduces known features of tidal tails and stellar streams in the case of a circular GC orbit in an unevolving, spherical MW potential. While a vast simplification, this well-studied regime---standard for H\'{e}non's method---is ideal for a first test of our approach. Crucially, it also enables us to investigate several nuances that have yet to be explored in even this simple case, including chaotic behavior in the escape (survival) timescale of energetically unbound bodies within GCs, modification of this timescale by evolution of the GC potential, and impact of return trajectories. Note our modification does not generalize the internal dynamics of the Monte Carlo method to non-spherical potentials \citep[e.g.,][]{Vasiliev2015} or its escape prescriptions to evolving external tides \citep[e.g.,][]{Sollima2014, Rodriguez2023}. Such enhancements are complementary to our approach and the latter especially are similar to our intended future upgrades to \texttt{CMC} (see Section~\ref{S:discussion}).

As in W23, we shall often refer to escape before and after \textit{core collapse}, the GC's observable change from a flat (non-core-collapsed; NCC'd) to a steep (core-collapsed; CC'd) central surface brightness. This occurs upon ejection of the BH population, which weakens binary burning---hardening of binaries via encounters with passing bodies \citep[e.g.,][]{Heggie1975,Hills1975}. The potential energy released by the binaries heats the bodies involved, supporting the GC's core against collapse. BH binaries are strong heat sources due to their mass, but GCs born especially dense quickly harden and eject them; binary burning then relies on (less massive) white dwarfs, reducing heating and allowing the core to observably collapse \citep[e.g.,][]{Chatterjee2013a,Kremer2019a,CMCCatalog,Kremer2021,Rui2021a}. We recap how this affects GC escape mechanisms in Section~\ref{S:results_energy_distribution}, but see also W23.

The paper is organized as follows. We review relevant theory in Section~\ref{S:theory} and describe our GC simulations in Section~\ref{S:models}. We explain how we integrate (in post-processing) the trajectories of unbound bodies removed from the simulations in Section~\ref{S:escaper_evolution}. In Section~\ref{S:results}, we first examine the escape energies and escape timescale from typical NCC'd and CC'd GCs. We then examine how GC mass loss affects the escape timescale and how return trajectories impact stream morphology and kinematics. In Section~\ref{S:discussion}, we discuss observational implications, additional complexities and caveats, and potential upgrades to escape in \texttt{CMC}. We conclude with a summary of our findings and planned future work in Section~\ref{S:summary}.

\section{Theory} \label{S:theory}
We examine escape in the simplified case of an \textit{evolving} spherical GC potential $\phiclus$ circularly orbiting within an \textit{unevolving} spherical Galactic potential $\phigal$. We assume that once a cluster member becomes a \textit{potential escaper} (PE) by gaining enough energy to eventually escape it ceases to be influenced by scattering with other stellar bodies. This collisionless approximation allows the PE mass distribution to follow the non-spherical tidal field despite the spherical $\phiclus$ in \texttt{CMC}. The PE trajectories in this case are an example of the circular restricted three-body problem \citep[CR3BP, e.g.,][]{Szebehely1967, Marchal1990, Henon1997, ValtonenKarttunen2006, Koon2011}. In this Section, we review the elements of the CR3BP most essential to our new results on the escape timescale and return trajectories, leaving details of our algorithmic implementation to Section~\ref{S:models}. While the classic CR3BP features unevolving $\phiclus$ and $\phigal$, much of its conceptual framework remains useful in our case, and we comment where relevant on the impact of $\phiclus$'s time-dependence.

\subsection{Problem Setup} \label{S:theory_problem_setup}
We study trajectories of stellar bodies in the gravitational fields of their birth GC---of slowly evaporating mass $m(t)$---and the MW, each on instantaneously circular orbits about their mutual center-of-mass at a constant angular speed $\omega=v_g/\rgc$, where $v_g$ is the GC's circular speed at Galactocentric distance $\rgc$. To satisfy Kepler's third law at $t=0$, we define the MW's mass enclosed within $\rgc$ to be a constant $M\equiv\rgc v_g^2/G - m(0)$. Technically, a constant $M$, $\omega$, $v_g$, and $\rgc$ is mildly inconsistent with a decreasing $m(t)$ that would transfer about half of its lost mass to the volume within $\rgc$, but these assumptions are very accurate since $m/M\sim 10^{-6}$ for a MWGC and since the circular speed profile in the MW halo is very flat. Defining coordinates in the center-of-mass frame $XYZ$ in units of $\rgc$, the MW and GC centers instantaneously trace circles in the $XY$-plane with respective dimensionless radii $[\mu(t),1-\mu(t)]$,
where $\mu(t) \equiv m(t) / [M+m(t)]$.\footnote{The Galactic nucleus' high density means MWGCs have $\mu\ll 1$ even at low $\rgc$, so accounting for $\mu$ is pedantic in our context. We do so to be precise and mindful of related contexts where $\mu$ is non-negligible, e.g., GCs in dwarf galaxies, which we may explore later in this series.}

We also define a clustercentric frame $xyz$, in units of $\rgc$, that co-rotates with the GC as it orbits the center-of-mass. In this frame, the GC's center is the origin ($\vect{r_c}=\vect{0}$) with velocity $\vect{v_c}=[1-\mu(t)]v_g\vect{\hat{y}}$ while the MW's center has constant position $\vect{r_g}=-\vect{\hat{x}}$. The effective potential at position $\vect{r}$ in this frame is
\begin{equation} \label{Eq:phieff}
\phieff(\vect{r},t) \equiv \phiclus(\vect{r},t) + \phigal(\vect{r}) - \frac{v_g^2}{2} \left\{\left[x+1-\mu(t)\right]^2+y^2\right \},
\end{equation}
where the last term is centrifugal. $\phieff$ sets the tidal boundary's shape and the minimum energy a body in the GC must have to escape. If $\phieff$ were static and the cluster dynamics collisionless, then the motion of each body with speed $v$ in the $xyz$ frame would conserve a quantity known as the Jacobi energy (in our case time-dependent):
\begin{equation} \label{Eq:EJacobi}
\Ejacobi(t) \equiv \frac{v(t)^2}{2} + \phieff(\vect{r},t) - A(t).
\end{equation}
Here $A(t)=\max\left[\phieff(z=0,t)\right]$, achieved at the L4/L5 Euler-Lagrange points, tips of both in-plane equilateral triangles whose base vertices are $\vect{r_c}$ and $\vect{r_g}$. The convention of subtracting $A$ \citep[e.g.,][]{Spitzer1987,FukushigeHeggie2000} has no impact on trajectories but affects normalized energy definitions like $\Etilde$ in Equation~(\ref{Eq:raw_criterion}), so is essential when later comparing to the latter study, hereafter referred to as FH00.

\begin{figure*} 
\centering
\includegraphics[width=0.85\linewidth]{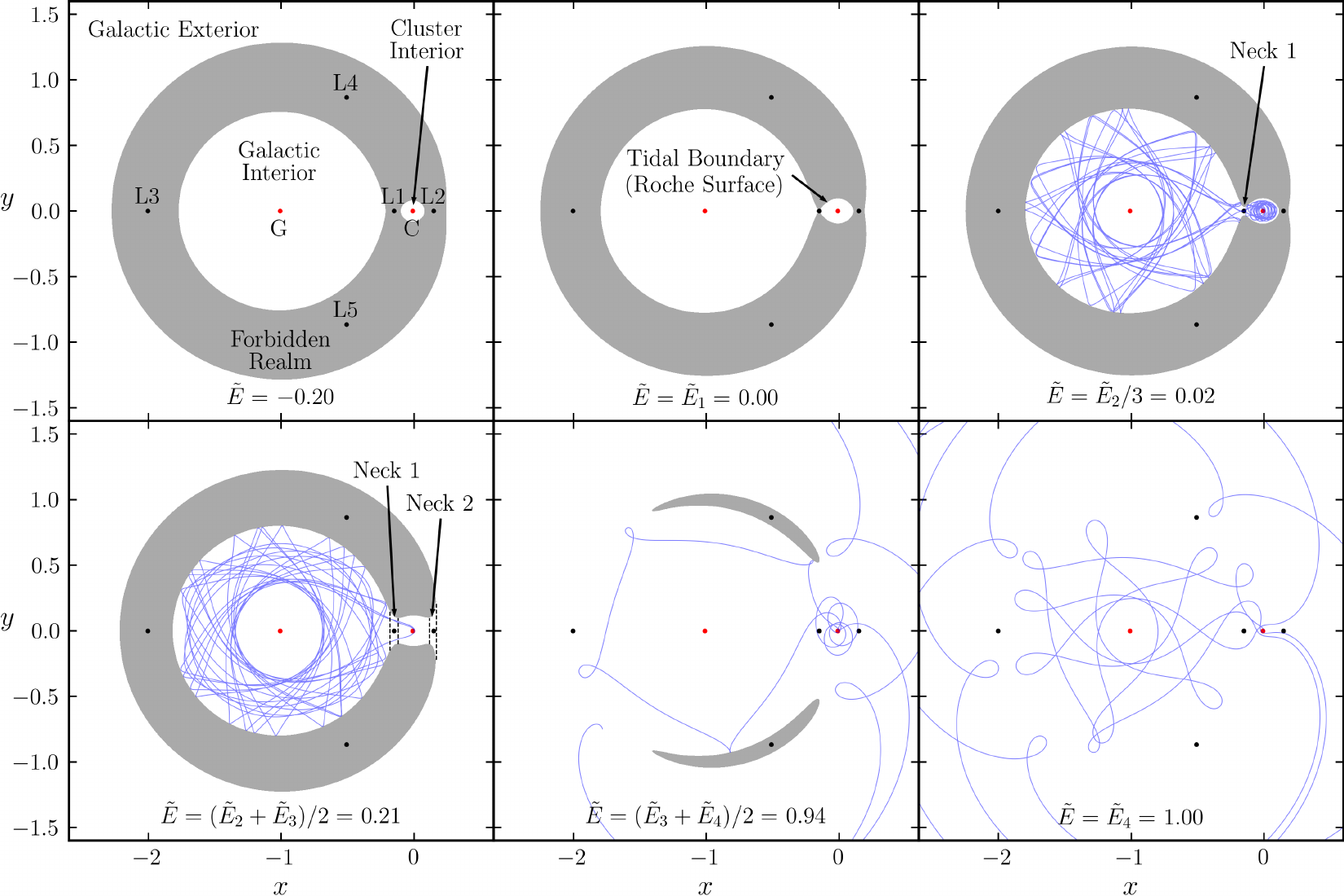}
\caption{Slices in the $xy$-plane of the rotating, clustercentric frame, illustrating features of the CR3BP for the case of Keplerian $\phigal$ and $\phiclus$, $\mu=1/101$, and in units where $G=\omega=1$. The panels distinguish different choices of excess relative energy $\Etilde$, in terms of the values $\Etilde_i$ at each Euler-Lagrange point (L$i$) when $v=0$. These points and the centers of the Galaxy (G) and cluster (C) are labeled in the first panel but appear in all. The forbidden realm (gray) is bounded by the zero-velocity surface and cannot be entered by bodies with the given $\Etilde$ or lower. The blue curves in the last four panels exemplify trajectories at the given $\Etilde$ initiated with $(x,y,z)=(\mu,0,0)$, velocity $\vect{v_0} \parallel \hat{y}$, and integrated for 20 full orbits of the cluster in the Galaxy. Upper left: $\Etilde<0$, so bodies cannot transit between the cluster interior, Galactic interior (within $\rgc$), or Galactic exterior (beyond $\rgc$). Upper center: $\Etilde=0$, so the zero-velocity surface can expand no further before opening a neck between the cluster and Galactic interiors at L1. The closed portion of the surface passing through L1 is the tidal boundary. Upper right: $\Etilde_1<\Etilde<\Etilde_2$ opens the neck at L1. The example trajectory---nearly periodic in the rotating frame---transits back-and-forth through the cluster/Galactic interiors and illustrates an escaper taking many $t_c$ to escape before eventually returning to the cluster. Lower left: $\Etilde_2<\Etilde<\Etilde_3$ opens a second neck at L2, connecting all three domains. The example trajectory illustrates an escaper that immediately finds the neck to the Galactic interior and briefly passes again through the cluster after nearly 20 full Galactocentric orbits. Lower center: $\Etilde_3<\Etilde<\Etilde_4$ opens a third neck at L3. The example trajectory transits between all three domains, with the escaper temporarily returning to the cluster after only four orbits about the Galaxy (to the cluster's three in that time). Lower right: $\Etilde=\Etilde_4=1$, at which point the entire $xy$-plane is available to escapers, though the zero-velocity surface still exists out-of-plane. The example trajectory alternates between orbiting in the Galactic interior/exterior, passing back through the cluster during each transit. Though at higher-$\Etilde$ and not nearly as regular, this trajectory is loosely analogous to those of some Jovian comets.}
\label{fig:hill-surface-diagram}
\end{figure*}

Since $v^2\geq 0$, Equation~(\ref{Eq:EJacobi}) implies that bodies with energy $\Ejacobi$ cannot enter regions where $\phieff(\vect{r}) > \Ejacobi+A$. Figure~\ref{fig:hill-surface-diagram} shows the $xy$-plane's intersection with these `forbidden realms' (gray) for several $\Ejacobi$ (in its normalized form $\Etilde$; see below). As $\Ejacobi$ increases, the \textit{zero-velocity surface} bounding these realms ($\phieff = \Ejacobi+A$) ranges from separately enclosing the GC and MW centers (upper left/middle panels), to allowing passage through one or two openings directly toward and away from the latter (upper right--lower center) to disappearing entirely from the $xy$-plane (lower right).

Slowly shrinking due to global mass loss, the GC's tidal boundary lies on the highest-$\Ejacobi$ zero-velocity surface that (at the time) fully encloses the GC (upper center). This boundary terminates nearest to the Galactic center at a saddle point of $\phieff({\vect{r}})$ known as the L1 Euler-Lagrange point and furthest just shy of another saddle, L2.\footnote{Various texts label the Euler-Lagrange points differently. We follow the aerospace convention numbered by increasing $\phieff$ \citep[e.g.,][]{Koon2011}.} Their respective locations, $\vect{r_1}(t)$ and $\vect{r_2}(t)$, are numerically solvable from the definition of a critical point, $\nabla\phieff(\vect{r},t)=0$, but for MWGCs ($\mu\sim 10^{-6}$), a 2nd-order expansion about $\mu\rightarrow0$ is very accurate---to about one part in $4\mu^{-1/3}$ \citep{Szebehely1967}. In this limit $\lvert\vect{r_2}-\vect{r_c}\rvert \rightarrow \lvert\vect{r_1}-\vect{r_c}\rvert$ and $\phieff(\vect{r_2})\rightarrow\phieff(\vect{r_1})$, so the tidal boundary is symmetric, terminating exactly at both saddles. Yet for generality we define the \textit{tidal radius} as the maximum clustercentric distance to the tidal boundary, $r_t \equiv \lvert \vect{r_1}-\vect{r_c} \rvert\rgc$. For Keplerian $\phiclus$ and $\phigal$, the expansion under $\mu\rightarrow0$ yields $r_{\rm t}/\rgc \approx (\mu/3)^{1/3}$, becoming $r_{\rm t}/\rgc \approx (\mu/2)^{1/3}$ for a logarithmic $\phigal$ closer to that of the MW halo \citep[e.g.,][]{Spitzer1987}.

Bodies within the tidal boundary can only pass beyond it if they have $\Ejacobi(t) > \Ecrit(t)$, where $\Ecrit(t) \equiv \phieff(\vect{r_1},t)-A(t)$. We shall refer to this escape criterion as the \textit{raw criterion} to distinguish it from more complex modeling alternatives discussed later. Following FH00 and to emphasize the trajectories' extreme sensitivity to $\Ejacobi$ as $\Ejacobi\rightarrow\Ecrit^{+}$, we henceforth express this criterion in normalized form as
\begin{equation} \label{Eq:raw_criterion} \Etilde(t) \equiv \frac{\Ejacobi(t)-\Ecrit(t)}{\lvert\Ecrit(t)\rvert} > 0, \end{equation}
This definition has the benefit that $\Etilde=\Etilde_4=1$ when $\phieff=A$. Here and elsewhere, the $\Etilde$ subscript $i$ indicates it is the value of $\Etilde$ at the Euler-Lagrange point L$i$ for $v=0$.

The space accessible to PEs (white space in Figure~\ref{fig:hill-surface-diagram}) varies with $\Etilde$ and reduces to three domains: the cluster, and the Galactic interior and exterior (beyond the tidal boundary at Galactocentric distances ${<}\rgc$ and ${>}\rgc$, respectively). For $\Etilde\leq 0$, transit between domains is energetically disallowed, but once $\Etilde>\Etilde_1=0$, a neck in the zero-velocity surface opens at L1 to allow transit between the cluster/Galactic interior. A similar neck at L2 allows transit between the cluster/Galactic exterior once $\Etilde>\Etilde_2$. This is the most relevant geometry to tidal tail formation from MWGCs, for which $\Etilde_1\approx\Etilde_2$. Only once $\Etilde>\Etilde_3$ (corresponding to L3, the third and final saddle point of $\phieff$) is direct transit between the Galactic interior/exterior possible. The forbidden realm disappears entirely from the $xy$-plane once $\Etilde\geq\Etilde_4=1$, though it still exists at $z \neq 0$, receding away from the $xy$-plane as $\Etilde$ grows. So escape at high $\Etilde$ is nearly unconstrained while escapers at low $\Etilde$ must pass near L1/L2.

\subsection{Tidal Tail and Stellar Stream Formation}\label{S:theory_trajectories}
The equations of motion in the CR3BP in the $xyz$ frame are
\begin{align} \label{Eq:equations_of_motion}
\ddot{\vect{r}} = -\nabla\phieff(\vect{r}) - 2\vect{\omega}\times\dot{\vect{r}} = -\nabla\phieff(\vect{r}) + 2\omega(\dot{y}\vect{\hat{x}}-\dot{x}\vect{\hat{y}}).
\end{align}
The second term in each equality, the Coriolis acceleration, drives much of the trajectories' behavior, including production of tidal tails from low-$\Etilde$ PEs. Since these must escape near L1/L2, their velocities in the necks are biased to be parallel to $\mp\vect{\hat{x}}$. The Coriolis effect then bends these trajectories into tails leading/trailing the GC, respectively, and induces epicycles in each trajectory's projection onto the $xy$-plane (Figure~\ref{fig:hill-surface-diagram}). The epicycles' characteristic size depends on $\mu$ and $\Etilde$, but for $(\mu,\Etilde)\ll 1$, the Coriolis effect keeps escapers near to the GC's orbital path, forming elongated stellar streams along it. Since each epicycle has a turning point minimizing the velocity parallel to the stream, there are periodic over-densities in the streams near these points, spaced ${\sim}10r_t$ apart for $\Etilde\ll 1$ \cite[e.g.,][]{CapuzzoDolcetta2005,Kupper2008,Kupper2010,Kupper2012,Just2009}. The Coriolis effect also stabilizes trajectories retrograde to the GC orbit, slowing escape of bodies on retrograde orbits and allowing those with $\Etilde>0$ within and even \textit{beyond} the tidal boundary to remain near the GC for an arbitrarily long time (\citealt{Henon1970}; FH00; \citealt{Read2006}; \citealt{Ernst2008}).

\subsection{Dynamical Systems Theory and Return Trajectories} \label{S:theory_escape_and_return_trajectories}

Stellar escape from GCs can be given extensive mathematical formalism from a standpoint of dynamical systems theory (e.g., FH00; \citealt{Ernst2008,TanikawaFukushige2010,deAssisTerra2014,Zotos2015a,Zotos2015b,ZotosJung2017}). For each neck, twin, infinitely winding/branching tubes in phase space known as invariant manifolds enclose all possible transit trajectories into or out of the GC. Trajectories outside the invariant manifolds cannot transit between the CR3BP's three domains, implying that some portion of even bodies with $\Etilde>0$ would never escape without the aid of perturbations from other bodies or evolution in $\phieff$ (e.g., via GC mass loss, an evolving $\phigal$, an eccentric GC orbit, or passage near MW substructure). These effects induce additional phase space diffusion, enabling bodies otherwise stuck forever on non-transiting orbits outside the invariant manifolds to move to transiting orbits within them. Reframed energetically, the internal gravitational scattering and evolution of $\phieff$ cause diffusion in $\Etilde$. Since the shapes of the manifolds change with $\Etilde$, this effectively smears them out to encompass more phase space, promoting escape.

Trajectory families known as \textit{heteroclinic orbits} asymptotically connect the saddle points to each other. Through the union of such orbits, one can design \textit{heteroclinic chains} that, e.g., periodically alternate between orbiting in the interior and exterior Galactic domains, crossing (and temporarily orbiting within) the GC during each transit. The intentional design of such itineraries in the solar system context is essential to many space missions \citep[e.g.,][]{Koon2011} but these trajectories exist naturally, too. For example, some comets, including Oterma and Gehrels 3 \citep[e.g.,][]{BelbrunoMarsden1997,Koon2000}, periodically transition from heliocentric orbits outside to inside the orbit of Jupiter and vice versa. During each transition, the comets become temporarily bound to Jupiter. The existence of such orbits (with the Sun--Jupiter system analagous to the MW--GC system) means that GCs can re-capture past members that previously escaped. In GC models, the recapture rate is typically assumed negligible due to a time-dependent and/or asymmetric $\phigal$ and diffusion of the return trajectories by MW substructure. Yet the rate may be significant in the case of a nearly circular GC orbit in an unevolving $\phigal$. Indeed, we demonstrate that robust `returning tidal tails' resulting from such trajectories can form in this case in Section~\ref{S:results_tails_and_streams}. Though idealized, this case is a reasonable starting point since the impact of returning bodies on stellar streams has not been studied before. The idealization also enables us to determine upper bounds on the impact of returners on stream observables (e.g., surface density and velocity dispersion) under more realistic conditions.

\subsection{Escape Timescale}  \label{S:theory_escape_timescale}
Weak, diffusive two-body relaxation has long been known to dominate energy transport in GCs \citep[e.g.,][]{Ambartsumian1937, Spitzer1940, Chandrasekhar1942}. Despite some early caveats in the idealized context of isolated GCs \citep[e.g.,][]{Henon1960,Henon1969}, relaxation therefore naturally dominates escape from GCs in a tidal field (e.g., \citealt{SpitzerShapiro1972}; W23). So while various strong encounters propel some bodies to $\Etilde\gg 1$, ejecting them on the crossing timescale $t_c$, most bodies first acquire sufficient energy to escape by gradually random-walking in energy to $\Etilde>0$, remaining at $\Etilde\ll 1$ for potentially many relaxation times.

Since the necks about L1/L2 are narrow for $\Etilde\ll 1$, relaxation-driven PEs may take many $t_c$ to `find' and escape through these necks, a fact highlighted in the seminal work by FH00 \citep[see also][]{TanikawaFukushige2010}. Via direct $N$-body modeling with analytical support, they showed that the typical \textit{escape timescale} $\Delta t_{\rm esc}$---between first satisfying $\Etilde>0$ and crossing beyond $r_t$---is of order an entire Hubble time for MWGCs in the idealized case of a static $\phiclus$. They also found that for $\Etilde\ll 1$ and static $\phiclus$, then $\Delta t_{\rm esc}\propto\Etilde^{-2}$. Yet $\Delta t_{\rm esc}$ has attracted little further study, likely due to its limited relevance to direct $N$-body models \citep[though see][]{Baumgardt2001}, which typically only remove bodies once they pass beyond several $r_t$ anyway. $\Delta t_{\rm esc}$ \textit{is} important for models that use energy-based escape criteria, however, like most Fokker-Planck or Monte Carlo codes, including \texttt{CMC} and \texttt{MOCCA}. For reasons we discuss in the next Section, such codes usually remove bodies immediately once they satsify $\Etilde>0$ or a similar energy-based criterion, effectively assuming $\Delta t_{\rm esc}=0$ \citep[e.g.,][]{SpitzerShull1975,LeeGoodman1995,Giersz2008,Chatterjee2010}. Especially since the \texttt{MOCCA} code now implements delayed escape based on FH00's idealized findings  \citep{Giersz2013}, $\Delta t_{\rm esc}$'s extreme length merits revisiting. In particular, $\Delta t_{\rm esc}$ in a realistically evolving $\phiclus$---a highly practical scenario that should hasten escape due to GC mass loss---has yet to be explored at all, motivating our later attention to this case.

\section{Globular Cluster Simulations} \label{S:models}
We employ the latest public version of the Cluster Monte Carlo code \citep[\texttt{CMC};][]{CMCRelease} to simulate GCs. \texttt{CMC} includes numerous physical processes essential to GC evolution, including internal stellar evolution with \texttt{COSMIC} \citep{Breivik2020}---an updated version of \texttt{SSE}/\texttt{BSE} \citep{Hurley2000, Hurley2002}---two-body relaxation \citep{Joshi2000,Pattabiraman2013}, Galactic tides \citep{Joshi2001, Chatterjee2010}, strong binary encounters and physical collisions \citep{Fregeau2003,Fregeau2007}, and three-body binary formation \citep{Morscher2013, Morscher2015}. \texttt{CMC} simulates strong binary encounters via the small $N$-body direct integrator \texttt{fewbody}, which accounts for post-Newtonian dynamics \citep{Fregeau2004, Antognini2014, AmaroSeoane2016, Rodriguez2016, Rodriguez2018a, Rodriguez2018b}. \texttt{CMC} also allows for two-body binary formation through gravitational wave dissipation and tidal capture \citep{Kremer2021, Ye2022}, but for simplicity we only allow the former in this study.

\begin{deluxetable}{lccl}[ht!]
\tabletypesize{}
\tablewidth{\linewidth}
\tablecaption{Simulations
\label{table:sims}}
\tablehead{
    \colhead{\#} &
    \colhead{$r_v/\pc$} &
    \colhead{Status at $12\,\Gyr$} &
    \colhead{Escape Criterion}}
    
\startdata
1 & 2   & NCC'd & \multirow[c]{2}{*}{Raw: Equation~(\ref{Eq:raw_criterion2})} \\
2 & 0.5 &  CC'd & \\ \hline
3 & 2   & NCC'd & \multirow[c]{2}{*}{$\alpha$: Equation~(\ref{Eq:Giersz_criterion})} \\
4 & 0.5 &  CC'd & \\
\enddata
\tablecomments{\footnotesize For all, $N_i=8\times10^5$, $\rgc=8\,\kpc$, and $Z=10^{-1}\,Z_\odot$.}
\end{deluxetable}
\vspace{-42pt}

We study escape in four cases (Table~\ref{table:sims}): an archetypal non-core-collapsed (NCC'd) and core-collapsed (CC'd) MWGC, each under two distinct escape criteria. The corresponding GCs from the \texttt{CMC Cluster Catalog} \citep{CMCCatalog} are numbered 2 and 8 in W23. Other than the escape criteria (see below) these simulations differ only in the initial virial radius $r_v=0.5\,\pc$ (for the CC'd GC) and $2\,\pc$ (for the NCC'd GC). All other initial parameters are identical, such as the initial total number of particles (singles plus binaries) $N_i=8\times10^5$, Galactocentric distance $\rgc=8\,\kpc$, and metallicity $Z/Z_\odot=0.1$. As in W23, stellar masses (primary mass, in the case of a binary) draw from the standard \cite{Kroupa2001} stellar initial mass function from $0.08$--$150\,\Msun$. We also assume that all neutron stars born in core-collapse (electron-capture) supernovae receive natal kicks drawn from a Maxwellian with dispersion $\sigma=265\,{\rm km\,s}^{-1}$ ($20\,{\rm km\,s}^{-1}$). Natal kicks for BHs share the core-collapse kick distribution of the neutron stars, but reduced in magnitude by the fraction of the stellar envelope's mass that falls back onto the BH via the prescription from \cite{Hobbs2005}---see also \cite{Fryer2012}. For any further details or justifications regarding the models \textit{other} than the two updates from W23 mentioned below, see W23 and references therein.

For completeness, we note two minor differences from the earlier versions of the models used in W23. Both changes are purely to bolster consistency with concurrent and future works using \texttt{CMC}. First, instead of a uniform initial binary fraction $f_b=5\%$, we now set $f_b=5\%$ for stars born with mass $M<15\,M_\odot$ \citep[in accord with \textit{present-day} MWGC observations, e.g.,][]{Milone2012} and $f_b=f_{b,{\rm high}}=50\%$ for those born with $M\geq 15\,M_\odot$ \citep[more in line with observations in young star clusters, e.g.,][]{Sana2009,Sana2012,MoeDiStefano2017}. Though still simplistic, this two-component initial binary fraction is more consistent with models of binary formation in star clusters embedded in molecular clouds \citep[e.g.,][]{Cournoyer-Cloutier2021,Guszejnov2023}, in which $f_b$ increases steeply with stellar mass. The update has minimal impact on the GC evaporation rate or distribution of escape mechanisms and energies (evident by comparing later figures to those in W23). We also now utilize the \cite{Fryer2012} `delayed' supernova prescription---\texttt{CMC}'s new default---to compute compact remnant masses since this treatment produces a robust remnant population in the BH lower-mass gap ($2$--$5\,\Msun$), in line with observations from the third Gravitational Wave Transient Catalog \citep{GWTC3}. This was not the case under the \cite{Fryer2012} `rapid' supernova prescription used in the \texttt{CMC Cluster Catalog} models used in W23. This change has negligible dynamical impact but expands the population of bodies labeled as BHs and so explains an increase in BH ejections seen in later figures relative to those in W23.

\subsection{Escape Criteria in \texttt{CMC}}
\label{S:escape_criteria}

While \texttt{CMC} allows for an arbitrary time-varying tidal field \citep[specified by a tidal tensor; see][]{Rodriguez2023}, we explore \texttt{CMC}'s default tidal scenario---the time-dependent CR3BP from Section~\ref{S:theory} with a logarithmic Galactic potential and $\omega\rgc = v_g = 220\,\kms$, typical of MWGCs \citep[e.g.,][]{Spitzer1987,BinneyTremaine2008}. \texttt{CMC}'s tidal radius is then
\begin{equation} \label{Eq:rtCMC}
r_t^{\rm CMC}(t) \equiv [\mu(t)/2]^{1/3} \rgc = [Gm(t)/]^{1/3}\omega^{-2/3}.
\end{equation}
Note the time-dependence due to GC mass-loss via escape and stellar evolution. The GC's spherical potential $\phiclus^{\rm CMC}(r,t)$ is similarly time-dependent, computed at each \texttt{CMC} timestep from the positions of all bodies. Unless otherwise noted, this time-dependence carries over to our integration of escape trajectories in post-processing. Its main impact is to slowly raise (make less negative) $\phiclus$ and thereby $\Etilde$, promoting escape.

Since the raw escape criterion $\Etilde>0$ involves $\phieff$, \texttt{CMC}'s spherical version of this criterion is approximate, becoming
\begin{equation} \label{Eq:raw_criterion2} \tilde{E}^{\rm CMC}(t) \equiv \frac{E^{\rm CMC} - E_{\rm crit}^{\rm CMC}}{\lvert E_{\rm crit}^{\rm CMC} \rvert} > 0 \end{equation}
where
\begin{equation} \label{Eq:E_cmc} E^{\rm CMC}(t) \equiv \frac{v(t)^2}{2} + \phiclus^{\rm CMC}(r,t),\end{equation}
and
\begin{equation} \label{Eq:Ecrit_cmc} E_{\rm crit}^{\rm CMC}(t) \equiv \frac{3}{2} \phiclus^{\rm CMC}\left[r_t^{\rm CMC}(t),t\right].\end{equation}
As explained in W23's Appendix~A.2, a term ${\approx}-(r/r_t)^2/B$ (where $B=12$ for logarithmic $\phigal$ or $9$ for Keplerian $\phigal$) is omitted from the coefficient of Equation~(\ref{Eq:Ecrit_cmc}) since most bodies first satisfy the escape criterion with $r/r_t\ll 1$.

The raw energy criterion for escape has never strictly been used in \texttt{CMC}. Instead, \texttt{CMC}'s current default is to remove bodies in a two-step process at every timestep: first any bodies with clustercentric apocenter distance $r_a>r_t^{\rm CMC}$, and then any remaining bodies satisfying a modified version of the raw criterion we call the \textit{$\alpha$ criterion} \citep{Giersz2008}:
\begin{equation} \label{Eq:Giersz_criterion}
E^{\rm CMC}(t) > \alpha \phiclus^{\rm CMC}\left[r_t^{\rm CMC}(t),t\right],
\end{equation}
\noindent where
\begin{equation} \label{Eq:Giersz_alpha}
\alpha \equiv 1.5 - 3 \left( \frac{\ln\Lambda}{N_i} \right)^{1/4}.
\end{equation}
\noindent In the Coulomb logarithm, $\ln\Lambda = \ln\left(\gamma N_i\right)$, we use $\gamma=0.01$, appropriate for GCs with realistic stellar initial mass functions \citep[e.g.,][]{Freitag2006b,Rodriguez2018c,CMCRelease}.
The first step's purpose is simply to ease comparison to GC models using apocenter-based criteria (e.g., \texttt{CMC} before 2010 or most direct $N$-body codes).\footnote{\texttt{CMC} no longer defaults to the $r_a>r_t$ criterion because this was found to significantly underpredict the evaporation rate compared to direct $N$-body codes \citep{Giersz2008,Chatterjee2010}. This occurs because $r_t$ is the \textit{maximum} clustercentric distance to the tidal boundary, so bodies can satisfy $\Etilde>0$ without satisfying $r_a>r_t$. See also W23's Appendix~A.2.} Physically, the two steps reduce to solely the $\alpha$ criterion, which all bodies with $r_a>r_t^{\rm CMC}$ must satisfy for $\alpha>1$ \citep[$N_i\gtrsim 5\times10^3$, below which H\'{e}non's method is unreliable anyway;][]{Aarseth2008}.

The $\alpha$ criterion's purpose is to make escape slightly harder to account for scattering of PEs back to $\Etilde<0$ \cite[e.g.,][]{Chandrasekhar1942,King1959,Baumgardt2001}.
This would occur if the PEs were to continue participating in collisional dynamics on their way out of the GC, since weak encounters induce a random walk in $\Etilde$. Over the escape timescale (typically long; FH00), this could either increase or decrease $\Etilde$---in the latter case, potentially delaying escape. The effect's strength scales inversely with $N_i$; specifically, \cite{Baumgardt2001} found that averaged over the half-mass time of a GC, the fraction of bodies within $r_t$ that are PEs scales roughly as ${\sim}[\ln(\Lambda)/N_i]^{1/4}$. Using this factor as a measure of the strength of back-scattering, \cite{Giersz2008} tuned the second coefficient in Equation~(\ref{Eq:Giersz_alpha}) so that H\'{e}non-type codes like \texttt{MOCCA} and \texttt{CMC} best match the evaporation rate from direct $N$-body codes. Since the only intent was to improve accuracy for \textit{internal} GC evolution, the impact on tidal tails (e.g., their velocity dispersion and corresponding width) was not considered and remains untested.

It is unclear if the $\alpha$ criterion should lead to more accurate tidal tails than the raw criterion. By roughly accounting for the impact of ongoing collisional dynamics on PEs, the former may better reproduce the escape timescale for PEs, and hence the evaporation rate, as it slightly raises the $\Etilde$ at which removal from \texttt{CMC} (promotion to `PE status') occurs. This should hasten escape of PEs, qualitatively matching the expectation if \texttt{CMC} were to allow continued relaxation for PEs rather than removing them immediately. But by raising $\Etilde$, the $\alpha$ criterion may artificially increase the velocity dispersion (width) in the tidal tails. We therefore study and compare escape under both criteria. Carefully doing so now may aid future efforts to improve escape physics in \texttt{CMC} by providing a rough sense of how much accounting for the impact of collisional dynamics on PEs changes the morphology and kinematics of tidal tails.

\section{Escape Trajectory Integration} \label{S:escaper_evolution}
We now describe how we evolve PEs removed from \texttt{CMC} with the galactic dynamics code \texttt{Gala} \citep{GalaCodeRelease}. All of these steps are currently executed in post-processing of \texttt{CMC} output using Python scripts; the functionality is not yet embedded into \texttt{CMC}'s base code, though see Section~\ref{S:discussion_monte_carlo_method} for a discussion of such potential enhancements in the future.

\subsection{Coordinate Systems and Trajectory Initialization} \label{S:coordinates}
Since \texttt{CMC} imposes spherical symmetry, the output phase space coordinates of removed bodies are $(r_{\rm rmv},v_{r,{\rm rmv}},v_{t,{\rm rmv}})$, the radial position, and radial and tangential velocities, respectively---all in the rotating clustercentric frame $xyz$ at removal time $t_{\rm rmv}$. (To tidy notation, we forgo subscript `rmv' in the rest of this subsection.) Yet trajectories in the true non-spherical $\phieff$ require full phase space coordinates, so we isotropically project the PEs' positions/velocities at removal into full 6-dimensional phase space. We first orient each PE in the $xyz$ frame with position $\vect{r}=(0,0,r)$ and velocity $\vect{v}=(v_t,0,v_r)$. To distribute $v_t$ isotropically relative to $v_r$, we rotate $\vect{v}$ about $\vect{\hat{z}}$ by angle $\psi=U(0,2\pi)$, where $U(a,b)$ indicates a random sample from the uniform distribution between $a$ and $b$. To isotropically distribute $\vect{r}$ and $\vect{v}$, we then rotate each about $\vect{\hat{y}}$ by angle $\theta = \arccos[U(-1,1)]$ and again about $\vect{\hat{z}}$ by angle $\phi = U(0,2\pi)$, yielding positions $[x,y,z] = r[\sin\theta\cos\phi,\sin\theta\sin\phi,\cos\theta]$ and velocities
\begin{align} \label{Eq:6D_projection}
\begin{bmatrix} \dot{x} \\ \dot{y} \\ \dot{z} \end{bmatrix} = \begin{bmatrix}
\left(v_r\sin\theta+v_t\cos\theta\cos\psi\right)\cos\phi - v_t\sin\psi\sin\phi \\
\left(v_r\sin\theta+v_t\cos\theta\cos\psi\right)\sin\phi + v_t\sin\psi\cos\phi \\
v_r\cos\theta - v_t\cos\psi\sin\theta \end{bmatrix}.
\end{align}
Whenever we convert to the non-rotating $XYZ$ coordinates we also need to specify a phase, so we define the cluster to be located at $(X,Y,Z)=(1-\mu,0,0)$ at time $t=0$. The coordinate transformation from $xyz$ to $XYZ$ at any time is then
\begin{align} \label{E:cluster_position}
\begin{bmatrix} X(t) \\ Y(t) \\ Z(t) \end{bmatrix} = \begin{bmatrix}
\left[x(t)+1-\mu(t)\right]\cos(\omega t) - y(t)\sin(\omega t) \\
\left[x(t)+1-\mu(t)\right]\sin(\omega t) + y(t)\cos(\omega t) \\
z(t) \end{bmatrix}.
\end{align}

\subsection{Trajectory Integration} \label{S:orbits}
We use the galactic dynamics code \texttt{Gala} \citep{GalaCodeRelease} to integrate the PE trajectories in the true $\phieff$ from the time and location of removal from \texttt{CMC} (usually deep within the GC; W23). The trajectory integration takes place in the rotating frame and uses \texttt{Gala}'s wrapper for the DOP853 \texttt{SciPy} integrator, an implementation of the Dormand-Prince method---within the Runge-Kutta family---of order 8(5,3). This implementation internally yields positional errors less than one part in $10^{10}$ per output interval, which we set to be every $\Myr$. When solving the trajectories using the truly time-dependent $\phiclus$, we do so up to our simulations' final age $14\,\Gyr$ (i.e., an integration time of $14\,\Gyr-t_{\rm rmv}$) or until the first time in the above output that the PE's Galactocentric distance exceeds $2\rgc$, whichever comes first. This limits the computational burden while enabling resolution of full stellar streams and return trajectories. When solving the trajectories using a constant $\phiclus(t_{\rm rmv})$ to compare directly to FH00, we use a full $14\,\Gyr$ integration time, regardless of $t_{\rm rmv}$.

\begin{figure*}[ht!]
\centering
\includegraphics[width=0.99\linewidth]{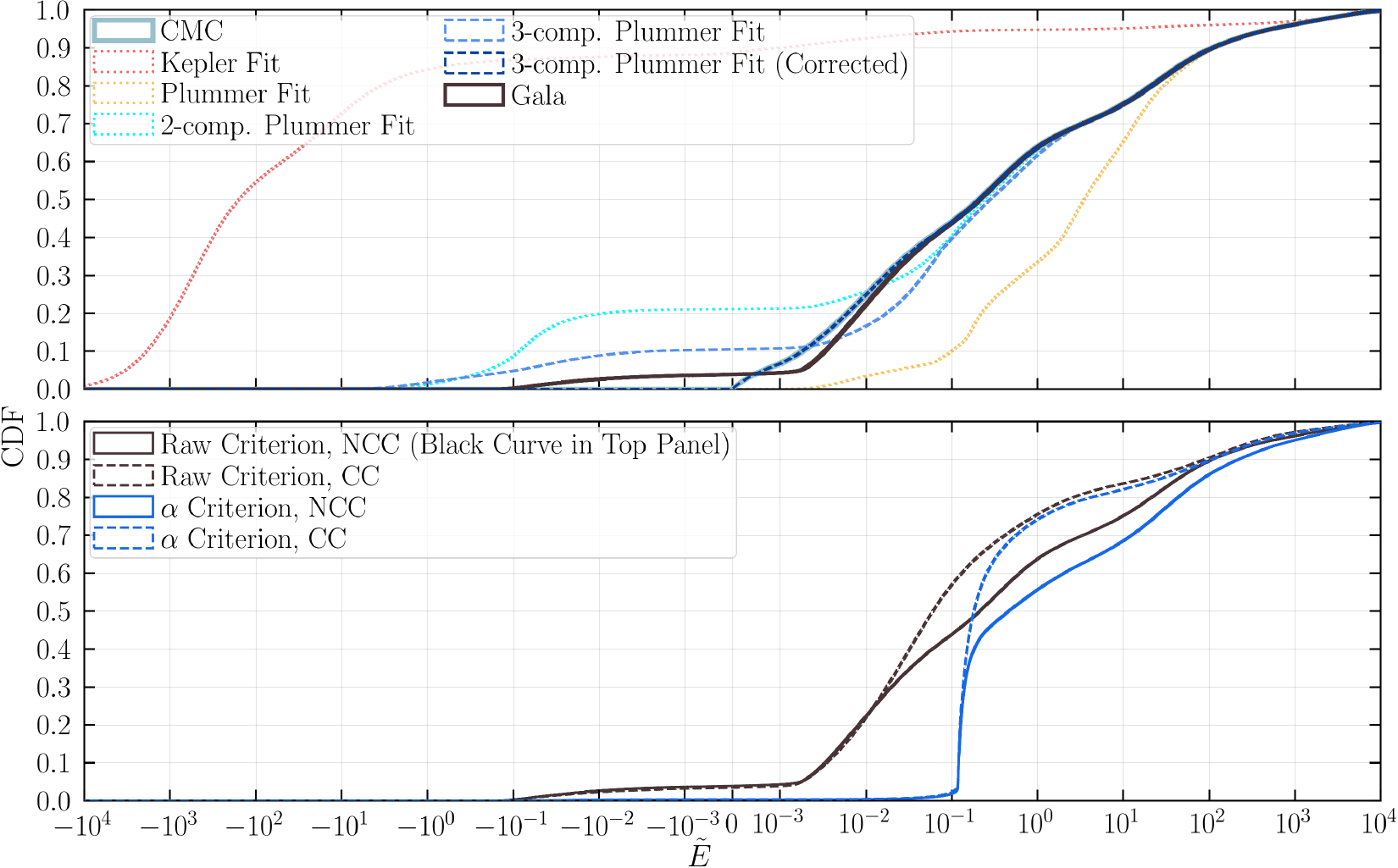}
\caption{Upper panel: The cumulative density function (CDF) for $\Etilde$ at key steps transforming \texttt{CMC}'s $\Etilde$ (solid gray; approximate due to the assumption of spherical symmetry) into \texttt{Gala}'s $\Etilde$ (solid black; the \textit{true} $\Etilde$ in the full, asymmetric tidal field). Since \texttt{Gala} uses analytically-defined potentials, deviation from a perfect analytic fit to the numerical \texttt{CMC} potential can dramatically bias the $\Etilde$ fed into \texttt{Gala}. For illustrative purposes, the dotted red, yellow, and teal curves exemplify how such biases arise, with decreasing severity, from fitting Kepler, Plummer, and 2-component Plummer potentials, respectively. Even fitting a 3-component Plummer potential (our final choice; dashed light blue) still biases $\Etilde$ to higher values. We undo this fitting-induced bias by minutely changing the velocities of each escaper via Equation~(\ref{Eq:velocity_correction}), yielding the dashed dark blue curve---a perfect realignment of the post-fit $\Etilde$ to match the original \texttt{CMC} distribution. Finally, the projection of radially-symmetric \texttt{CMC} positions and velocities into full 6-dimensional phase space (and precise numerical computation of $\Ecrit$, rather than the expansion as in the \texttt{CMC} definition) smears out the low end of the $\Etilde$ distribution when accounting for the asymmetry of the true tidal field in \texttt{Gala} (black curve). Lower panel: The final $\Etilde$ CDF as computed in the \texttt{Gala} potential---the black curve from the upper panel---but shown for four different models: the archetypal NCC'd and CC'd GCs under both energy criteria---Equations~(\ref{Eq:raw_criterion2}) and (\ref{Eq:Giersz_criterion}). The $\alpha$ criterion results in significantly higher $\Etilde$.}
\label{fig:Etilde}
\end{figure*}

\subsection{Ensuring Consistency in $\Etilde$ Between \texttt{CMC} and \texttt{Gala}} \label{S:potential_fitting}. To smoothly transition PEs removed from \texttt{CMC} into an orbit integration in \texttt{Gala}, we must ensure consistency between the sphericalized $\Etilde$ estimated by \texttt{CMC}---Equation~(\ref{Eq:raw_criterion2})---and $\Etilde$ numerically computed in \texttt{Gala} from $\phiclus$, $\phigal$, and the PE's position and velocity at $t_{\rm rmv}$. In \texttt{Gala}, we use the logarithmic Galactic potential $\phigal(R)=v_g^2\ln(R)$ with $v_g=220\,\kms$. This is identical to the $\phigal$ assumed by \texttt{CMC} when computing $r_t^{\rm CMC}$ and thereby $\Ecrit^{\rm CMC}$---Equations~(\ref{Eq:rtCMC}) and (\ref{Eq:Ecrit_cmc}), respectively. It thus ensures that $\phigal$'s contribution to $\Etilde$ via $\Ecrit$ is consistent between \texttt{CMC} and \texttt{Gala}. Consistency in the contribution to $\Etilde$ from the cluster potential $\phiclus$ is harder to ensure. While a recent \texttt{Gala} upgrade allows users to input cylindrical spline potentials built from \texttt{CMC}-like lists of masses and positions, this functionality is still being optimized; at present, the computation is orders of magnitude faster for analytic potentials.

An analytic fit to $\phiclus^{\rm CMC}$ must be chosen carefully. While $\phigal$'s functional form has a greater impact on the tidal boundary,\footnote{Because the GC's enclosed mass (and thus $\nabla \phiclus$) near the tidal boundary is very flat compared to that of the MW (and $\nabla \phigal$).} $\phiclus$ more strongly affects $\Etilde$ since PEs typically first satisfy the escape criterion in the GC's core (W23). So even slight inconsistencies in $\phiclus$ between \texttt{CMC} and \texttt{Gala} greatly affect $\Etilde$ (Figure~\ref{fig:Etilde}), and thereby the escape timescale.
For example, a Keplerian fit $\phiclus^{\rm fit}$ parameterized by the GC mass $m(t)$ is too steep in the core relative to the true $\phiclus^{\rm CMC}$. This causes \texttt{Gala} to underestimate $\Etilde(t_{\rm rmv})$, enough to prevent all but the most energetic PEs from crossing beyond $r_t(t)$ within a Hubble time. Meanwhile, a \cite{Plummer1911} $\phiclus^{\rm fit}$ set by $m(t)$ and the GC half-mass radius causes the opposite issue; this choice is too shallow in the core and escape occurs far too rapidly, on the crossing timescale. Finally, though our simulations sample initial positions/velocities from a \cite{King1966} profile, which also fits well $\phiclus$ after core collapse, we find it is not an ideal match prior to collapse (most ages; explained shortly). In all cases, these discrepancies are far more visually apparent in the logarithmic enclosed mass profile as it magnifies changes in $\nabla \phiclus$ at small $r$.

We find a three-component Plummer potential---Equation~(\ref{Eq:PlummerPotential})---provides a generally excellent fit to $\phiclus^{\rm CMC}$ (and the GC's enclosed mass profile) across all $t$ and $r$. We determined this via trial and error upon observing that once BHs segregate to the core, but  before their eventual loss precipitates core collapse, the enclosed mass profile in \texttt{CMC} deviates strongly from a typical Plummer or King profile. Specifically, the mass density profile becomes bimodal due to the BHs forming a denser sub-cluster in the deep core, naturally leading to a two-component profile (one for BHs, one for stars). That a three-component fit also performs slightly better in Figure~\ref{fig:Etilde} is due more to the additional degrees of freedom. We use \texttt{SciPy}'s curve-fitting functionality to fit such a potential to each simulation snapshot, linearly interpolating the fitted parameters both spatially and in time to yield $\phiclus^{\rm fit}(r,t)$ for the \texttt{Gala} integration:
\begin{equation} \label{Eq:PlummerPotential}
\phiclus^{\rm fit}(r,t) = -\frac{Gm(t)}{\sum_{i=1}^{3}m_i(t)} \sum_{i=1}^{3}\left[\frac{m_i(t)}{\sqrt{r^2+b_i(t)^2}}\right].
\end{equation}
The six fitted parameters $m_i(t)$ and $b_i(t)$ are the characteristic masses and Plummer scale lengths, respectively, for each piece of $\phiclus^{\rm fit}$, interpolated to time $t$. This definition guarantees the GC's enclosed mass tends to $m(t)=\sum_{i=1}^{3}m_i(t)$ as $r \rightarrow \infty$.

While a vast improvement over the above alternatives, the interpolated three-component Plummer fit still allows small inconsistencies between $\phiclus^{\rm CMC}$ and $\phiclus^{\rm fit}$, enough to affect the $\Etilde$ distribution in Figure~\ref{fig:Etilde}. So we add a final step to correct for this. The equivalent of Equations~(\ref{Eq:raw_criterion2})--(\ref{Eq:Ecrit_cmc}) after the fit/interpolation are:
\begin{equation} \label{Eq:Etilde_fit} \tilde{E}^{\rm fit} \equiv \frac{E^{\rm fit} - E_{\rm crit}^{\rm fit}}{\lvert E_{\rm crit}^{\rm fit} \rvert}, \end{equation}
\begin{equation} \label{Eq:E_fit} E^{\rm fit} \equiv \frac{v_{\rm rmv}^2}{2} + \phiclus^{\rm fit}\left(r_{\rm rmv}\right), \end{equation}
and
\begin{equation} \label{Eq:Ecrit_fit} E_{\rm crit}^{\rm fit} \equiv \frac{3}{2} \phiclus^{\rm fit}\left[r_t^{\rm CMC}(t_{\rm rmv}),t_{\rm rmv}\right]. \end{equation}

\noindent To correct the inconsistency in $\Etilde$ from fitting/interpolating $\phiclus$, we then slightly adjust $v$ so that $\tilde{E}^{\rm fit} = \tilde{E}_{\rm rmv}^{\rm CMC} \equiv \tilde{E}^{\rm CMC}(t_{\rm rmv})$. The corrected speed that achieves this is
\begin{equation} \label{Eq:velocity_correction} v_{\rm corr} \equiv \sqrt{v_{\rm rmv}^2 + 2\left[\left(1-\tilde{E}_{\rm rmv}^{\rm CMC}\right)E_{\rm crit}^{\rm fit} - E^{\rm fit}\right]}. \end{equation}
The resulting $\Etilde$ is $\Etilde^{\rm corr}\equiv E^{\rm corr}-\Ecrit^{\rm fit}/| \Ecrit^{\rm fit}|$, where $E^{\rm corr}\equiv E^{\rm fit} + (v_{\rm corr}^2 - v_{\rm rmv}^2)/2$. In the few cases (${\lesssim}1$ in $10^4$ PEs) this results in imaginary $v_{\rm corr}$, we instead use $v_{\rm corr}=v_{\rm rmv}$. The (10th, 50th, 90th) percentiles of $v_{\rm corr}/v_{\rm rmv}$ are ${\approx}(0.99,1.00,1.01)$. While small, this correction does impact $\Etilde$ in Figure~\ref{fig:Etilde}; in the top panel, the uncorrected $\Etilde^{\rm fit}$ (dashed light blue) deviates significantly from $\Etilde^{\rm CMC}$ (solid gray). The correction works as intended, bringing $\Etilde^{\rm corr}$ (dashed dark blue) in line with $\Etilde^{\rm CMC}$.
For reference, we also show $\Etilde^{\rm fit}$ when instead setting $\phiclus^{\rm fit}$ to be Keplerian, Plummer, or two-component Plummer (dotted red, yellow, and cyan lines, respectively).

\begin{figure*}[ht!] 
\centering
\includegraphics[width=\linewidth]{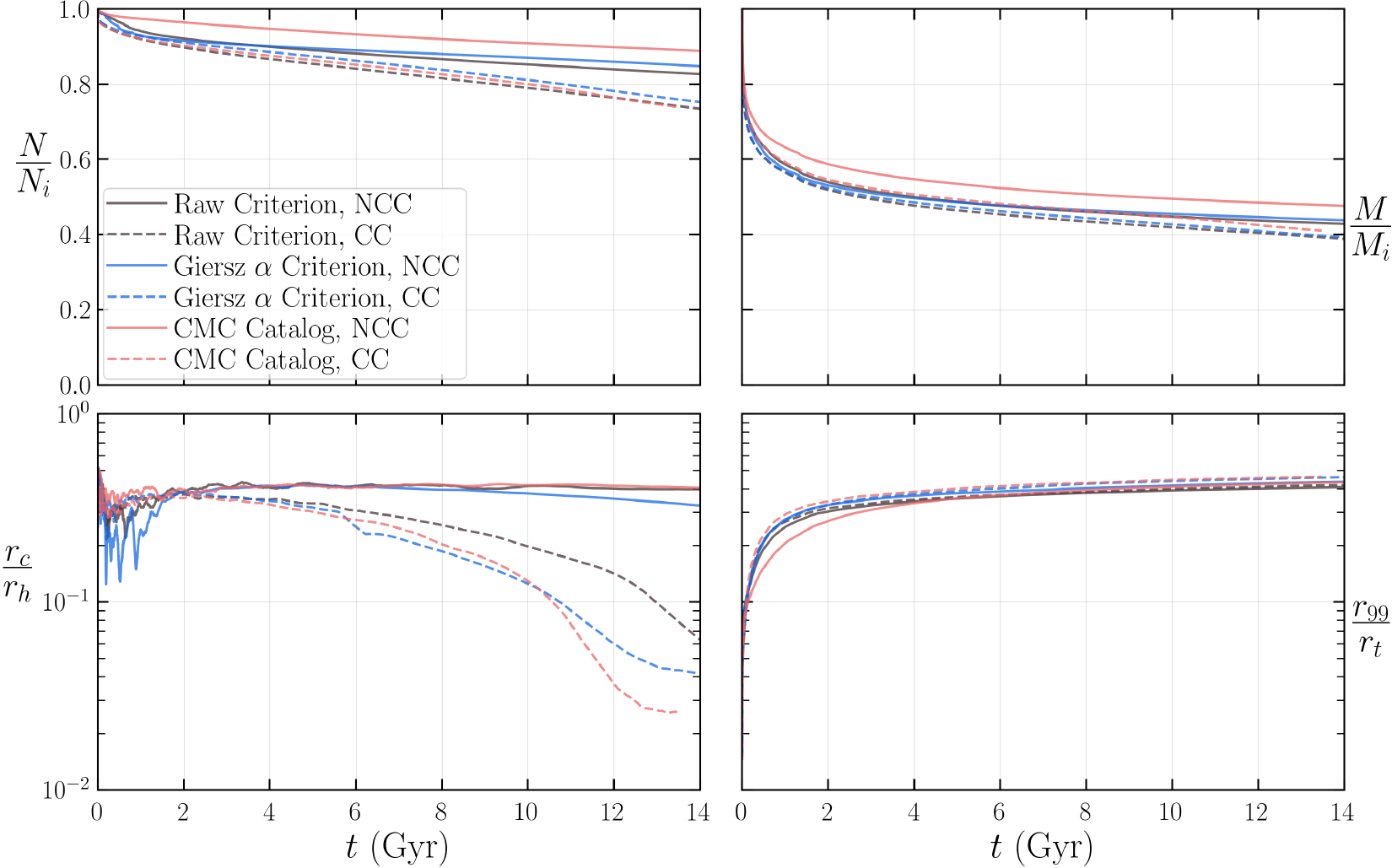}
\caption{Evolution of the GC models over time. Top left: number of particles relative to their initial number. Top right: cluster mass relative to the initial mass. Bottom left: core radius over half-mass radius (rolling average). Bottom right: 99\% Lagrange radius over tidal radius (rolling average). In all panels, solid/dashed curves correspond to the NCC'd/CC'd GCs, black/blue to the raw/$\alpha$ escape criterion, and red to the near-identical simulations in W23 (which used the $\alpha$ criterion).}
\label{fig:macro_evolution}
\end{figure*}

Technically the above correction only accounts for inconsistency in $\Etilde$ between \texttt{CMC} and \texttt{Gala} directly attributable to inaccuracy in fitting/interpolating $\phiclus^{\rm CMC}$. There are two other minor discrepancies. First, the raw criterion in \texttt{CMC} is only a spherical approximation, whereas the true $\Etilde$---defined in \texttt{Gala} as in Equation~(\ref{Eq:raw_criterion})---depends on the PE's full positional coordinates. So projecting the PEs into full six-dimensional phase space for initialization in \texttt{Gala} changes $\Etilde$ from what \texttt{CMC} measured. The inconsistency has a small impact, causing ${\sim}5\%$ of \texttt{CMC}'s `PEs' to actually have $\Etilde<0$ when measured under the true raw criterion in \texttt{Gala} (solid black). These mostly have low $|z(t_{\rm rmv})|$ and high $|x(t_{\rm rmv})|$---see, e.g., W23's Equation~(A4). To avoid bias, we exclude them from our analysis of the escape timescale in Section~\ref{S:results}. Secondly, $r_t^{\rm CMC}$ and $E_{\rm crit}^{\rm CMC}$ are both first-order approximations in $\mu$ (albeit excellent ones) to the actual location and $\phieff$ of L1, so both criteria in \texttt{CMC}---Equations~(\ref{Eq:raw_criterion2}) and (\ref{Eq:Giersz_criterion})---are only approximate whereas Equation~(\ref{Eq:raw_criterion}) in \texttt{Gala} is exact to within machine precision. Other uncertainties in GC physics dwarf this tiny inconsistency, so, like all cluster-modeling codes we are aware of, we do not account for it in \texttt{CMC}.

\begin{figure*}
\centering
\includegraphics[width=\linewidth]{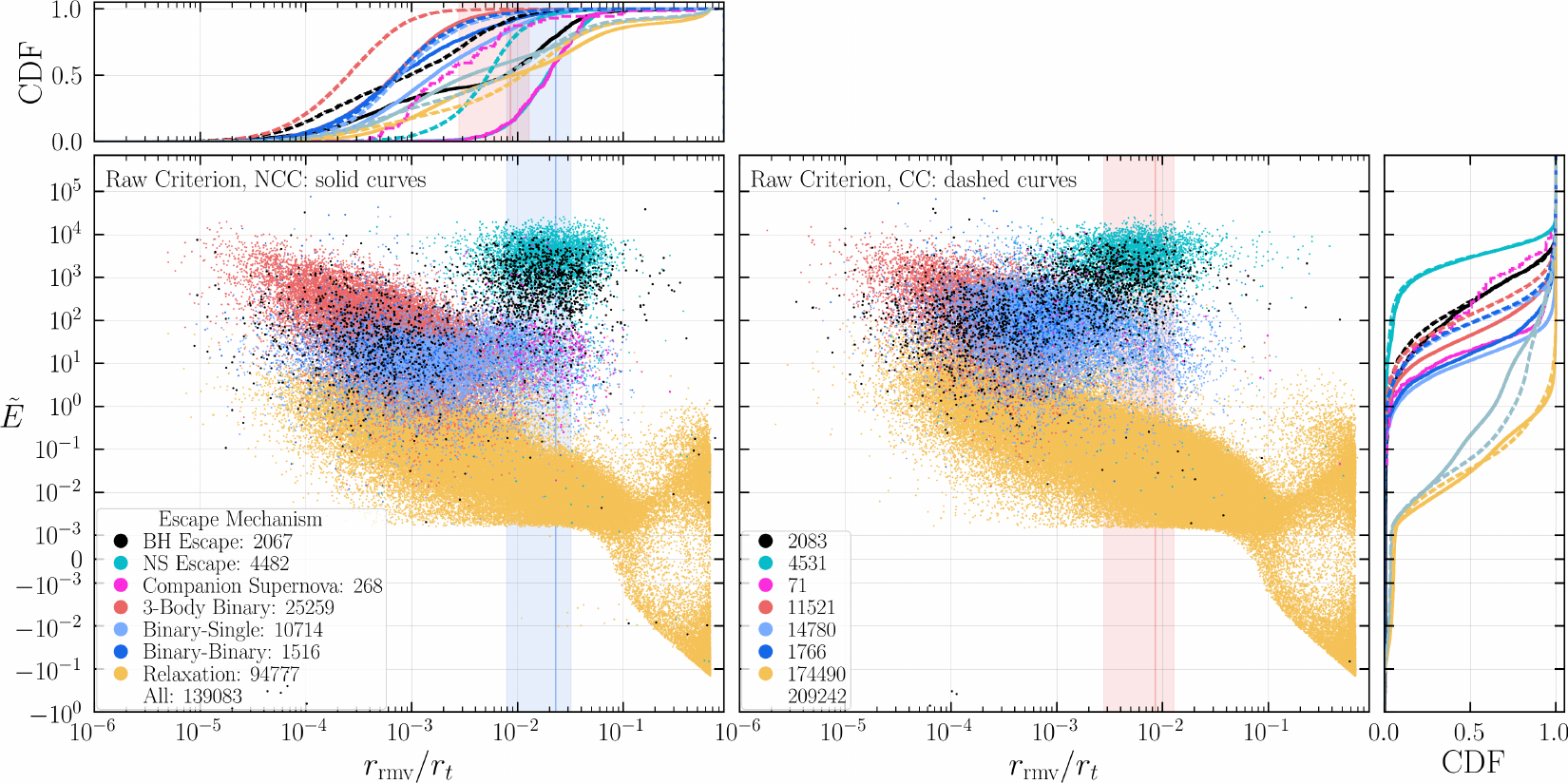}
\caption{Escapers from the archetypal NCC'd (lower left panel) and CC'd (central panel) GC models, distributed according to position $r_{\rm rmv}/r_t$ and excess relative energy $\Etilde$ upon removal from \texttt{CMC} ($t=t_{\rm rmv}$). The corner plots show the corresponding CDFs for $r_{\rm rmv}/r_t$ and $\Etilde$, with solid (dashed) curves corresponding to the NCC'd (CC'd) model. The gray curves in the CDFs include all escapers while other colors distinguish different escape mechanisms. Regardless of mechanism, escapers (single \textit{or} binary) containing a BH (black) or neutron star (but no BH; teal) are shown separately. All other escapers are categorized by mechanism: those caused by the induced kick from a binary companion's supernova (magenta), three-body binary formation (from three singles; red), binary--single (light blue) and binary--binary (dark blue) strong encounters, and two-body relaxation (yellow). The legends display the total number of escapers and subtotals for each category. The vertical lines and surrounding shaded intervals indicate the median and 10th--90th percentile range of the theoretical density-weighted core radius $r_c(t_{\rm rmv})$ from \cite{CasertanoHut1985}, normalized by $r_t(t_{\rm rmv})$.}
\label{fig:rejrtej_Etilde}
\end{figure*}

\begin{figure*}
\centering
\includegraphics[width=\linewidth]{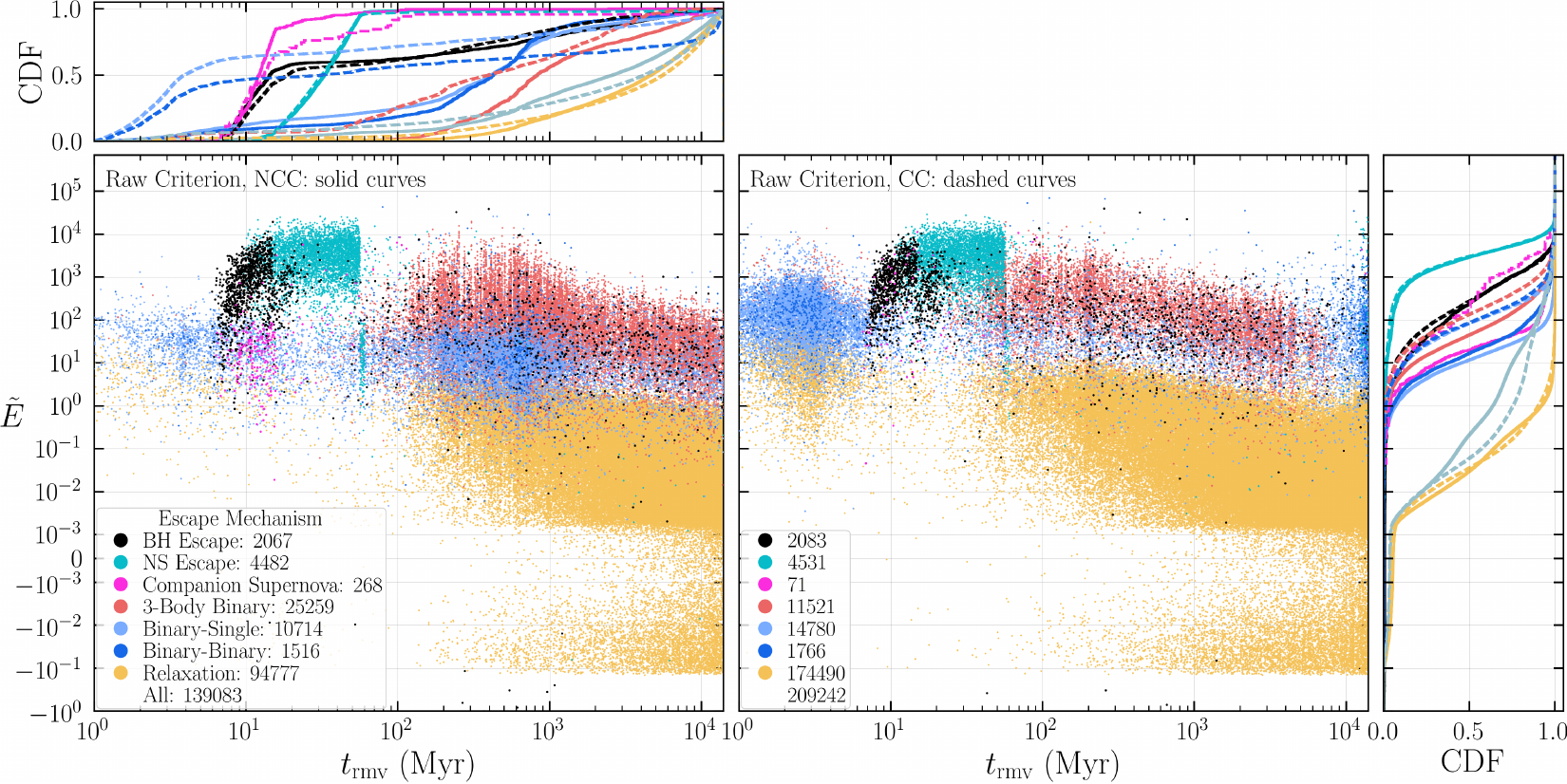}
\caption{Same as Figure~\ref{fig:rejrtej_Etilde} but the horizontal axis is now the time of removal from \texttt{CMC}, $t_{\rm rmv}$.}
\label{fig:tej_Etilde}
\end{figure*}

\section{Results} \label{S:results}
Figure~\ref{fig:macro_evolution} shows the evolution of our archetypal NCC'd and CC'd GCs (solid and dashed curves, respectively) under both the raw (black) and $\alpha$ (blue) escape criteria. For reference, we also include the nearly-identical models from W23 (red), which used the $\alpha$ criterion. The top panels show the retention fractions of the initial number of particles $N(t)/N_i$ (left) and cluster mass $M(t)/M_i$ (right). The CC'd GCs ($r_v=0.5\,\pc$) evaporate faster than the NCC'd GCs ($r_v=2\,\pc$) due to their higher density and correspondingly shorter dynamical and relaxation timescales. As intended, the $\alpha$ criterion lowers the evaporation rate by raising the escape threshold---to $\Etilde\gtrsim 0.1$ (Figure~\ref{fig:Etilde}). Yet the impact on the evaporation rate is small---less than the typical stochastic variation between separate statistical realizations of \texttt{CMC} models. This is unsurprising given that \cite{Giersz2008} saw relatively modest changes for $N_i=10^4$, 80 times smaller than our $N_i$. Since $\alpha$ scales inversely with $N_i$ in Equation~(\ref{Eq:Giersz_criterion}) the two criteria are much closer in our case. Together, Figures~\ref{fig:Etilde} and \ref{fig:macro_evolution} suggest that changing from the raw to $\alpha$ criterion primarily affects $\Etilde$, and thereby the escape timescale and escaper velocities, rather than the evaporation rate. So tuning the escape criterion to most accurately capture the latter (due to its greater relevance to \textit{internal} GC evolution) may not be ideal for Galactic archeology applications, where the former are also important.

The lower left panel of Figure~\ref{fig:macro_evolution} shows a rolling average of the theoretical core radius $r_c$, expressed as a ratio to the half-mass radius $r_h$. The steep drop in $r_c/r_h$ between $8$ and $13\,\Gyr$ for the CC'd GCs demonstrate how their cores indeed collapse within a Hubble time, accompanying the transition from a centrally flat to a centrally steep surface brightness (an observationally CC'd state) upon loss of most BHs \cite[e.g.,][]{CMCCatalog,Kremer2021,Rui2021b}. The lower right panel shows the evolution of $r_{99}/r_t$, the normalized radius enclosing 99\% of the GC's mass. Each GC significantly underfills its tidal boundary at birth but tends toward a tidally-filling state where $r_{99}/r_t\lesssim 0.626$---the minimum clustercentric distance to the tidal boundary for a logarithmic Galactic potential \citep[in the $z$-direction;][]{Claydon2017}.

\subsection{Escaper Energy Distribution} \label{S:results_energy_distribution}
Figure~\ref{fig:rejrtej_Etilde} shows the distribution of $\Etilde$ versus the clustercentric position $r_{\rm rmv}/r_t$ when each PE first satisfies $\Etilde>0$. Corner plots show the cumulative density functions (CDFs) in $\Etilde$ and $r_{\rm rmv}/r_t$ while the lower left and central panels show the corresponding scatter plots for the NCC'd and CC'd GCs (solid and dashed curves in the CDFs, respectively). Each use the raw criterion since the equivalent plot under the $\alpha$ criterion simply shifts PEs with $\Etilde\lesssim 0.1$ to $\Etilde\gtrsim 0.1$. Colors distinguish escape mechanisms, indicated in the legends and caption. As a reproduction of Figure~2 in W23, with only slightly updated models and the vertical axis now $\Etilde$ instead of velocity, we only highlight the Figure's key features. For more detailed discussion, including the algorithmic definitions of escape mechanisms, see W23.

First, two-body relaxation (yellow) dominates overall, producing escapers with $\Etilde\ll 1$. About half of escapers from relaxation originate within the typical core radius at removal $r_c(t_{\rm rmv})$, indicated by the vertical line and shaded interval in each scatter plot. This reflects that even bodies with $\Etilde$ just below $0$ in the GC halo typically first cross to $\Etilde>0$ only after first plunging back through the core, where the higher density greatly enhances relaxation's efficiency \citep[e.g.,][]{SpitzerShapiro1972}. Strong fewbody encounters dominate escape at high $\Etilde$ and from the deep core (though relaxation still dominates in the core overall). Strong encounters are especially prolific in the CC'd GC (dashed curves), which also features several times more escapers from strong binary--single (light blue) and binary--binary (dark blue) interactions, as well as two-body relaxation. These reflect the increased density and correspondingly faster dynamics. Meanwhile, the faster loss of BHs in the CC'd clusters quenches three-body binary formation (from three singles; red) due its steep mass dependence (see W23). This mechanism dominates high-$\Etilde$ escape prior to observable core collapse, corresponding to the present in most MWGCs \citep{Trager1995}.

The expression of energy as a fractional difference from $\Ecrit$ is an important qualitative difference from the similar figure in W23, since it emphasizes behavior at $\Etilde\ll 1$. This reveals that two-body relaxation in \texttt{CMC} applies stronger kicks at higher density (smaller $r$). This arises because the average squared velocity kick applied to each body per (spatially uniform) timestep in \texttt{CMC}'s relaxation algorithm is proportional to the local density---see Equation (9) of \cite{CMCRelease}. This discretization introduces some uncertainty to $\Etilde$ from relaxation since it is truly a continuous diffusive process. $\Etilde$ from three-body binary formation also decreases with increasing $r$ because \texttt{CMC} limits the semi-major axis of the newly formed binary to a minimum $a_{\rm min} \equiv G m_1 m_2 / (50 \langle m \rangle \sigma^2)$, where $m_1$ and $m_2$ are the binary's component masses, $\langle m \rangle$ is the local average mass, and $\sigma$ is the local velocity dispersion \citep[see Section 2.3.1 of][]{CMCRelease}. So the maximum potential energy released in binding the binary is $25\langle m \rangle \sigma^2$, and both quantities scale inversely with $r$.

Figure~\ref{fig:tej_Etilde} complements Figure~\ref{fig:rejrtej_Etilde} by showing the distribution of $\Etilde$ versus removal time $t_{\rm rmv}$. The first ejections occur primarily from strong binary-mediated scattering (blues) in both the NCC'd and CC'd GCs, but especially in the latter due to its higher density. Bursts of BH and neutron star ejections (black and teal, respectively) follow from $10$--$100\,\Myr$, mostly due to supernovae. The smaller secondary burst of BH ejections arises from the BHs ejected by the supernova of a neutron star companion. A similar secondary burst of neutron star ejections at ${\approx}60\,\Myr$ is a result of electron-capture supernovae, which occur a bit later and at lower ejection speeds than the dominant core-collapse supernovae. By $t\sim100\,\Myr$, escape occurs through a mix of two-body relaxation, three-body binary formation, and strong binary encounters, with many bursts of escapers at common times (vertical streaks) visible due to gravothermal oscillations, a mathematically chaotic phenomenon in which the core sharply contracts and re-expands at frequent irregular intervals throughout the GC's life \citep[e.g.,][]{HeggieHut2003}. These density spikes especially promote three-body binary formation since the rate for this process scales with the density cubed. A burst of strong binary-mediated ejections from $0.2\lesssim t/\Gyr \lesssim 1$ occurs in the NCC'd GC due its unusually long and deep early core contraction (Figure~\ref{fig:macro_evolution}) reversed by BH binary burning. The loss of almost all BHs in the CC'd GC curtails three-body binary formation around $t\sim 10\,\Gyr$ and induces core collapse, promoting strong binary-mediated ejections instead. 

\begin{figure*}
\centering
\includegraphics[width=0.95\linewidth]{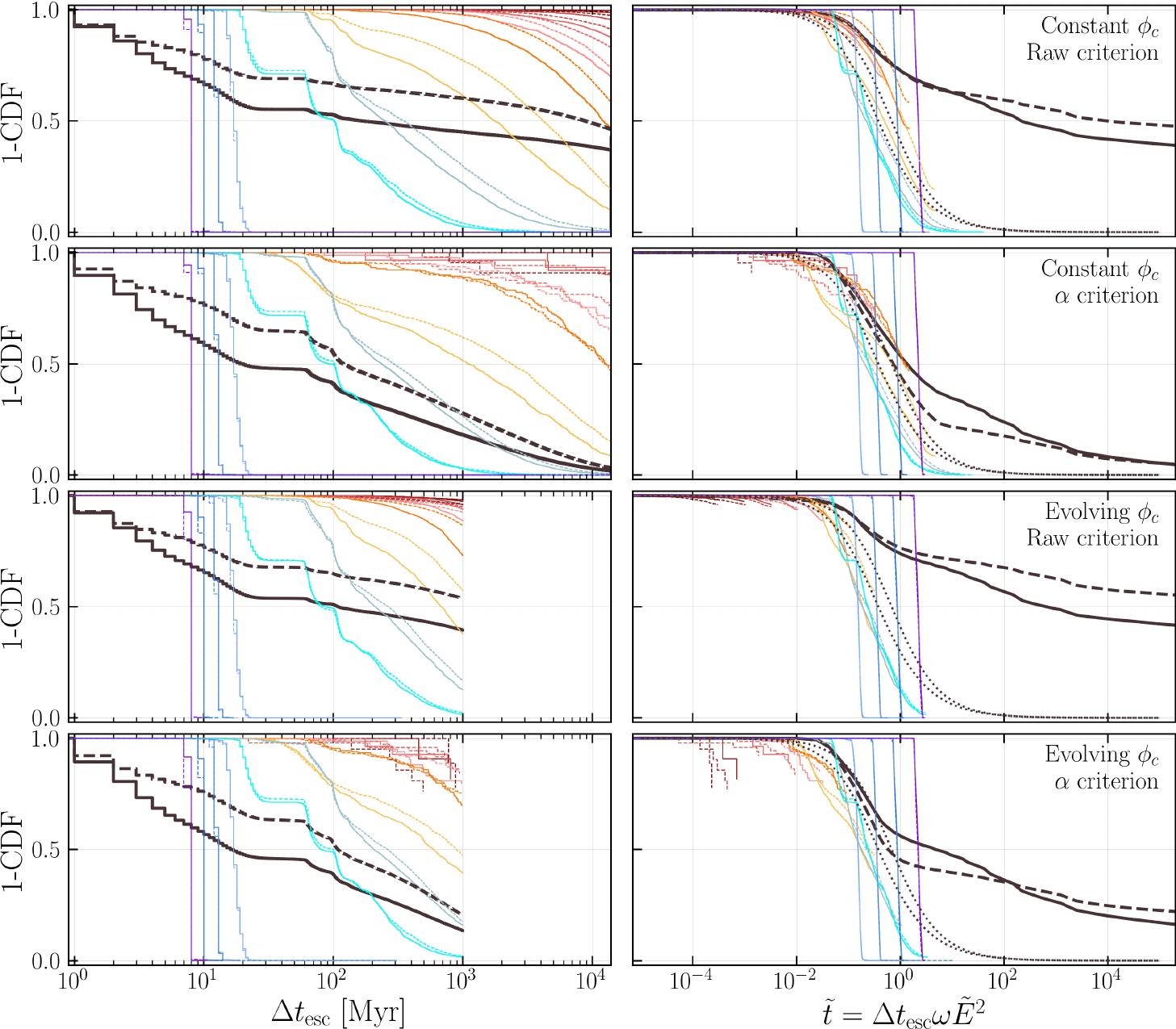}
\caption{Survival functions (1-CDF) for the escape timescale $\Delta t_{\rm esc}$ (left panels) and its normalized equivalent $\tilde{t}=\Delta t_{\rm esc}\omega\Etilde^2$ (right panels). In the top two rows, we integrate each PE's trajectory for a full $14\,\Gyr$ beyond the time $t_{\rm rmv}$ that it first satisfies the raw (top) or $\alpha$ (2nd from top) escape criteria, assuming constant $\phiclus=\phiclus(t_{\rm rmv})$. In the bottom two rows (again, one for each criterion), we instead integrate all PEs with $t_{\rm rmv} < 13\,\Gyr$ for only $1\,\Gyr$ beyond $t_{\rm rmv}$ in \texttt{CMC}'s truly \textit{time-dependent} $\phiclus(t)$---see explanation in text. Again, solid curves denote the NCC'd GC and dashed the CC'd GC. The thick black curve shows the survival function across \textit{all} escapers while the remaining curves show the contributions from each of 13 thin bins in $\Etilde$. These are colored in rainbow order (right to left in the left panels) with bin centers $\Etilde_{\rm cen}$ uniformly spaced in log-scale $1/4$ dex apart $\Etilde_{\rm cen}=10^{-2.5}$ (dark red) to $\Etilde_{\rm cen}=10^{0.5}$ (violet). Each bin's lower/upper bound is defined narrowly as $(\Etilde_{\rm low},\Etilde_{\rm upp})=(0.9,1.1)\times\Etilde_{\rm cen}$. Truncation of curves correspond to exclusion of PEs that do not cross beyond $r_t^{\rm CMC}$ within the integration time. Finally, we show for comparison the fitted $\tilde{t}$ distributions (dotted black) from Figure~9/Table~1 of FH00 (see text).}
\label{fig:tesc}
\end{figure*}

The typical $\Etilde$ decreases over time---most notably for three-body binary formation since the average mass of bodies in the core drops as the GC ejects its BHs---but peaks again after core collapse in the CC'd GC due to the increased core density. Note in the right corner plot (identical between Figures~\ref{fig:rejrtej_Etilde}--\ref{fig:tej_Etilde}) that $\Etilde$ is cumulatively higher across all times and escape mechanisms in the NCC'd GC (solid gray) but the CC'd GC (dashed gray) has a comparable typical $\Etilde$ when restricted to ages near a Hubble time. This is attributable to its high post-collapse core density and the correspondingly stronger relaxation kicks and late burst of strong encounters.

\subsection{Escape Timescale} \label{S:results_escape_timescale}
We now examine the distribution of escape times, $\Delta t_{\rm esc}\equiv t_{\rm esc}-t_{\rm rmv}$, between removal from \texttt{CMC} (becoming a PE) and first passage beyond $r_t$, at which point we may say the body has `escaped.' While such `escapers' can and often \textit{do} circulate back within the GC's tidal boundary at least once before the GC's eventual dissolution, we examine the ramifications of these return trajectories later, focusing here on the timescale to cross beyond $r_t$ for the \textit{first} time (e.g., FH00; \citealt{Ernst2008, TanikawaFukushige2010, deAssisTerra2014, Zotos2015b, Zotos2016, ZotosJung2017}).

Figure~\ref{fig:tesc} shows survival functions for $\Delta t_{\rm esc}$ (left panels) and its normalized form from FH00, $\tilde{t} = \Delta t_{\rm esc}\omega\Etilde^2$ (right panels). In the top two rows, we separately evolve each PE's trajectory for $14\,\Gyr$ past the time $t_{\rm rmv}$ that it first satisfies the raw (top) or $\alpha$ (2nd from top) escape criteria, assuming constant $\phiclus=\phiclus(t_{\rm rmv})$. This guarantees conservation of $\Etilde$ while making escape take longer (by neglecting GC mass loss) and allows direct comparison to the results of FH00, who also assumed a constant $\phiclus$. In the bottom two rows (again, one for each criterion), we instead evolve each trajectory with $t_{\rm rmv}<13\,\Gyr$ for only $1\,\Gyr$ each in the true \textit{evolving} \texttt{CMC} $\phiclus(t)$. In this case, we do not know $\phiclus(t)$ beyond the simulation end time ($14\,\Gyr$), so we impose the $1\,\Gyr$ cutoff to avoid biasing the $\Delta t_{\rm esc}$ distribution to smaller values (since many PEs removed at large $t_{\rm rmv}$ will not have had time to escape). Again, solid curves denote the NCC'd GC and dashed the CC'd GC. The thick black curves represent the total survival across all $\Etilde$ while the thinner rainbow curves each correspond to the partial contribution from each of several thin bands in $\Etilde$ spaced $1/4$ dex apart from $\log_{10}(\Etilde)\in[-2.5,0.5]$ (dark red to violet, respectively; see caption for more detail). Curves are truncated because of the many escapers that do not cross beyond $r_t^{\rm CMC}$ within the integration time.

\begin{figure*} 
\centering
\includegraphics[width=\linewidth]{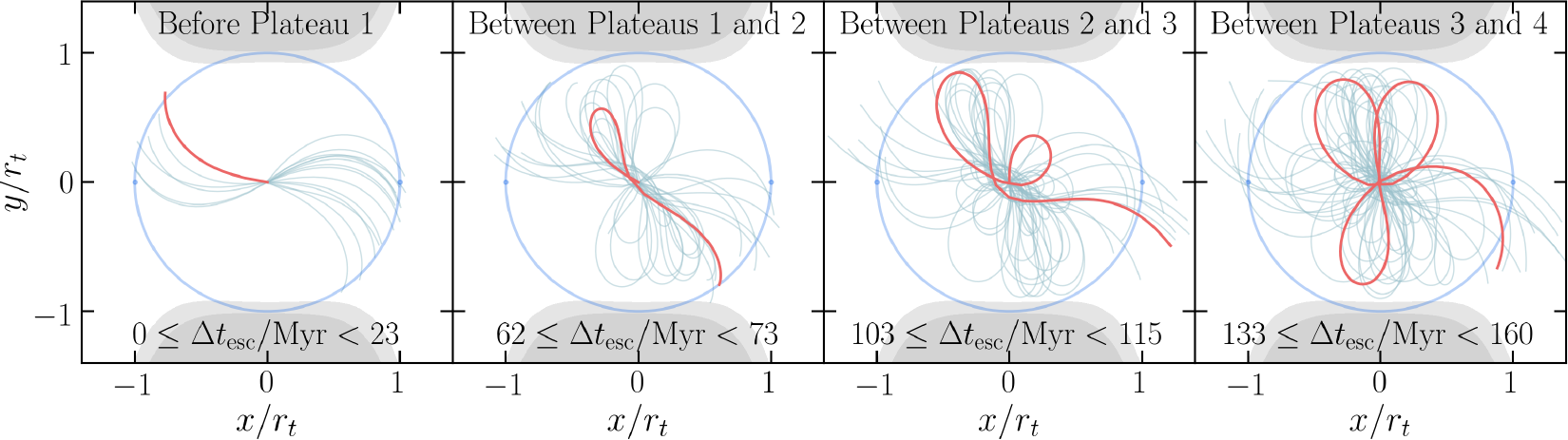}
\caption{Escaper trajectories selected from the solid teal curve in the upper left panel of Figure~\ref{fig:tesc}---$\Etilde\in[0.9,1.1]\times10^{-1/2}$---assuming a constant $\phiclus=\phiclus(t_{\rm rmv})$ from the removal time $t_{\rm rmv}$ of each escaper. Each is projected into the $xy$-plane and integrated for time $1.1\Delta t_{\rm esc}$, just beyond when they cross $r_t$ (blue circle; note the projection only makes it \textit{appear} some of these do not cross $r_t$). Each panel shows 20 such escapers, one highlighted in red for clarity, belonging to a different $\Delta t_{\rm esc}$ interval (see labels), selected to be immediately before/between the first four plateaus visible in the $\Delta t_{\rm esc}$ distribution at this $\Etilde$. Trajectories immediately before the $\mathcal{N}$th plateau typically loop back $\mathcal{N}-1$ times before escape. We show in blue the L1 (left) and L2 (right) Euler-Lagrange points and in gray the two forbidden realms (of the 80 across all shown trajectories) enclosing the GC the least (dark gray) and most (light gray). This spread is due primarily to the finite $\Etilde$ bin width when selecting the displayed escapers and secondarily to the variation in $\mu(t_{\rm rmv})$ across these, since they are removed from \texttt{CMC} at different ages.}
\label{fig:plateau_trajectories}
\end{figure*}

Numerous interesting results are apparent in Figure~\ref{fig:tesc}, so we start by comparing the overall $\Delta t_{\rm esc}$ distributions (black curves; left panels). As pointed out by FH00, the dominance of two-body relaxation (low $\Etilde$) causes a large fraction ($\gtrsim 40\%$) of PEs under the raw criterion to escape on the Hubble timescale $t_{\rm H}$, at least when integrated in constant $\phiclus$ (top row). On its own, this would appear problematic for GC modeling, which often neglects phenomena occurring on $t_{\rm H}$---e.g., evolution of the Galactic potential. But removal under the raw criterion neglects the impact of ongoing weak two-body encounters during $\Delta t_{\rm esc}$. By raising the $\Etilde$ necessary to become a PE in an attempt to roughly account for these interactions, the $\alpha$ criterion dramatically reduces the probability of $\Delta t_{\rm esc}\gtrsim t_H$ (2nd row from top). Further accounting for the impact of GC mass loss by integrating PEs in \texttt{CMC}'s true evolving $\phiclus(t)$ also hastens escape (lower two rows) but is less influential than the escape criterion because the initial dissolution timescale for most MWGCs surviving today is $\gtrsim t_{H}$. For all four combinations of escape criterion and $\phiclus$ assumption, $\Delta t_{\rm esc}$ overall is significantly shorter for the NCC'd GCs (solid) than the CC'd GCs (dashed). This is unsurprising since Figure~\ref{fig:Etilde} showed the former have higher $\Etilde$ overall (due to more strong ejections from BH-driven three-body binary formation; Figures~\ref{fig:rejrtej_Etilde}--\ref{fig:tej_Etilde} and W23).

A visually striking feature of Figure~\ref{fig:tesc} is the appearance of successive plateaus in the $\Delta t_{\rm esc}$ distribution, especially at intermediate energies---the yellow, gray, and teal curves with respective $\Etilde\approx(0.1,0.18,0.32)$. These features are not evident in the results of FH00, \cite{Ernst2008}, or \cite{TanikawaFukushige2010}, perhaps in part due to their focus on low $\Etilde$, where $\Delta t_{\rm esc}$ is so long that these plateaus are minute. Yet both FH00 and \cite{TanikawaFukushige2010} show $\Delta t_{\rm esc}$ for similar bands in $\Etilde$ as high as 0.24, so the phenomenon may be unapparent in their results simply because they cut out low $\Delta t_{\rm esc}$, where the plateaus are most noticeable. Regardless, the plateaus are a true physical feature arising from the geometry of the zero-velocity surface. For $0<\Etilde < 0.1$ (dark red to orange curves), the necks are so narrow that PEs typically loop back many times through the GC before finally `finding' and escaping through a neck. For $\Etilde\gtrsim 0.5$ (light blue to violet), the necks are so large that most PEs cross directly beyond $r_t$. But for $0.1\lesssim\Etilde\lesssim 0.5$, especially near the center of that interval, escape often requires a small number of additional crossings before the body finally passes through either neck. We show in Figure~\ref{fig:plateau_trajectories} that bodies escaping immediately before the $\mathcal{N}$th plateau in the $\Delta t_{\rm esc}$ distribution typically correspond to $\mathcal{N}-1$ such additional crossings.

That the plateaus become less obvious as $\mathcal{N}$ and $\Delta t_{\rm esc}$ increase relates to chaotic scattering theory, in which the basins (regions) of phase space leading to escape through different necks in an underlying conservative potential are separated from each other by a non-attracting fractal boundary known as a \textit{chaotic saddle} \citep[see, e.g.,][and references within]{OttTel1993,Ott2002,TelGruiz2006,Ernst2008,Seoane2013}. Such fractals also appear in the phase space basins corresponding to different bins in the $\Delta t_{\rm esc}$ distribution \citep[e.g.,][]{deAssisTerra2014, Zotos2015a, Zotos2015b, Zotos2016, ZotosJung2017}. In general, though there are regions of phase space where trajectories are regular and $\Delta t_{\rm esc}$ is effectively infinite (especially retrograde orbits in the CR3BP, e.g., FH00), $\mathcal{N}$ and $\Delta t_{\rm esc}$ increase where the phase space winds into finer (more chaotic) regions of the fractal boundary. Infinitely many locally smooth patches of each basin (for escape through either neck) exist in the saddle, each with a different characteristic $\Delta t_{\rm esc}$ that, when exceeded, induces a sharp drop off in the overall $\Delta t_{\rm esc}$ distribution as PEs from the patch exit the GC in a burst, leaving behind a plateau. The shrinking size and winding of these patches in the chaotic saddle as $\mathcal{N}$ and $\Delta t_{\rm esc}$ increase diminishes the bursts and blurs them together until, for $\Delta t_{\rm esc}\rightarrow\infty$, the surviving fraction of PEs decays roughly exponentially with $\Delta t_{\rm esc}$. For an excellent discussion with graphics in the context of the H\'{e}non-Heiles potential see \cite{Aguirre2001}, especially their Section~IV.A/Figure~9 relating to the plateaus.

Figure~\ref{fig:tesc} also shows that for either constant or evolving $\phiclus$, changing the escape criterion negligibly alters the $\Delta t_{\rm esc}$ distributions for almost all specific $\Etilde$ (excluding $\Etilde\approx 0.1$ in yellow). This is reasonable; while exact phase space details affect $\Delta t_{\rm esc}$, $\Etilde$ controls the width of the zero-velocity surface's openings. We see more significant differences when comparing $\Delta t_{\rm esc}$ at specific $\Etilde$ between the CC'd and NCC'd GCs. $\Delta t_{\rm esc}$ is higher in the former at all $\Etilde$ shown (especially $\Etilde\approx 0.1$ in yellow) \textit{except} $\Etilde\lesssim 0.02$ (red curves), where the opposite is true. These discrepancies at identical $\Etilde$ must relate to differences in the initial phase space distributions of PEs. Figure~\ref{fig:rejrtej_Etilde} showed $\Etilde\approx 0.1$ corresponds mostly to relaxation deep in the core ($r\lesssim r_c/10$) while $\Etilde\lesssim 0.02$ corresponds mostly to relaxation in the halo ($r\gtrsim r_c$). So relative to the NCC'd GCs, relaxation in the CC'd GC yields slower escape from the deep core and faster escape from the halo.

The faster escape in the halo of the CC'd GC may result from radial velocity anisotropy (bias to $|v_r|>v_t$), which develops in the halos of GCs born centrally dense, especially near core collapse, but not in GCs born more diffuse \citep[e.g.,][]{GierszHeggie1997,Takahashi1997,TakahashiLee2000,BaumgardtMakino2003,Tiongco2016a,Zocchi2016b,Claydon2017}. PEs on such elongated orbits have long been known to escape more easily for $0<\Etilde\ll 1$ since they probe the full tidal boundary, precessing to eventually find either neck. The stability of near-circular retrograde orbits within or near the tidal boundary also promotes preferential escape of radial orbits. Meanwhile, the \textit{slower} escape from the deep core of the CC'd GC may result from the faster loss of BHs, an important driving source of strong kicks, which increase the effective orbital eccentricity of the kicked body. So the early loss of BHs in the CC'd GC may dampen radial velocity anisotropy in the deep core, slowing escape, but firmer explanation of such subtleties will require further study.

Figure~\ref{fig:tesc}'s right column shows the $\Delta t_{\rm esc}$ distribution in normalized form, $\tilde{t}\equiv\Delta t_{\rm esc}\omega\Etilde^2$. This is motivated by the approximate scaling relation $\Delta t_{\rm esc}\propto\omega^{-1}\Etilde^{-2}$ derived by FH00 for the most common case $\Etilde\ll 1$ \citep[see also][]{TanikawaFukushige2010,Renaud2011}. If this scaling is correct, $\tilde{t}$ distributions at different $\Etilde$ in this limit should be nearly identical. They should also roughly match the fitted $\tilde{t}$ distributions from Figure~9/Table~1 of FH00 (dotted black) since $\Etilde$ is independent of $N$---their Equation~(9)---and weakly dependent on the central concentration or assumed $\phigal$ \citep{TanikawaFukushige2010}. But cumulatively, there are significant differences between FH00's setup and our own; the former assumed a Keplerian (instead of logarithmic) $\phigal$ and sampled PEs from a static \cite{King1966} $\phiclus$ while we sample PEs from (and for the lower panels, even evolve trajectories within) \texttt{CMC}'s time-dependent $\phiclus$. This $\phiclus$ features large density fluctuations (gravothermal oscillations), velocity anisotropy, and $\Etilde$ drawing from numerous escape mechanisms.

Given the above differences, the relatively close match in the $\tilde{t}$ distribution for $\Etilde<0.1$ under constant $\phiclus$ in Figure~\ref{fig:tesc} is encouraging. For these curves (top two rows, red through orange), our $\tilde{t}$ distributions converge and do not show a monotonic trend in the $\tilde{t}$ space (e.g., gradual drift to lower $\tilde{t}$ as $\Etilde$ decreases). This supports the $\Etilde^{-2}$ scaling of $\Delta t_{\rm esc}$ under constant $\phiclus$ in FH00. Yet this is contrary to how $\tilde{t}$ increases monotonically with $\Etilde$ at $\Etilde \gtrsim 1$ (blue and violet). The FH00 scaling relation, based on the phase space flow rate near L1/L2, is not valid at such high $\Etilde$ since the corrresponding flow is no longer restricted to the vicinity of L1/L2. This also largely explains the discrepancy between our overall $\tilde{t}$ distributions in black and those of FH00. The tail at high $\tilde{t}$ in the former (literally a sum over the continuous procession of the nearly vertical blue curves) arises from our inclusion of many escapers from strong encounters. The fits to FH00 are worse in the upper panels because many PEs that have not yet escaped are excluded in our curves. If our simulations were run longer, new `escapers' beyond $r_t$ would be added at the top left in each panel, pushing the tail at high $\tilde{t}$ down until it appears more like the thick black curves in the right panel second from the top.

\begin{figure*}
\centering
\includegraphics[width=\linewidth]{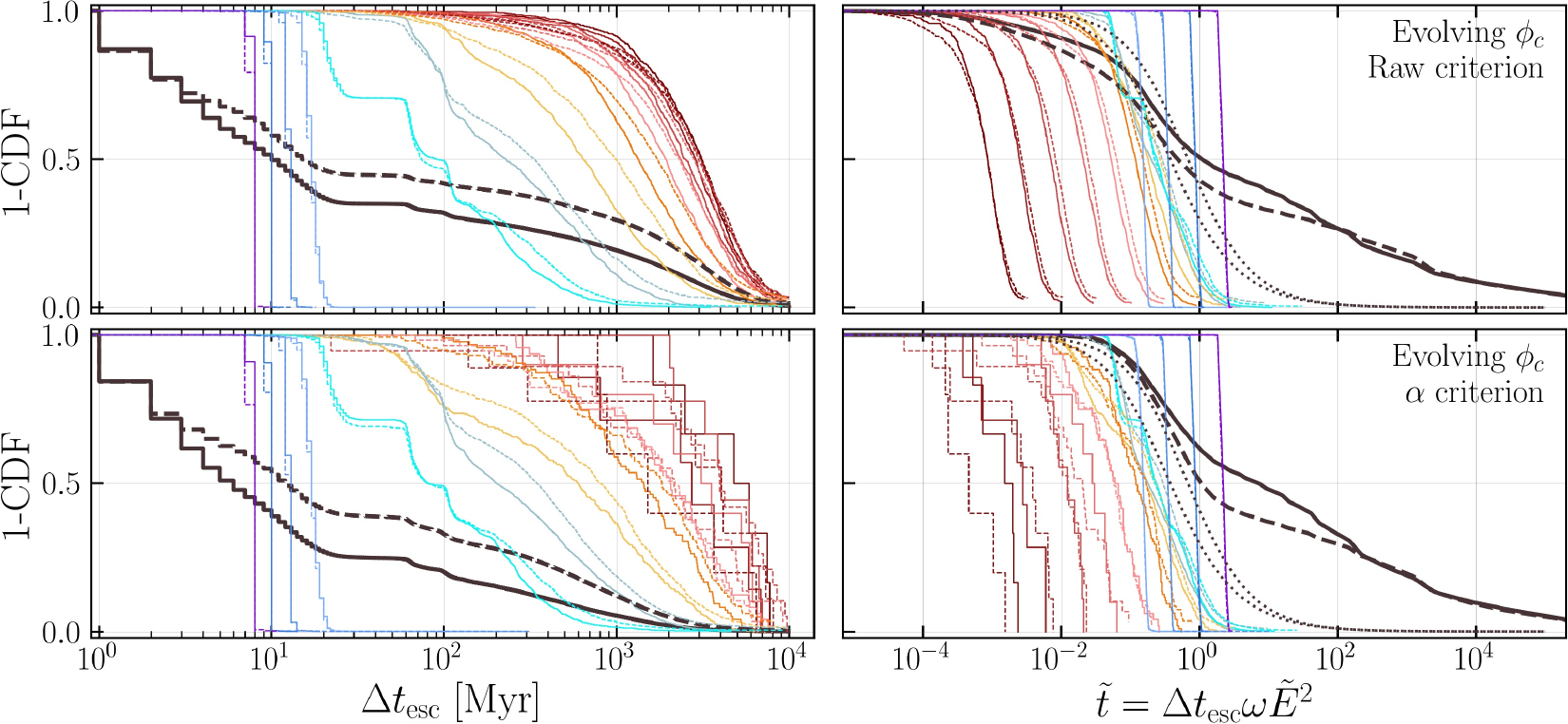}
\caption{Same as the bottom two rows of Figure~\ref{fig:tesc}, but we instead integrate all PEs with $t_{\rm rmv} < 4\,\Gyr$ for $10\,\Gyr$ beyond $t_{\rm rmv}$.}
\label{fig:tesc4Gyr}
\end{figure*}

\begin{figure}
\centering
\includegraphics[width=\linewidth]{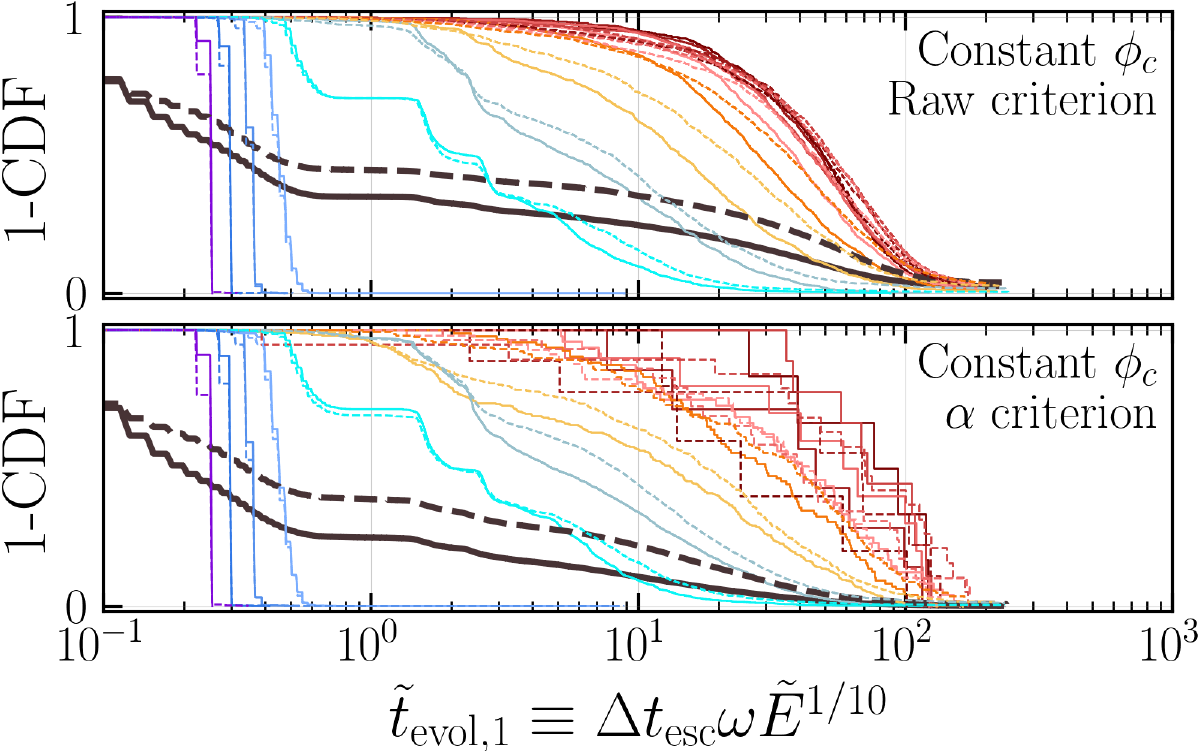}
\caption{Same as the right panels in Figure~\ref{fig:tesc4Gyr} but with $\tilde{t}$ redefined from Equation~(\ref{eq:tesc_evolving_phi}) for low $\Etilde\lesssim 0.03$, the red curves. These converge much better to a common $\tilde{t}$ distribution.}
\label{fig:ttilde_evol1}
\end{figure}

\begin{figure}
\centering
\includegraphics[width=\linewidth]{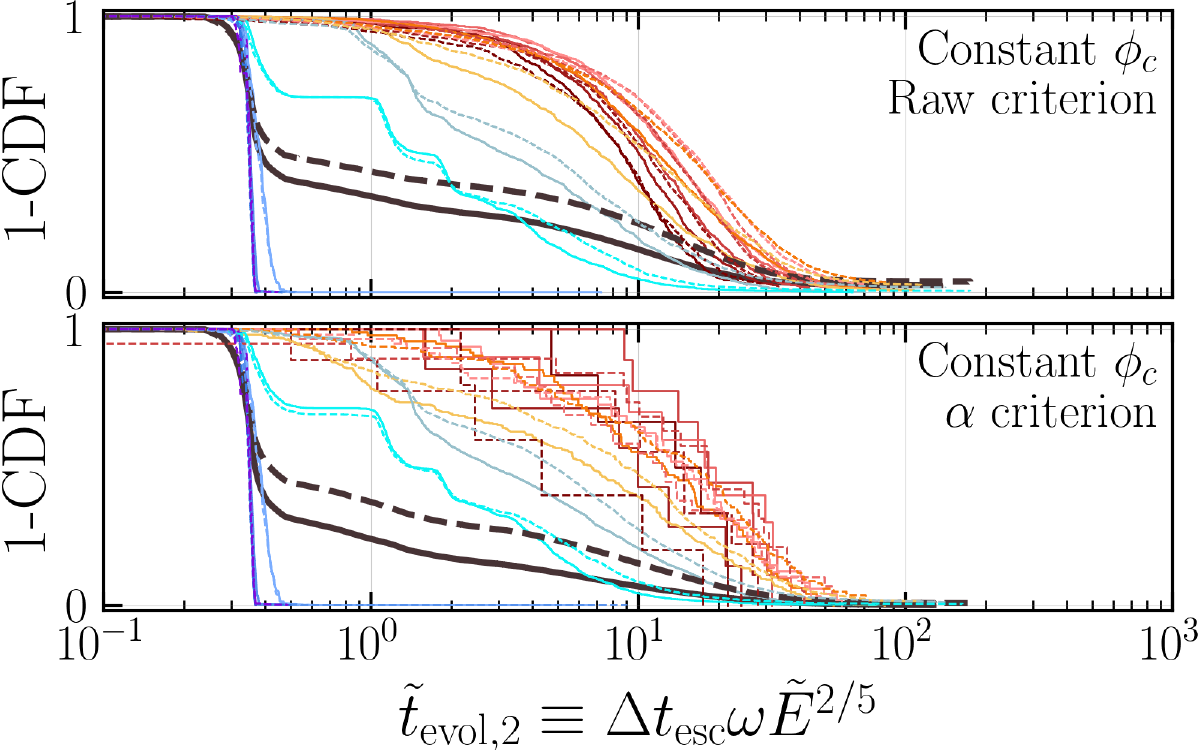}
\caption{Same as Figure~\ref{fig:tesc4Gyr} but with $\tilde{t}$ redefined from Equation~(\ref{eq:tesc_evolving_phi}) for high $\Etilde\gtrsim 0.56$, the blue and violet curves. These, too, converge much better to a common $\tilde{t}$ distribution.}
\label{fig:ttilde_evol2}
\end{figure}

\subsection{An Empirical Escape Timescale for an Evolving Cluster Potential} \label{S:results_new_escape_timescale}
When $\phiclus(t)$ is allowed to evolve during the trajectory integration (bottom two rows of Figure~\ref{fig:tesc}), the $\tilde{t}$ distributions for different $\Etilde \ll 1$ no longer overlap, indicating $\Delta t_{\rm esc}$ no longer scales as $\Etilde^{-2}$ in this more complex case. This is hard to see in Figure~\ref{fig:tesc} because the truncation at $1\,\Gyr$ cuts out nearly the entire $\tilde{t}$ distribution for each of the red and orange curves, but it is at least clear that for such low energies, $\tilde{t}$ monotonically decreases with $\Etilde$. This occurs because the initial dissolution timescale of old MWGCs is ${\sim}t_H$, so has negligible impact for moderate to high $\Etilde$, which have $\Delta t_{\rm esc}\ll t_H$. But as $\Etilde$ decreases and $\Delta t_{\rm esc}$ lengthens, GC mass loss grows more relevant, limiting $\Delta t_{\rm esc}$. The pattern is much more evident in Figure~\ref{fig:tesc4Gyr}, a duplicate of the lower two rows of Figure~\ref{fig:tesc} that only shows escapers from $t_{\rm rmv}<4\,\Gyr$, allowing us to truncate without bias the $\Delta t_{\rm esc}$ distribution at $10\,\Gyr$. Figure~\ref{fig:tesc4Gyr} shows how GC mass loss causes the $\Delta t_{\rm esc}$ distribution at low $\Etilde$ to converge; for $\Etilde$ low enough that escape proceeds on $t_H$, GC evaporation itself limits $\Delta t_{\rm esc}$.

Please note however that the curves in Figure~\ref{fig:tesc4Gyr} are \textit{not} directly comparable to Figure~\ref{fig:tesc}. Except for ages after core collapse, lowering the maximum removal time from $13$ to $4\,\Gyr$ increases the typical $\Etilde$, greatly reducing overall $\Delta t_{\rm esc}$ in black. Due mostly to stellar winds in young massive stars and supernovae, GC mass loss is faster at early times (Figure~\ref{fig:macro_evolution}), so more rapidly increases $\Etilde(t)$ of PEs as they find their way out of the GC. This truncates the $\Delta t_{\rm esc}$ curves more sharply in Figure~\ref{fig:tesc4Gyr} than in Figure~\ref{fig:tesc} for any specific, but sufficiently low, $\Etilde$. The distributions at higher $\Etilde$ change less between these Figures since the correspondingly faster escape reduces the impact of GC mass loss on $\Delta t_{\rm esc}$.

Figure~\ref{fig:tesc4Gyr} shows a very nearly uniform ${\approx}0.47$ dex gap between each of our red curves at different $\Etilde\lesssim10^{-1.5}\approx0.03$ and a ${\approx}0.40$ dex gap between each of our blue/violet curves at $\Etilde\gtrsim10^{-0.25}\approx0.56$. Since the curves are $1/4$ dex apart in $\Etilde$, these trends suggest that for a realistically evolving $\phiclus$, $\tilde{t}\propto\Etilde^{1.9}$ for $0<\Etilde \lesssim 0.03$ and $\tilde{t}\propto\Etilde^{1.6}$ for $\Etilde \gtrsim 0.56$. So
\begin{equation} \label{eq:tesc_evolving_phi}
\Delta t_{\rm esc} \propto
    \begin{cases}
        \omega^{-1} \Etilde^{-0.1} & \text{if } 0 < \Etilde \lesssim 0.03 \\
        \omega^{-1} \Etilde^{-0.4} & \text{if }     \Etilde \gtrsim  0.56.
    \end{cases}
\end{equation}
We provide no expression for $0.03\lesssim \Etilde \lesssim 0.56$ because it is apparent in Figures~\ref{fig:tesc} and \ref{fig:tesc4Gyr} that there is no clean power-law scaling due to the elevated importance of the specific initial phase space coordinates at these energies and resulting appearance of more complex features like the plateaus in both the $\Delta t_{\rm esc}$ and $\tilde{t}$ distributions.

From the above expressions, we can redefine the normalized escape timescale for the case of an evolving $\phiclus$ as
\begin{equation} \label{eq:ttilde_evolving_phi}
\tilde{t}_{\rm evol} \equiv
    \begin{cases}
        \Delta t_{\rm esc} \omega \Etilde^{0.1} & \text{if } 0 < \Etilde \lesssim 0.03 \\
        \Delta t_{\rm esc} \omega \Etilde^{0.4} & \text{if }     \Etilde \gtrsim  0.56.
    \end{cases}
\end{equation}

\begin{figure*}[h!]
\centering
\includegraphics[width=\linewidth]{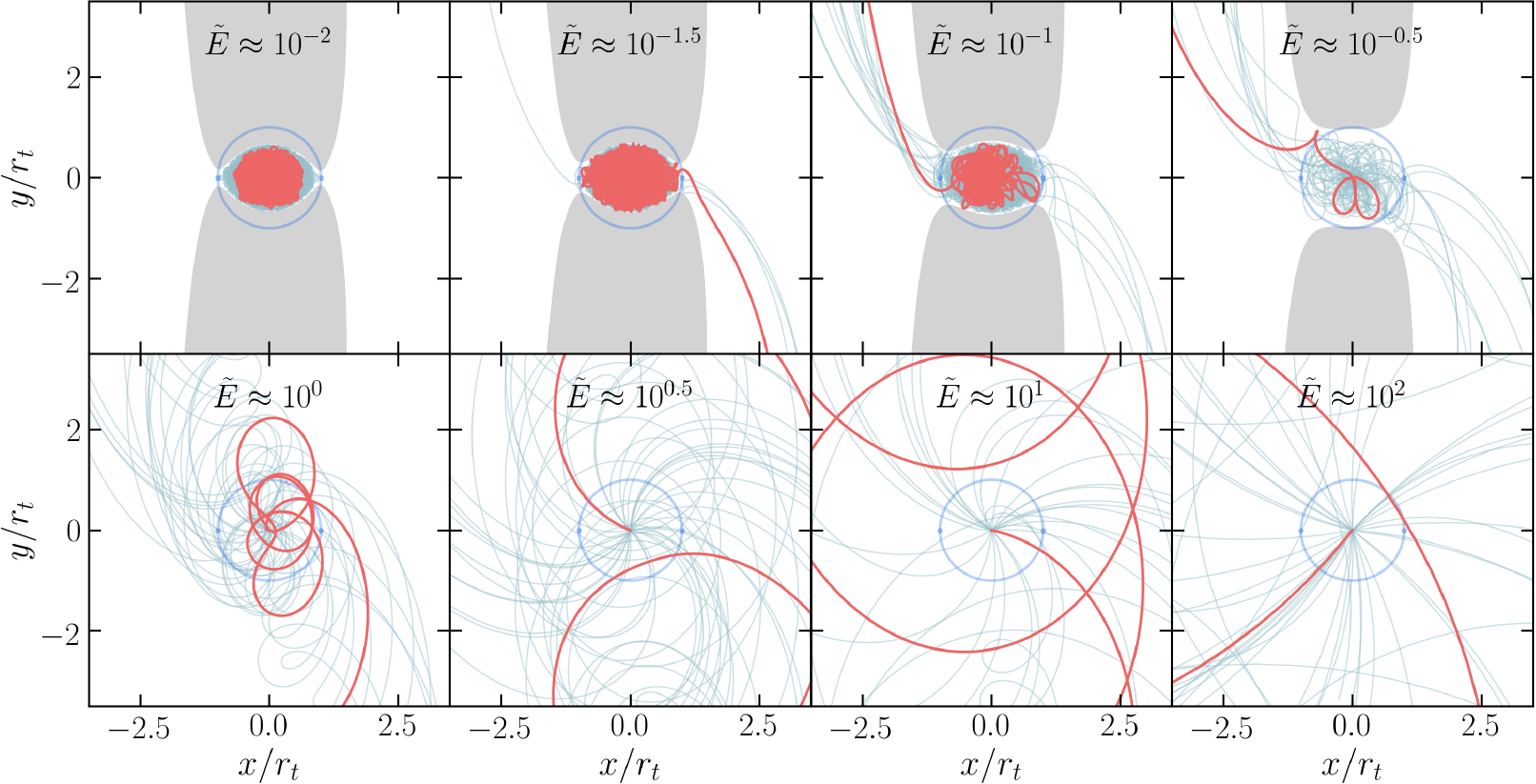}
\caption{As in Figure~\ref{fig:plateau_trajectories}, but the panels now distinguish subsets (of 50 escaper trajectories each) belonging to 8 bins in $\Etilde\in[10^{-2},10^2]$ from the archetypal NCC'd GC under the raw criterion and integrated to age $t_{\rm rmv}+\Delta t_{\rm esc}+2\,\Gyr$. Except in the last panel (due to its high $\Etilde$), this time limit excludes the portions of the trajectories that return to the GC after entirely circumnavigating the Galaxy in the rotating frame, cleaning up the Figure considerably (see also Figure~\ref{fig:trajectories_galactocentric}, identical to this one but in the inertial center-of-mass frame). Unlike in Figure~\ref{fig:plateau_trajectories}, there is no additional filter for escapers belonging to specific windows in $\Delta t_{\rm esc}$, so we make the bins ten times narrower---between 0.99 and 1.01 the $\Etilde$ specified in each panel. This makes the difference between the least and most enclosing forbidden realm in each panel nearly imperceptible.}
\label{fig:trajectories_clustercentric}
\end{figure*}

\begin{figure*}[h!]
\centering
\includegraphics[width=\linewidth]{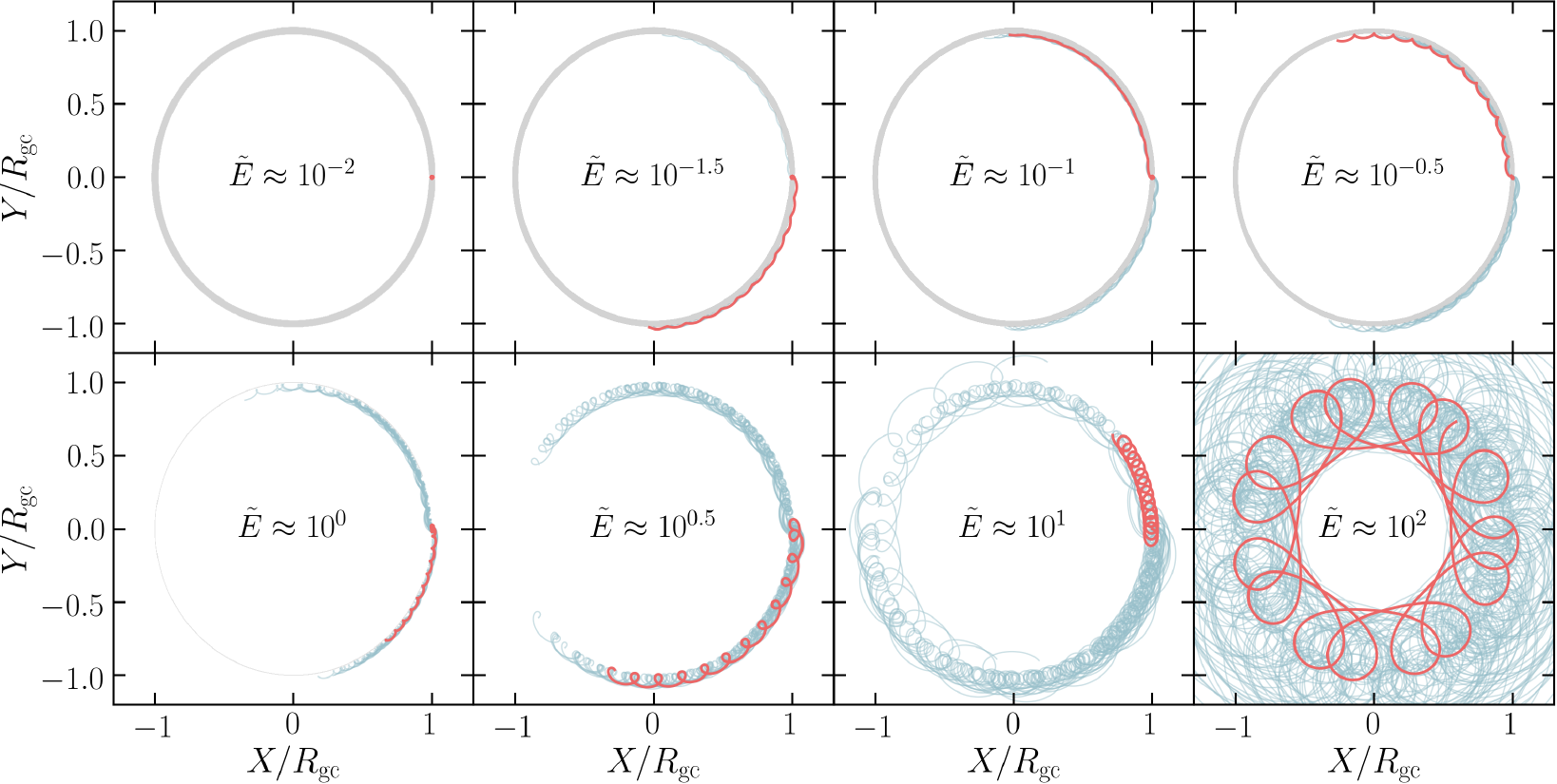}
\caption{As Figure~\ref{fig:trajectories_clustercentric}, but shown in the rotating center-of-mass frame of the Galaxy and GC.}
\label{fig:trajectories_galactocentric}
\end{figure*}

\begin{figure*} 
\centering
\includegraphics[width=\linewidth]{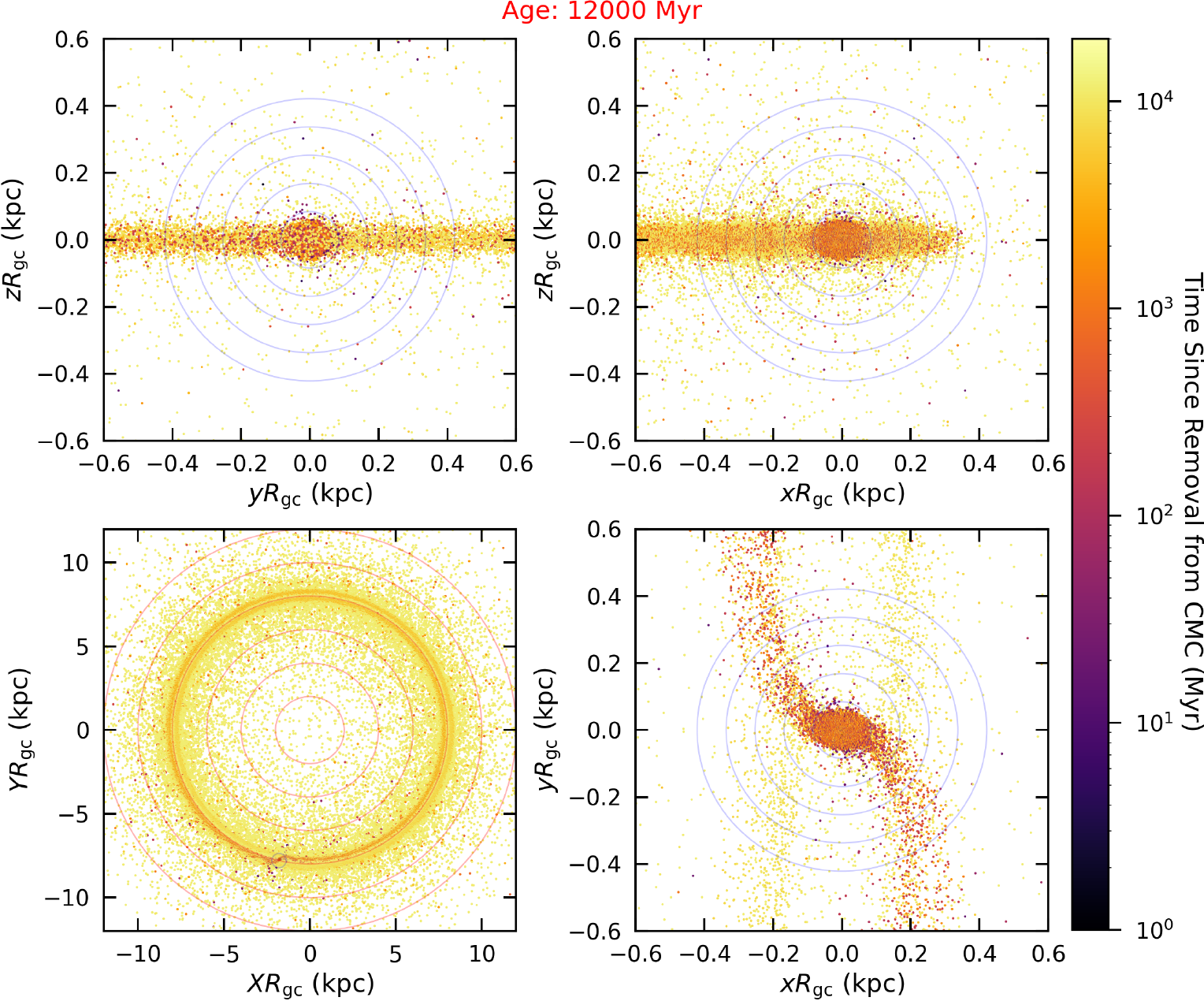}
\caption{Two-dimensionally projected positions of PEs, colored by time since removal from \texttt{CMC}, at age $t=12\,\Gyr$ (but see a full $14\,\Gyr$ movie accessible via \dataset[DOI: 10.5281/zenodo.10714860]{https://doi.org/10.5281/zenodo.10714860} and \href{https://www.youtube.com/watch?v=zJKCvAf6U3E}{youtu.be/zJKCvAf6U3E}). We show the denser CC'd GC under the $\alpha$ criterion simply to optimize visualization, as the higher density produces more PEs and the lower $\Delta t_{\rm esc}$ under the $\alpha$ criterion amplifies the color contrast. The lower left panel shows the projected positions in the static center-of-mass coordinates (Section~\ref{S:coordinates}), a face-on view of the GC orbit. The other panels show the three orthographic projections along each of the cardinal directions in the rotating clustercentric coordinates. The views in the upper row are edge-on to the GC's orbit, looking along (upper left) and perpendicular to (upper right) the ray connecting the cluster center to the Galactic center, while the view in the lower right panel is face-on to the cluster orbit. In the three orthographic panels, the blue circles have radii $r/r_t=[1,2,3,4,5]$. In the lower left panel, the red circles have radii $R/\kpc=[2,4,6,8,10,12]$, and the blue circle radius $r/r_t=5$ with guiding center at $R=\rgc$. The true cluster center's slight offset from this circle's center illustrates the subtle decrease in the distance between the GC and center-of-mass as the GC loses mass.}
\label{fig:tails_and_streams_movie_snapshot}
\end{figure*}

We verify these empirical relations in Figures~\ref{fig:ttilde_evol1} and \ref{fig:ttilde_evol2}, which show that $\tilde{t}_{\rm evol}$ in Equation~(\ref{eq:ttilde_evolving_phi}) leads to much better convergence in the normalized escape time distributions at low and high $\Etilde$, respectively. We also note that we have tested power law exponents in the vicinity of the ones specified above and found them to indeed provide the narrowest convergence (especially the one for high $\Etilde$).

\subsection{Formation of Tidal Tails and Stellar Streams} \label{S:results_tails_and_streams}
Having empirically re-examined the escape timescale, we turn our attention to the tidal tails and stellar streams from our simulated GCs. Future upgrades to \texttt{CMC}'s escape physics will involve tuning via careful comparison to direct $N$-body models, so for now we simply demonstrate macroscopic features (including new returning tails) and show that escapers from \texttt{CMC} can already reasonably reproduce established tidal phenomena for circular GC orbits in a static, spherical MW potential, so long as PE trajectories are evolved collisionlessly.

Figure~\ref{fig:trajectories_clustercentric} projects into the clustercentric $xy$-plane sample PE trajectories (from the raw criterion's NCC'd GC) integrated under constant $\phiclus=\phiclus(t_{\rm rmv})$, distinct for each PE. Each panel shows 50 trajectories, one highlighted in red as a visual aid, belonging to different narrow bins spaced $1/2$ dex apart in the range $\Etilde\in[10^{-2},10^2]$---see caption. To give a sense of the timescale and avoid clutter, we cut off each trajectory $2\,\Gyr$ after it first crosses to $r>r_t$. The PEs with low $\Etilde$ (top left) escape through the necks in their forbidden realm (gray) near L1/L2, the points on the blue circle indicating $r_t(t)$. Increasing $\Etilde$ expands the necks, allowing PEs to cross beyond $r_t$ with higher $\dot{y}$ or $\dot{z}$ (in/out of the page). The lower panels show how the forbidden realm's retreat from the $xy$-plane at $\Etilde\geq 1$ enables more immediate return to the GC. At $\Etilde\approx 1$ (lower left), in particular, the Coriolis effect causes ${\sim}10\%$ of escapers to temporarily return to $r<r_t$---sometimes many times, akin to periodic extratidal orbits \citep[e.g.,][]{Henon1969}---before moving beyond several $r_t$ from the GC.

In the rotating center-of-mass frame of the GC and MW (Figure~\ref{fig:trajectories_galactocentric}) the trajectories with $\Etilde\ll 1$ complete a single epicycle every ${\sim}10r_t$ \citep[as expected from, e.g.,][]{Kupper2008,Just2009}---longer for higher $\Etilde$. Many of the escapers with $\Etilde{\sim}10^2$ (lower right panel) entirely circumnavigate the MW in this frame within $2\,\Gyr$. Even the trajectories at $\Etilde \ll 1$ do so in ${\approx}8\,\Gyr$, well within the dissolution timescale of most MWGCs. This, too, agrees well with the epicyclic approximation for low $\Etilde$, where escapers drift away from the GC along the tails at a speed $v_d\approx 2\omega r_t$ for a logarithmic $\phigal$---e.g., Equation~(18) of \cite{Kupper2010}. In our case, $\omega\rgc\approx 220\,\pc\,\Myr^{-1}$ and $\rgc/r_t\approx 80$, so $v_d\approx 5.5\,\pc\,\Myr^{-1}$. Since the distance traveled per full orbit about the MW is $2\pi\rgc\approx 5\times10^4\,\pc$, the drift period in the tails (timescale to return to the GC) is $T_d\approx 9\,\Gyr$.

The potential impact of return trajectories is apparent in Figure~\ref{fig:tails_and_streams_movie_snapshot}, containing several views of the PEs from the CC'd GC under the $\alpha$ criterion at age $12\,\Gyr$ (a snapshot from a full $14\,\Gyr$-long movie linked in the caption). The lower right panel shows the projected positions in the orbital ($XY$) plane of the inertial center-of-mass frame, and the other panels the three orthographic projections along each axis of the rotating clustercentric frame (see caption). Since weak two-body relaxation dominates escape at $t=12\,\Gyr$, the streams closely follow the GC's circular orbit. This contrasts with ages ${\lesssim}3\,\Gyr$, when a clumpier, more energetic $\Etilde$ distribution (Figure~\ref{fig:tej_Etilde}) leads to more irregular, branching streams (see the full movie). But perhaps the most notable feature---novel in the context of star cluster literature---is the appearance (lower right) of robust `returning tidal tails', which form an X-like structure with the usual outgoing tails. This structure only appears at ages ${\gtrsim}T_d$, so coloring the PEs by $t_{\rm rmv}$ emphasizes the large age difference between the outgoing (mostly red) and returning (yellow) tails. Crucially, however, the robustness of the latter is largely due to the assumptions of a circular GC orbit in a spherical, unevolving $\phigal$. In reality, perturbations from MW substructure and an asymmetric, evolving $\phigal$ likely disperse such tails to much lower density. We further discuss these caveats in Section~\ref{S:discussion_return_tails}.

\noprint{\figsetstart}
\noprint{\figsetnum{14}}
\noprint{\figsettitle{Tidal Tail Surface Densities}}
\figsetgrpstart
\figsetgrpnum{14.4}
\figsetgrptitle{Simulation 4: CC'd, $\alpha$ Criterion}
\figsetplot{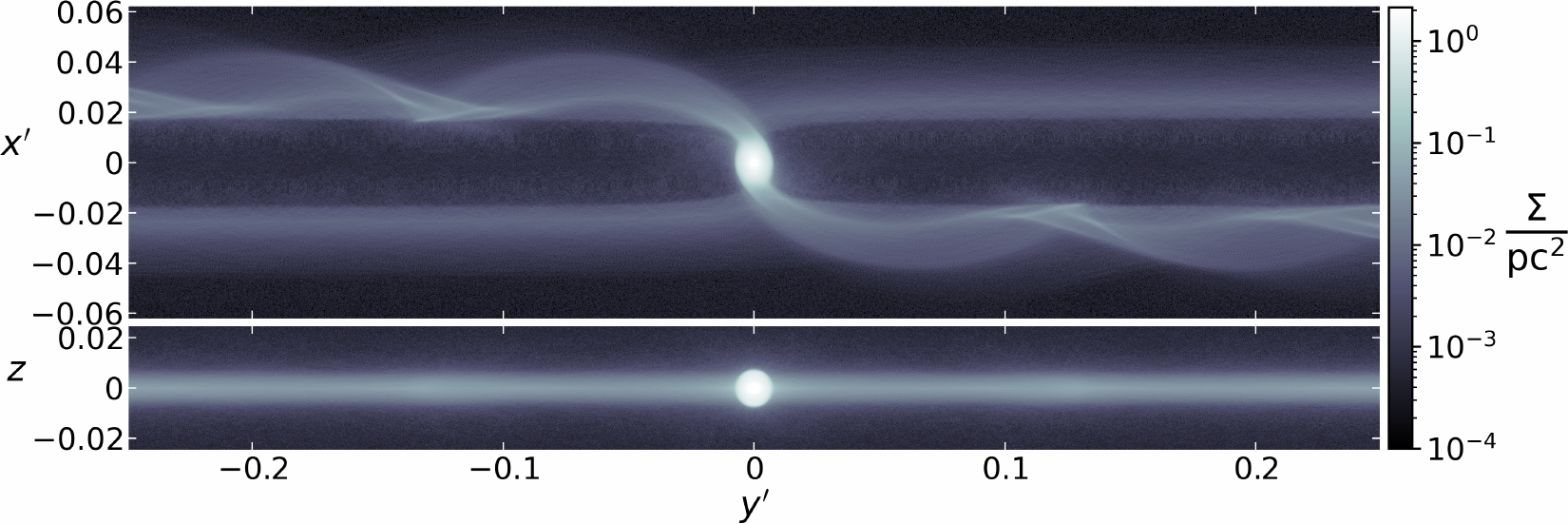}
\figsetgrpnote{Projected surface number density $\Sigma$ of PEs from the CC'd GC under the $\alpha$ criterion. $\Sigma$ is time-averaged between ages $t/\Gyr\in[11,13]$, achieved by stacking $2001$ snapshots of PEs in that interval, finely binning PEs by location, and dividing each bin count by $2001$ and the bin area in ${\rm pc}^2$. As described in the text, the new clustercentric coordinates $(x',y')$, still in units of $\rgc$, are flattened to map the full circular orbit to the line segment $y'\in[-\pi,\pi]$, where $(x',y',z')=0$ corresponds to the GC center. So the GC's velocity is to the right. The upper panel is a face-on view to the GC orbit (with the Galactic center down the page at $x'=-1$) while the lower panel is a panoramic edge-on view from the Galactic center to any point along the GC orbit.}
\figsetgrpend
\figsetend

\begin{figure*} 
\centering
\includegraphics[width=\linewidth]{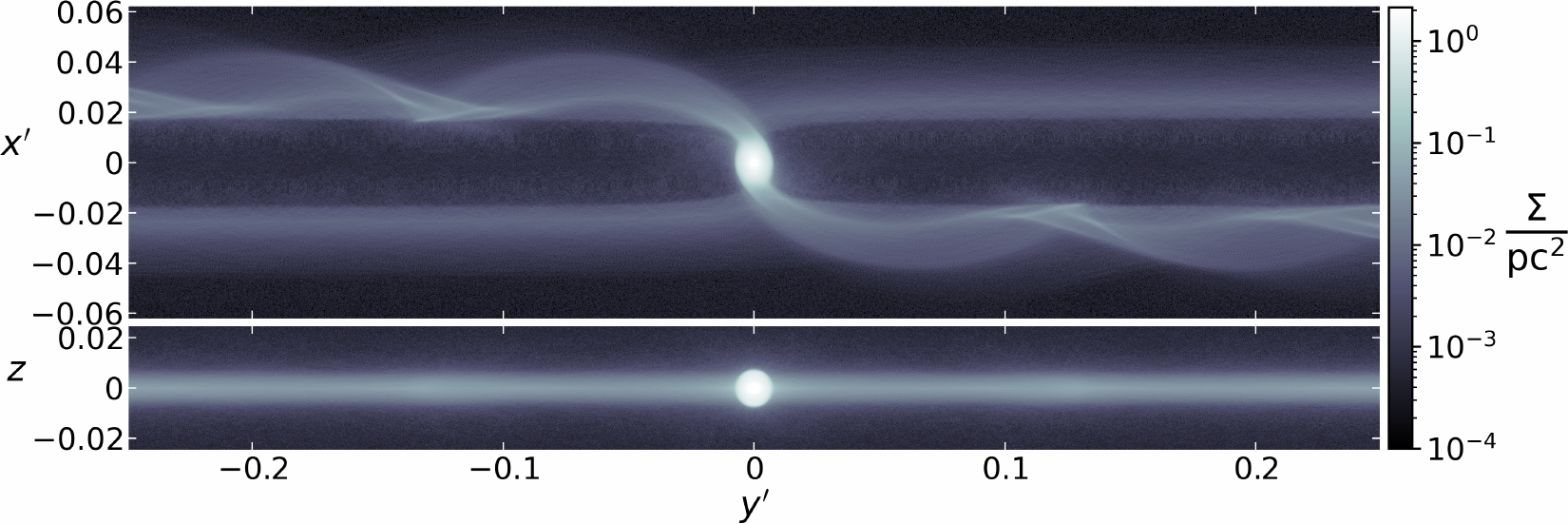}
\caption{Projected surface number density $\Sigma$ of PEs from the CC'd GC under the $\alpha$ criterion (see the full Figure set in the online journal for the other simulations). $\Sigma$ is time-averaged between ages $t/\Gyr\in[11,13]$, achieved by stacking $2001$ snapshots of PEs in that interval, finely binning PEs by location, and dividing each bin count by $2001$ and the bin area in ${\rm pc}^2$. As described in the text, the new clustercentric coordinates $(x',y')$, still in units of $\rgc$, are flattened to map the full circular orbit to the line segment $y'\in[-\pi,\pi]$, where $(x',y',z')=0$ corresponds to the GC center. So the GC's velocity is to the right. The upper panel is a face-on view to the GC orbit (with the Galactic center down the page at $x'=-1$) while the lower panel is a panoramic edge-on view from the Galactic center to any point along the GC orbit.}
\label{fig:fancy_tail}
\end{figure*}

\noprint{\figsetstart}
\noprint{\figsetnum{15}}
\noprint{\figsettitle{Full Stellar Stream Surface Densities}}
\figsetgrpstart
\figsetgrpnum{15.4}
\figsetgrptitle{Simulation 4: CC'd, $\alpha$ Criterion}
\figsetplot{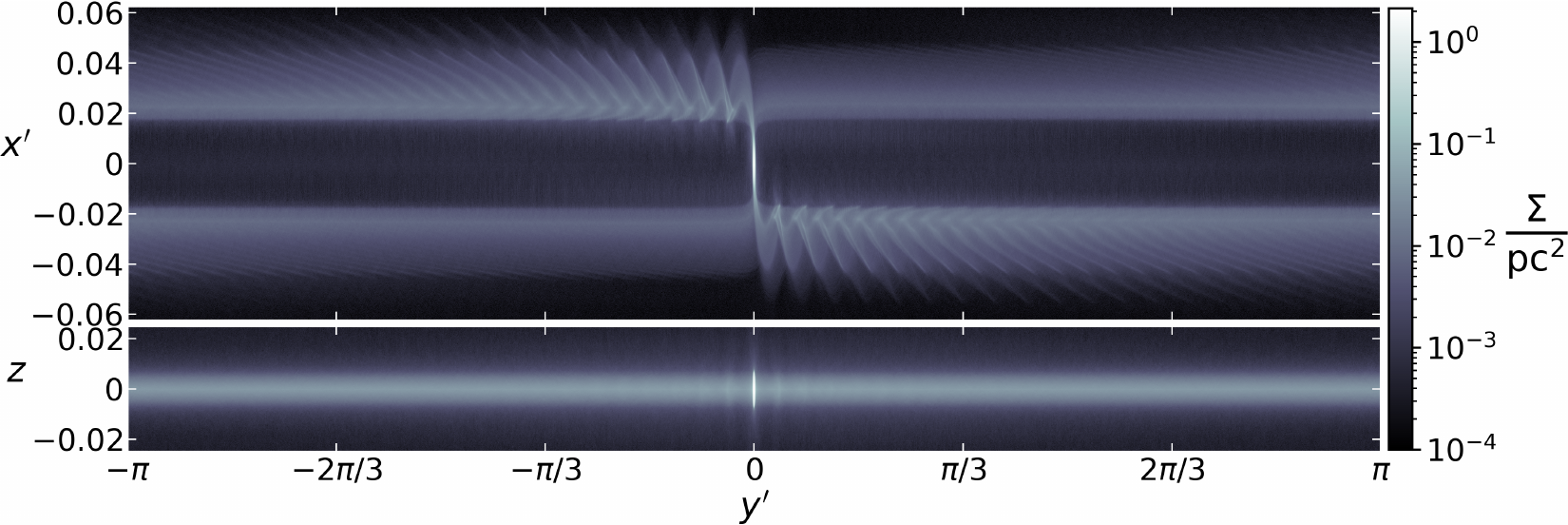}
\figsetgrpnote{As Figure~\ref{fig:fancy_tail} (for the CC'd GC under the $\alpha$ criterion) but with the horizontal axis compressed to fully span the entire GC orbit. The first several epicyclic overdensities are visible in the lower panel, spaced of order $10r_t$ apart. In the upper panel, the strip from $x'\rgc/r_t\in[-2,2]$ has much lower density than in the tails/streams due to the presence of the forbidden realm there for low $\Etilde$ (most escapers).}
\figsetgrpend
\figsetend

\begin{figure*} 
\centering
\includegraphics[width=\linewidth]{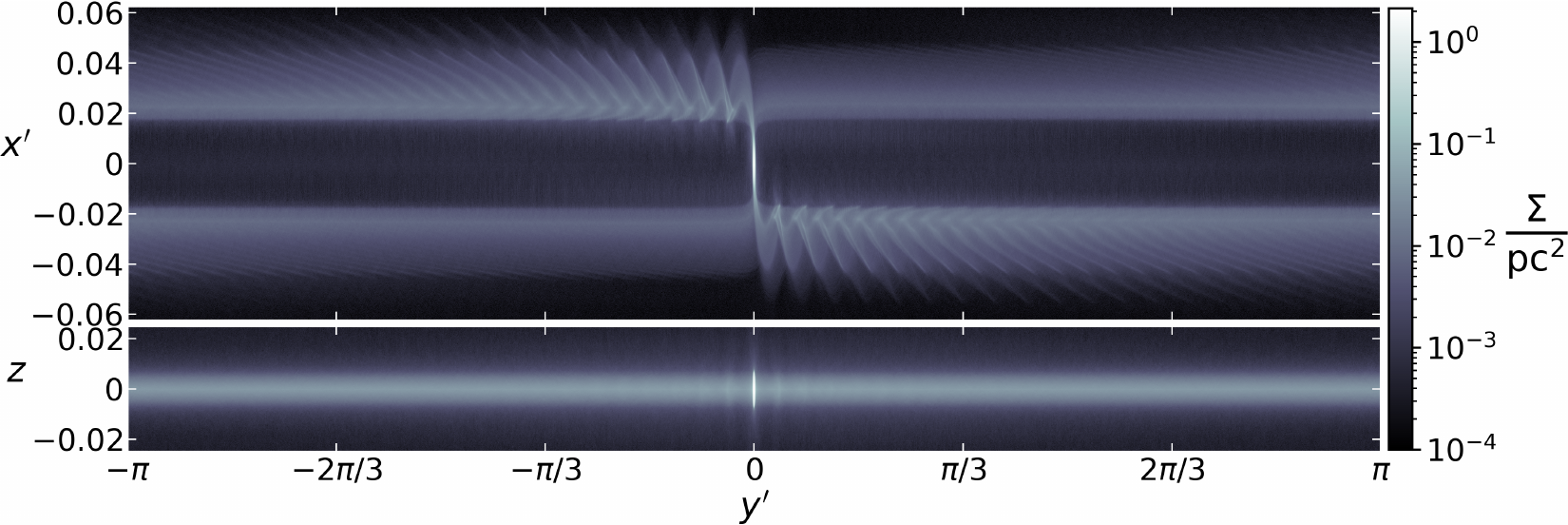}
\caption{As Figure~\ref{fig:fancy_tail} (including being a Figure set in the online journal) but with the horizontal axis compressed to fully span the entire GC orbit. The first several epicyclic overdensities are visible in the lower panel, spaced of order $10r_t$ apart. In the upper panel, the strip from $x'\rgc/r_t\in[-2,2]$ has much lower density than in the tails/streams due to the presence of the forbidden realm there for low $\Etilde$ (most escapers).}
\label{fig:fancy_stream}
\end{figure*}

\subsection{Stream Morphology} \label{S:stream_morphology}
To examine stream morphology in more detail, we show in Figures~\ref{fig:fancy_tail} and \ref{fig:fancy_stream} the surface number density $\Sigma$ of PEs near the tidal boundary and along the full stream, respectively, from the CC'd GC under the $\alpha$ criterion (see the online journal for the other simulations). $\Sigma$ is time-averaged between ages $t/\Gyr\in[11,13]$, achieved by stacking 2001 snapshots of the PEs in that interval, finely binning the PEs by position, and dividing each bin count by 2001 and the bin area in ${\rm pc}^2$. The new clustercentric coordinates $(x',y')$, still in units of $\rgc$, are flattened to map the full circular orbit to the line segment $y'\in[-\pi,\pi]$, where $(x',y')=0$ corresponds to the GC's center. Specifically, $x' \equiv [x +(1-\mu)(1-\cos\theta)]/\cos\theta$ and $y' \equiv (1-\mu) \theta$, where $\theta \equiv {\rm arctan2}[y/(x+1-\mu)]$. So in each Figure the GC's velocity points to the right and the upper panel is a face-on view to the orbit (with the MW's center down the page at $x'=-1$) while the lower panel is a panoramic edge-on view from the MW's center. The extreme sample size (${\gtrsim}10^8$ individual stellar positions) achieved from time-averaging escaper trajectories from such large GC simulations makes these Figures the highest-resolution tidal tail/stellar stream density profiles we were able to find in the literature.

The full stream consists of a leading and trailing tail sandwiching a low-density channel of width ${\approx}0.036\rgc\approx 3.4r_t$ (our GC models have average $r_t\approx 85\pc$ in the chosen age range). This channel closely follows the zero-velocity surface for $0<\Etilde\ll 1$, which has width $2\sqrt{3}r_t\approx 3.46 r_t$---see Equation~(17) of \cite{Just2009}. The width of each of the adjacent tails is ${\approx}2.3r_t$, in good agreement with the direct $N$-body models and epicyclic approximation presented by \cite{Just2009}; their Equations~(17) and (34) result in tail width ${\approx}2.1r_t$. Epicyclic overdensities are readily apparent, spaced ${\approx}12.2r_t$ apart. This is between the epicyclic approximations of $8.9r_t$ \citep{Kupper2010} and $15.4r_t$ \citep{Just2009}---their respective Equations~(20) and (22), given a logarithmic $\phigal$ has epicyclic frequency $\sqrt{2}\omega$. Their disagreement arises from their differing assumptions on the starting point for escapers' epicyclic trajectories: $x'=r_t$ or $x'=\sqrt{3}r_t$, respectively. Our results suggest an optimal approximation is intermediate to these extremes. Note the above distances in terms of $r_t$ are independent of GC mass $M$ and $\rgc$ in the MW halo \citep{Just2009}, where $\phigal$ remains logarithmic.

The Figures' exceptional resolution also reveals a subtle feature common to all four simulations: double-ridged density peaks in the tails near each epicyclic overdensity. These arise from a phase difference in the trajectories of low-$\Etilde$ escapers from opposing sides of each tail's neck (i.e., on either side of L1/L2); the corresponding overdensities streaking out from the tidal boundary are visible in the upper panel of Figure~\ref{fig:fancy_tail} (see also Appendix~\ref{S:tail_speed_maps}). This phenomenon disperses each epicyclic overdensity compared to the typical `particle spray' approximation  \citep[e.g.,][]{Kupper2008, Kupper2010, Kupper2012, Just2009} that escapers all exit the GC at exactly L1/L2 ($y'=0$).

While Figures~\ref{fig:fancy_tail} and \ref{fig:fancy_stream} display variations between simulations (see the online Figure Set), these are relatively minor. Most notably, the $\alpha$ criterion reduces the PE number density at clustercentric distances $0.6\lesssim r/r_t \leq 1$. This range, spanning the minimum and maximum $r$ to the tidal boundary, is where the GC density deviates most significantly from spherical. The disagreement, resulting from faster escape under the $\alpha$ criterion, means comparison to matching direct $N$-body models could determine which energy criterion best reproduces GC properties (e.g., density and velocities) within $0.6\lesssim r/r_t \leq 1$. But given the asymmetry here, it is likely that no spherically symmetric criterion \citep[even one based on energy \textit{and} angular momentum, e.g.,][]{Spurzem2005} will allow our collisionless PE approximation to reproduce all GC properties of interest here on its own. So development of more nuanced escape physics in \texttt{CMC}, perhaps using basis expansion to allow the Monte Carlo method to handle mildly asymmetric $\phiclus$ \citep[e.g.,][]{Vasiliev2015} may be worthwhile for this zone (see also Section~\ref{S:discussion_monte_carlo_method}).

Happily, changing between the $\alpha$ and raw criteria has lesser impact on features beyond the tidal boundary. For example, the former slightly widens the tail at the tidal boundary due to the higher typical $\Etilde$ and correspondingly larger necks about L1/L2. Further from the GC, this widening is most perceptible again at the first epicyclic overdensities. But this latter difference is minute, especially given our extreme time-averaged sample size is well beyond that achievable by observations, even stacking tails from many MWGCs. So, unlike for the asymmetric but still collisional region \textit{within} the tidal boundary, upgrades to escape in \texttt{CMC} should minimally impact the morphology of simulated stellar streams, in which the collisionless approximation is quite accurate.

\begin{figure*}
\centering
\includegraphics[width=\linewidth]{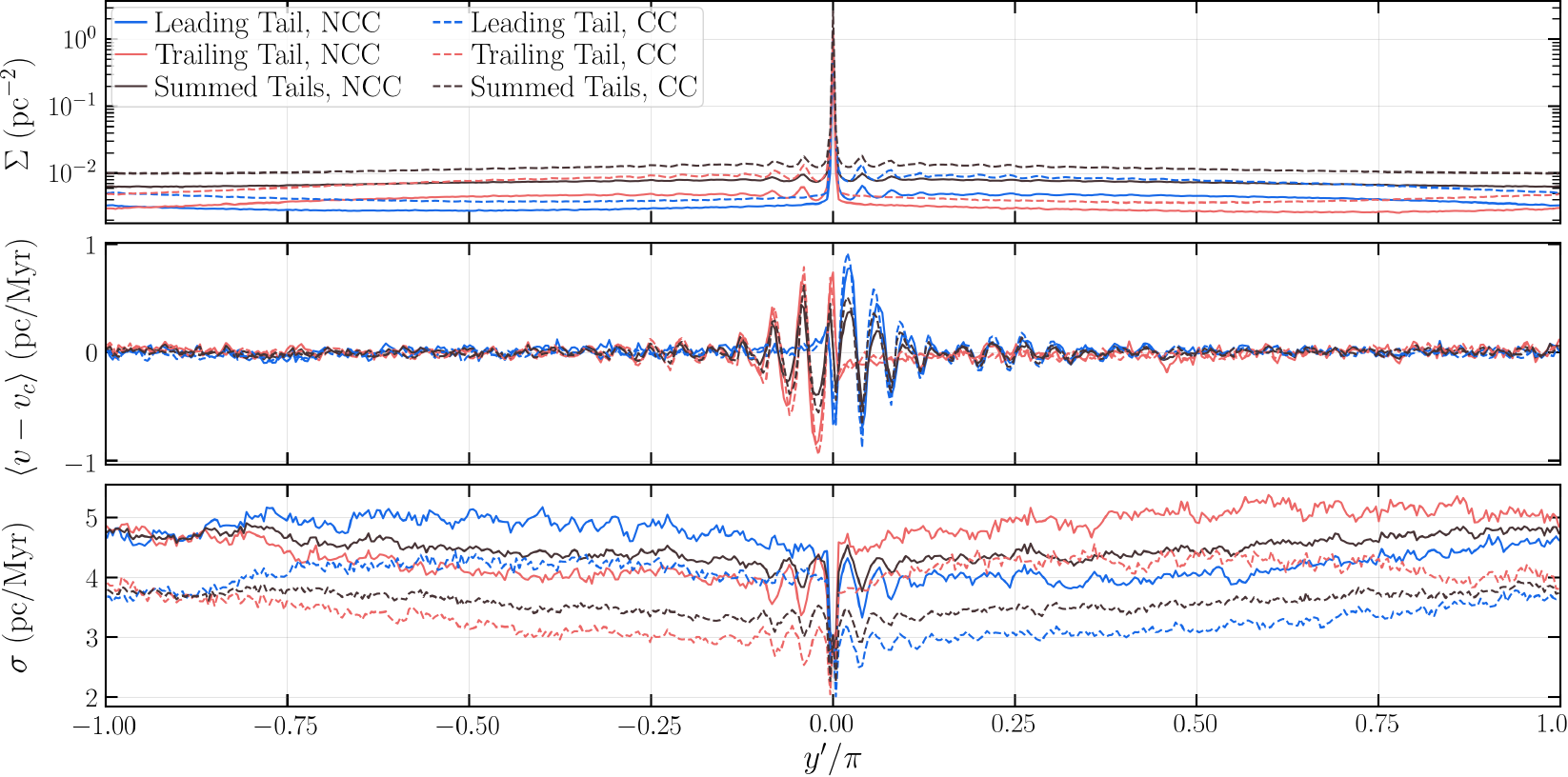}
\caption{The surface number density of PEs (viewed from the Galactic center; upper panel), their mean speed $\langle v-v_c\rangle$ relative to the GC's circular speed (middle panel), and the dispersion in this speed $\sigma \equiv \sqrt{\langle (v-v_c)^2\rangle - \langle v-v_{\rm c}\rangle^2}$ (lower panel), along the stellar stream's flattened $y'$-axis. The profiles are time-averaged across $t\in[11,13]\,\Gyr$ and split into 501 bins of uniform width ${\approx}100\,\pc$. To be counted in the tail/stream, a PE at time $t$ must have $|x'(t)|<0.05$ and $|z(t)|<r_t(t)/\rgc$. As usual, solid/dashed curves indicate the NCC'd/CC'd GCs---in this case, under the raw escape criterion. The blue/red curves correspond to the leading/trailing tails and the black curves to their sum.}
\label{fig:rhov_tail}
\end{figure*}

\begin{figure*}
\centering
\includegraphics[width=\linewidth]{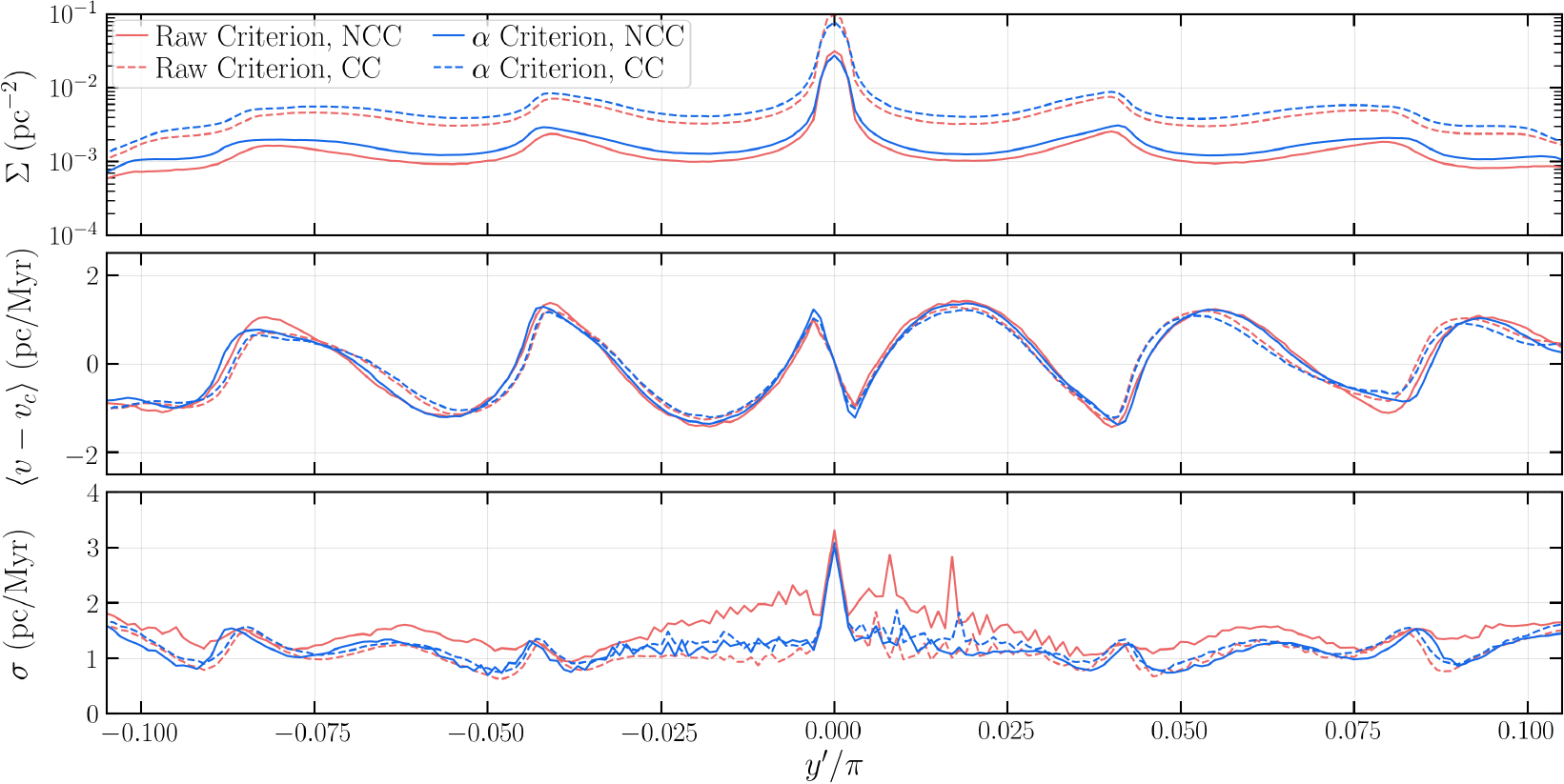}
\caption{As Figure~\ref{fig:rhov_tail} (including being a stack of 2001 snapshots spanning ages $11$--$13\,\Gyr$) except for the following changes. To simulate the impact of stream disruption, we now exclude the PEs in each snapshot for which ${>}500\,\Myr$ have elapsed since their `escape' (first crossing beyond $r>r_t$). We also only show the sum of the leading/trailing tails (i.e., the black curves from Figure~\ref{fig:rhov_tail}) and do so for all four simulations. The solid/dashed curves still correspond to the NCC'd/CC'd GCs, and the blue/red curves to the raw/$\alpha$ escape criteria. Finally, because of the limited tail length from the time cutoff, we use finer bins in $y'$---2001 across the entire GC orbit, corresponding to bin widths ${\approx}25\,\pc$.}
\label{fig:rhov_tail2}
\end{figure*}

\vspace{10pt}
\subsection{Stream Density and Internal Velocities} \label{S:stream_density_velocity}
While two-dimensional projections of stream properties are qualitatively revealing, it is easier to study variations between simulations and the impact of the returning tails via quantities averaged along the stream axis. As bulk properties, the stream density and internal velocity profiles are an ideal starting point. To this end, Figure~\ref{fig:rhov_tail} shows the surface number density $\Sigma$ of PEs (when viewed from the MW's center; top panel), their mean speed $\langle v-v_c\rangle$ with respect to the GC's circular speed (middle), and the dispersion in this speed $\sigma \equiv \sqrt{\langle (v-v_c)^2\rangle - \langle v-v_{\rm c}\rangle^2}$ (bottom), all from 501 uniform bins along the stellar stream's flattened $y'$-axis. Based on Figure~\ref{fig:fancy_stream}, we define the stream at any time $t$ as all PEs with $|x'(t)|<0.05$ and $|z(t)|<r_t(t)/\rgc$, including for reference the PEs within the tidal boundary. The results are again time-averaged across $t/\Gyr\in[11,13]$ and the solid/dashed curves correspond to the NCC'd/CC'd GCs under the raw criterion. The blue/red curves distinguish the contributions from the leading/trailing tails to their combined profile (black).

The epicyclic overdensities are again apparent and $\Sigma$ is nearly symmetric between the leading and trailing tails. The former is only ${\approx}5\%$ denser since the neck at L1 opens at slightly lower $\Etilde$ than the neck at L2. The density in both tails gradually decreases as they extend further from the GC before reaching a minimum and increasing again about a quarter orbit before returning to the GC. In the earliest portion of each tail's outgoing half---$y'>0$ (leading) and $y'<0$ (trailing)---$\Sigma$ is about twice as high from the CC'd GC than the NCC'd GC, consistent with their evaporation rates in Figure~\ref{fig:macro_evolution}. This factor shrinks closer to $3/2$ just before the tails return to the GC, consistent with steady leakage of PEs from the stream over time. Finally, since the circular speed $\omega R$ at any Galactocentric distance $R$ is constant in a logarithmic $\phigal$, then $\omega$ is slightly higher in the leading tail. This explains why the crossing point between the tail densities apparent at the far left of the top panel occurs at $y'/\pi$ just slightly ${>}-1$.

Under the epicyclic approximation, the mean speed in the stream relative to the GC's circular speed is $\langle v-v_c\rangle \approx [(2\omega/\kappa)^2-2]\omega r_t$, where $\kappa$ is the epicyclic frequency---see Equation~(21) of \cite{Kupper2010}. For logarithmic $\phigal$, $\kappa=\sqrt{2}\omega$, so $\langle v-v_c\rangle \approx 0$. The central panel of Figure~\ref{fig:rhov_tail} reproduces this expectation along the entire stream, and locally to within ${\sim}1\,\kms$. $\langle v-v_c\rangle$ locally peaks/troughs between the epicyclic overdensities in the leading/trailing tails since most (low-$\Etilde$) PEs here have a velocity near directly parallel/anti-parallel relative to the GC's velocity at the same $y'$. At each epicyclic overdensity, however, most PEs are briefly moving back toward the GC---or at least counter to the velocity of their epicyclic trajectory's guiding center, which is offset from the GC's orbit. This instead causes $\langle v-v_c\rangle$ to trough/peak, respectively. These features remain, albeit with lower magnitude, in the combined stream profile (black) since each tail's returning half---$y'<0$ (leading) and $y'>0$ (trailing)---is less dense than the outgoing half.

The amplitude in the $\langle v-v_c\rangle$ oscillations is slightly greater from the CC'd GC than the NCC'd GC. This is counter-intuitive since (to recap Section~\ref{S:results_energy_distribution}) the NCC'd GC has cumulatively higher ejection $\Etilde$ due to longer retention of BHs, which promote strong ejections from three-body binary formation (W23). Though we also noted the CC'd GC has comparable or greater typical $\Etilde$ when limited to $t\gtrsim 10\,\Gyr$, due to the higher post-collapse density's promotion of strong encounters and relaxation, this turns out \textit{not} to explain the CC'd GC's higher $\langle v-v_c\rangle$ amplitude in Figure~\ref{fig:rhov_tail}. (As we show shortly in Figure~\ref{fig:rhov_tail2}, restricting the stream to only PEs ejected after $10.5\,\Gyr$ instead causes the CC'd GC to have minutely \textit{lower} $\langle v-v_c\rangle$ amplitude than the NCC'd GC). Rather, the CC'd GC's faster mass loss---twice that of the NCC'd GG---is likely responsible, since the mass loss decreases $\lvert \phiclus \rvert$, causing the PEs to lose less speed while traveling `up' the GC's potential well to escape.

While the stream's $\langle v-v_c \rangle$ profile is very similar between the models, there is a larger difference in the profile of its velocity dispersion $\sigma$ (lower panel). Since $\sigma$ should correlate with the local tail width, it is unsurprising that $\sigma$ peaks between epicyclic overdensities and troughs within them (based on the overdensity locations in Figure~\ref{fig:fancy_stream}). Further along the tail, $\sigma$ both increases and oscillates less since small deviations in the typical epicycle period for different $\Etilde$ compound to randomize the velocity vector far from the GC (apparent in how the epicyclic overdensities distort and overlap far from the GC in Figure~\ref{fig:fancy_stream}). On average $\sigma$ is ${\approx}20\%$ higher from the NCC'd GCs than from the CC'd GCs. This is a direct result of including the returning tails, which allows bodies ejected many Gyr ago to contribute to Figures~\ref{fig:fancy_tail}--\ref{fig:rhov_tail}, despite only stacking snapshots of the stream in the age range $11$--$13\,\Gyr$. The NCC'd GC's wider spread in $\Etilde$ before ${\sim}10\,\Gyr$, due to more ejections from BH-driven three-body binary formation, thus inflates $\sigma$ in the NCC'd GC's stream even at late times.

Due to the impact of the returning tidal tails and their likely disruption by MW substructure in a realistic $\phigal$, we reproduce Figure~\ref{fig:rhov_tail}'s black curves (both tails combined) in Figure~\ref{fig:rhov_tail2}, this time showing all four simulations and eliminating the contribution of the returning tails. We do so by excluding from each snapshot in the Figure's $11$--$13\,\Gyr$ stack any escapers for which ${>}500\,\Myr$ have elapsed since their first crossing to $r>r_t$. This cutoff produces streams with angular span in Galactic longitude of ${\approx}40\deg$, roughly average for streams associated with MWGCs \citep[e.g.][]{Mateu2023}---hence the truncated horizontal axis.

Figure~\ref{fig:rhov_tail2}'s primary difference with respect to Figure~\ref{fig:rhov_tail} is that the amplitude in the $\langle v-v_c\rangle$ oscillations is ${\approx}2$ times higher while the dispersion $\sigma$ is ${\approx}3$ times lower. Eliminating the returning tail increases the former because there is no longer a returning flow opposing the outgoing tail (given our chosen viewpoint from the MW's center). That opposing flow significantly increases $\langle v-v_c\rangle$---compare, for example, the blue/red curves to the combined black curve in Figure~\ref{fig:rhov_tail}, or, for a helpful visual aid, see maps of the two-dimensional projected $\langle v-v_c\rangle$ in Appendix~\ref{S:tail_speed_maps}. These maps show that neglecting returning tails reduces $\sigma$ for the same reason, though the lower dispersion in $\Etilde$ at late times (Figure~\ref{fig:tej_Etilde}) likely helps.

Unlike when including the returning tails in Figure~\ref{fig:rhov_tail}, neglecting the returning tails in Figure~\ref{fig:rhov_tail2} results in there being little difference in $\sigma$ between the tails at $t\in [11,13]\,\Gyr$ from the NCC'd GCs versus those from the CC'd GCs. This is because the NCC'd and CC'd GCs have more similar $\Etilde$ distributions at $t\in [11,13]\,\Gyr$ (unlike their $\Etilde$ distributions \textit{cumulatively}). In particular, looking back at Figure~\ref{fig:tej_Etilde} for that age range, we see that though the NCC'd GC still has more high-$\Etilde$ ejections from three-body binary formation (due to longer BH retention; W23), the CC'd GC now has significantly more high-$\Etilde$ ejections from binary--single and binary--binary interactions (due to the higher density after core collapse). As a result, the total number of high-$\Etilde$ escapers contributing to $\sigma$ within this age window is roughly equivalent between the NCC'd and CC'd GCs, leading to a similar $\Etilde$ distribution and $\sigma$ within the tidal tails.

Together, Figures~\ref{fig:rhov_tail} and \ref{fig:rhov_tail2} show that the average speed of bodies in the tail $\langle v-v_c\rangle$ is not a reliable measure of a GC's state of core collapse and by extension its presently retained BH population. Yet $\langle v-v_c\rangle$ and its dispersion \textit{are} good indicators of the presence of a returning tidal tail, potentially making these quantities a useful constraint on the MW potential and the number and properties of massive perturbers---e.g., giant molecular clouds and dark matter subhalos---in the MW halo, since these control stream disruption.

\vspace{25pt}
\section{Discussion} \label{S:discussion}
\subsection{Implications and Caveats of Returning Tails} \label{S:discussion_return_tails}
A novel finding in this work is the appearance of returning tails in our simulations. When interpreting these features, it is crucial to recognize that our idealizations of a circular GC orbit in an unevolving, spherical MW potential are a \textit{best-case} scenario for the stream density. In reality, they should be much more diffuse as asymmetry, substructure, and time-dependence---including perturbations from the MW disk and bar, molecular clouds, and dark matter subhalos---should all naturally disrupt streams, blending them into the stellar background. Even detection of the denser \textit{outgoing} tails remains difficult; the number of GCs observed to have them remains low (${\approx}15$; \citealt{Piatti2020}). In none of these cases are returning tails apparent.

The MW is also not an especially accommodating place for returning tails as they need a drift period $T_d$ less than the GC's age to form. In the MW halo, where the circular speed $v_c$ is very flat (enclosed mass $M\propto\rgc$), $T_d\propto \rgc M^{1/3} v_c^{-1} \sim \rgc^{4/3}$ \citep[e.g.,][]{BinneyTremaine2008}. So $T_d$, already ${\approx}9\,\Gyr$ at $\rgc=8\,\kpc$, exceeds the age of many MWGCs altogether. Nor is it much shorter in the MW bulge since $v_c\propto\rgc$ there ($M\propto v_c^{2}$), yielding a sub-linear scaling $T_d\sim\rgc^{2/3}$. So bulge substructure, perhaps aided by a rotating bar \citep[e.g.,][]{Hattori2016,Pearson2017}, would have ample time to disrupt the streams. Eccentricity or even precession of the GC orbit---induced by, e.g., a triaxial $\phigal$ \citep{CapuzzoDolcetta2005}---could also misalign the bulk of the returning orbits from the GC at most phases of its orbit, preventing consistent returning tails at the tidal boundary. Even so, the sheer number of escapers (often ${>}10^5$) and their wide dispersal on the MW's crossing timescale (Figure~\ref{fig:tails_and_streams_movie_snapshot}), means that even major disruption or misalignment of the stream is unlikely to prevent \textit{some} escapers from returning to their original host GC after circumnavigating the MW. Tidal capture also requires low $\Etilde$ \citep{Koon2011}, so re-capture of past escapers from a dispersed stream may still exceed capture from the MW's stellar background. It is merely unclear whether more diffuse returning tails would be observable.

Yet detection of returning tails is not hopeless. Clearly an approach reliant on stellar surface density alone would be hindered by the background, and also the narrowness of the gap between the outgoing and returning tails; from many viewing angles no gap is apparent (e.g., the lower panels in Figures~\ref{fig:fancy_tail}--\ref{fig:fancy_stream}). But kinematic detection is more promising; thanks to kinematic (and chemodynamic) measurements, many MW streams are already known to be extremely elongated \citep{Mateu2023}. So while no streams to-date have been traced over more than a full orbit of the MW, this may be achievable with future surveys. Notably, we found in Section~\ref{S:stream_density_velocity} that the opposing flow of the outgoing and returning tails boosts the stream's velocity dispersion by ${\approx}3\,\kms$---substantial since the stream's dispersion without accounting for the returning tails is only ${\approx}1\,\kms$. This boost is similar in magnitude to that from perturbations by MW substructure, such as dark matter subhalos or the MW disk \citep[e.g.,][]{Carlberg2009,Carlberg2023}. It is also much greater than the likely inflation of $\sigma$ by unresolved binaries in streams. Multi-epoch radial velocity measurements of stars in streams \citep[][with the S5 Survey]{Li2019b} and in the MW halo \cite[][with the H3 Survey]{Conroy2019} indicate that systematic uncertainties from phenomena like binarity inflate observed $\sigma$ by an amount similar to the known uncertainties from the surveys' pipelines: ${\approx}0.5\,\kms$. So while the true impact of returning tails on $\sigma$ in streams should be substantially weaker than the $3\,\kms$ boost in our ideal case, returning tails may be relevant on a level comparable to observing biases such as unresolved binaries.

\subsection{Black Hole Mergers from Return Trajectories} \label{S:discussion_bh_mergers}
In Section~\ref{S:results_tails_and_streams}, we noted many escapers at moderate $\Etilde\sim 1$ promptly return to $r<r_t$, even multiple times, before moving more than several $r_t$ from the GC. An intriguing ramification is that BHs `ejected' from the GC may pass back through the core and undergo a strong encounter, potentially re-binding to the GC. In the case of BHs ejected by gravitational-wave merger kicks, this may enhance dynamical production of hierarchical BH mergers \citep[e.g.,][]{MillerHamilton2002,Rodriguez2019}. Checking all four simulations, we find that within the $14\,\Gyr$ simulation runtime, BH escapers pass from $r>r_t$ to $r\leq r_t$ an average of 200 times per GC---6 times when limited to BH merger remnants. These drop to 3.5 and 0.25 times, respectively, when considering only core passages from $r>r_c$ to $r\leq r_c$ \citep[the density-weighted core radius from][]{CasertanoHut1985}. Assuming a typical $r_c\sim1\,\pc$ containing stellar number density $10^3\,\pc^{-3}$ and a BH remnant with mass $30\,\Msun$ and speed $50\,\kms$, then the rate of strong encounters between returning BH merger remnants and typical stars in the core (capable of scattering the BHs back to $\Etilde < 0$) is only ${\sim}10^{-8}$ per GC per Hubble time. This suggests that returning BH merger remnants negligibly enhance hierarchical BH merger rates in GCs.

\subsection{How Cluster Density and Black Holes Affect Streams} \label{S:how_bhs_impact_tails}
While external tides dominate stellar stream morphology, factors internal to GCs are also relevant. In particular, the GC's initial number density $n$ (set by $r_v$ and $N$ in \texttt{CMC}) determines the timescales for both relaxation and stronger encounters \citep[e.g.,][]{Spitzer1987, HeggieHut2003, BinneyTremaine2008}. Along with $\rgc$, $n$ also determines how fully the GC's mass distribution fills its tidal boundary. These considerations have competing effects; higher $n$ leads to faster relaxation, \textit{hastening} evaporation, but also leads to a deeper central potential and so a larger gap between $\Ecrit$ and the typical $\Etilde$ in the GC. Since most bodies achieve $\Etilde>0$ in the GC's core this means higher $n$ can also hinder escape, \textit{slowing} evaporation. The competing effects can lead to seemingly divergent conclusions in the literature on $n$'s impact on evaporation. The speed-up of relaxation dominates in this study, so our denser (CC'd) models evaporate faster, yielding denser streams. This holds for typical \texttt{CMC} models, which have initial $n$, $N$, and $\rgc$ leading to evolved GCs consistent with most MWGCs today \citep{CMCCatalog}. But \textit{lower} density can lead to faster evaporation in \texttt{CMC} for GCs that more fully fill their tidal boundary (low $\rgc$ and high $r_v$ or low $N$; see W23's Figure~1). Relevant to diffuse, low-$N$ GCs in the MW halo, such as Palomar~5, this regime has been more strongly emphasized in recent direct $N$-body modeling \citep{Gieles2021, Gieles2023, Roberts2024}.

BH dynamics is similarly relevant as central heating from BH binary burning supports the GC against collapse, lowering the core density. As our models are tidally under-filling, our denser CC'd GCs evaporate faster and form denser streams despite ejecting their BHs earlier. But in the tidally-filling regime, BH heating increases the mass loss rate and stream density. \cite{Gieles2021} showed this may explain Palomar~5's unusually dense tidal tails, finding that low-$n$ direct $N$-body models could reproduce the tails if BHs make up ${\approx}20\%$ of Palomar~5's current mass. More broadly, variation in BH retention may help explain why only some MWGCs have observed tails \citep{Gieles2023, Roberts2024}. To holistically investigate this prospect and extrapolate to the full MWGC population, future studies of the impact of GC density and BH dynamics on tidal tails should consider both tidally-filling and underfilling GCs.

By extension, tidal tails and streams in the MW halo may be useful probes of not just BH populations in GCs but also more fundamental uncertainties regulating BH formation and retention, such as the stellar initial mass function (IMF) and supernova kick strengths. Notably, GCs born with top-heavy IMFs can produce so many BHs---and so much BH burning---that they rapidly dissolve in ${\sim}1\,\Gyr$ \citep[e.g.,][]{BanerjeeKroupa2011, Whitehead2013, Contenta2015, Chatterjee2017, Giersz2019, Weatherford2021}, potentially producing very dense tidal tails. That the $20\%$ BH mass fraction in Palomar~5 suggested by \cite{Gieles2021} is already higher than achievable from a canonical stellar IMF supports that tails may be a useful constraint on the IMF in low-density GCs. The accelerating evaporation rate near the end of such GCs' lives may further help explain why the density of some tidal tails (e.g., Palomar~5's) rises steeply with decreasing distance from the GC. We plan to investigate such prospects further in future work.

\subsection{Prognosis for Modeling Streams via the H\'{e}non Method} \label{S:discussion_monte_carlo_method}
\subsubsection{Collisional Dynamics for Potential Escapers} \label{S:potential_escaper_collisional_dynamics}
We have compared the raw energy criterion to the \cite{Giersz2008} $\alpha$ criterion, designed to account for scattering of PEs back to $\Etilde<0$ prior to escape. Though motivated by conservation of $\Etilde$ in the CR3BP, which does not hold for collisional dynamics, non-circular GC orbits, or evolving $\phiclus$ or $\phigal$, such energy-based escape criteria remain useful when the energy change is slow compared to the crossing timescale and $\phigal$ is not highly substructured near the GC. The intent of our comparison is not to identify which criterion yields more accurate streams, a judgement requiring direct $N$-body codes, but rather to gauge how much collisional dynamics for PEs during their escape may affect $\Delta t_{\rm esc}$ and stream properties.

Summarizing our findings, the $\alpha$ criterion raises the $\Etilde$ of PEs (Figure~\ref{fig:Etilde}), reducing $\Delta t_{\rm esc}$ (Figure~\ref{fig:tesc}) and lowering by several times the surface density of PEs in the portion of the GC's halo that is significantly non-spherical ($0.6\lesssim r/r_t \lesssim 1$; Figure Set~\ref{fig:fancy_tail}). So collisional dynamics for PEs may significantly alter the morphology and kinematics of the GC \textit{within} the tidal boundary. Yet the two criteria yield very similar evaporation rates $\dot{N}$ (Figure~\ref{fig:macro_evolution}) and \textit{extratidal} properties, including the stream density and velocity dispersion profiles. This is expected for slow, steady evaporation ($\Delta t_{\rm esc}\ll N/\dot{N}$), as in our models representative of typical MWGCs. The small extratidal impact implies collisional dynamics can reasonably be neglected for PEs when simulating tails and streams from MWGCs, though perhaps not for $r<r_t$, or small $N$ and fast $\dot{N}$ (e.g., near GC dissolution).

To the extent that better accounting for PE collisional dynamics may still have small effects, keep in mind that weak two-body scattering is a competition between cooling (dynamical friction) and heating (relaxation)---e.g., Section~7.8 of \cite{BinneyTremaine2008}. By scattering some PEs back to $\Etilde<0$ before they can escape, cooling should reduce the evaporation rate, which the $\alpha$ criterion achieves by raising the threshold for removal from the GC dynamics to $\Etilde\gtrsim0.1$. But this does not account for how cooling would also lower the speed of the PEs that \textit{do} escape; by raising the threshold $\Etilde$, the criterion does the opposite. This may unintentionally help account for \textit{heating} of PEs during escape, but whether the $\alpha$ criterion does so accurately is unclear. The accuracy should at least vary with PE mass, since cooling dominates at high mass and heating at low mass (most stars).

A more physically-motivated alternative is delayed escape, in which PEs continue participating in collisional dynamics before removal from the GC simulation. This option has been implemented into \texttt{MOCCA} by \cite{Giersz2013}. \texttt{MOCCA} identifies PEs every timestep $\Delta T$ via the raw criterion. Based on FH00's results, it estimates the probability the PE will escape during $\Delta T$ as $P(\Delta T)\equiv1-(1+b\omega\Etilde^2\Delta T)^{-c}$, where $b\approx3$ and $c\approx0.8$ are tuned to best fit the mass loss rate from direct $N$-body codes. Random sampling with this probability determines which PEs to remove each $\Delta T$. Yet the scaling $\Delta t_{\rm esc}\propto\Etilde^{-2}$ from FH00 neglects GC dynamics and mass loss and only applies for $\Etilde\ll 1$. So $\Delta t_{\rm esc}$ from \texttt{MOCCA}'s algorithm does not necessarily scale properly with $\Etilde$. Our results accounting for GC evolution show that $\Delta t_{\rm esc}\sim(\Etilde^{-0.1},\Etilde^{-0.4})$ for PEs with (low, high) $\Etilde$, with some ambiguity in between due to chaotic scattering. In principle, these new scalings can be swapped with the $\Etilde^{-2}$ relation in \texttt{MOCCA}'s algorithm, and the coefficients $b$ and $c$ re-tuned against direct $N$-body codes.

While \texttt{MOCCA}'s delayed escape prescription could mildly improve accuracy for escaper kinematics, it has not been implemented into \texttt{CMC} since the impact on the GC mass loss rate is small. And like the simpler energy criteria with immediate escape, it still limits the H\'{e}non method's capacity to simulate more complex aspects of stellar stream morphology arising from evolution or substructure in the tidal field. More general alternatives are therefore necessary.

\subsubsection{Generalizing to Non-spherical Clusters and Evolving Tides} \label{S:generalizing_to_evolving_tides}
There are many sources of evolving tides that we do not account for in this work. The orbits of most MWGCs are eccentric and inclined relative to the MW disk \citep{Baumgardt2019}, traits that induce tidal shock heating from passage near the MW's center or through its disk, respectively. This hastens GC dissolution and causes the GC's bound mass to fluctuate \citep[e.g.,][]{GnedinOstriker1997, BaumgardtMakino2003, Webb2013, Webb2014a, Webb2014b, Madrid2014}. The eccentricity also causes the tidal tails to fan \citep[e.g.,][]{Kupper2008, Kupper2010} and more realistic, non-spherical MW potentials add further complexity. For instance, a triaxial $\phigal$ \citep[e.g.,][]{CapuzzoDolcetta2005} produces further stream gaps/overdensities and causes the GC orbit to precess, while a rotating Galactic bar induces further fanning and asymmetry in the lengths of the leading and trailing tails  \citep{Hattori2016,Pearson2017}. Modifications to gravity, an alternative to dark matter, can cause similar asymmetry \citep{Thomas2018,Kroupa2022}. Finally, as noted in Section~\ref{S:intro}, finer MW substructures such as dark matter subhalos, molecular clouds, other GCs, and infalling MW satellites can all heat and strip stars from streams, leaving behind significant gaps and kinks. These introduce fine time-dependence to the tides experienced by a GC, but tidal evolution on longer timescales is also likely, especially for the many GCs in the MW halo suspected to have been accreted from disrupted MW satellites \citep[e.g.,][]{Helmi2020}.

Our motivations for developing the H\'{e}non method for stream modeling are its detail for internal stream populations and its \textit{speed}---and thus its capacity, in principle, to generate large model grids of streams to explore the impacts of the many above competing factors. To realize this prospect, improvements to \texttt{CMC}'s escape criteria and tidal physics must be sufficiently general while not dramatically slowing \texttt{CMC}.

One option is to allow non-spherical $\phiclus$ and dynamics via basis expansion and position-dependent velocity diffusion coefficients \citep{Vasiliev2015}. While this only enables mild (factor of 2) asymmetries, so cannot on its own resolve tidal tails, it accounts for $\phiclus$'s asymmetry and allows PEs to continue participating in dynamics until escape. Collisionless trajectory integration could then follow as in our approach. But critical drawbacks accompany these advantages: (1) The method requires following each body's orbit and applying diffusion on the dynamical timescale, making it far slower than the H\'{e}non method. (2) The asymmetry makes assigning pairs of nearest neighbors highly non-trivial, hindering implementation of strong encounters. (3) Updating \texttt{CMC} to use the method or adding many years of \texttt{CMC} development (stellar evolution, strong encounters, escape) to Vasiliev's code would be time-consuming. The benefits for stream modeling are not likely enough to outweigh these drawbacks. As we argued above, explicit treatment of collisional dynamics for PEs should not substantially alter stream properties and $\phiclus$'s mild asymmetries in the GC halo are not as relevant to streams as external tides or dynamics in the core, where most diffusion to $\Etilde>0$ occurs (\citealt{SpitzerShapiro1972}; W23).

To generalize the H\'{e}non method to evolving tides, \cite{Sollima2014} developed an escape prescription for non-circular GC orbits in two distinct cases of a static, cylindrically symmetric MW potential: a point-mass $\phigal$ and $\phigal$ with bulge, disk, and halo components. At each timestep $\Delta T$, for each body in the GC, their algorithm isotropically samples a random escape vector, along which it computes the distance (and $\phieff$ at) the tidal boundary. It flags as PEs bodies with sufficient energy and angular momentum to cross the boundary during $\Delta T$, based on each body's orbital period in the GC. (Note this is mildly improper as it assumes escape occurs within several crossing times). The orbits of the GC and all PEs are then integrated within the external $\phigal$ for a full orbital period of the GC in the MW. PEs are only removed from the simulation if they pass beyond the maximum $r_t$ of the GC during this integration, allowing PEs to continue participating in dynamics prior to removal. Drawbacks are that the PE trajectories do not account for $\phiclus$ and that the above selection of PEs is based on laborious calculations (in their appendix) for their specific choices of $\phigal$, an approach hard to generalize to more substructured $\phigal$.

More recently, \cite{Rodriguez2023} developed new \texttt{CMC} escape prescriptions for non-circular GC orbits and applied them to an evolving, substructured $\phigal$ from a \texttt{FIRE}-2 MHD simulation of a MW-like galaxy. They first extracted time series data on the tidal tensor (spatial 2nd derivatives of the local $\phigal$) from the orbits of particles representing GCs in the galaxy simulation, then ran \texttt{CMC} computing $r_t$ every timestep based on that data (loaded as an input file). When loading tidal tensor data from \texttt{Gala} integrations of eccentric GC orbits in a static $\phigal$ (with bulge, disk, and halo components), \texttt{CMC} reproduced well the mass evolution of equivalent direct $N$-body models \citep{Webb2014a}---at least when using $r_{\rm apo}>r_t$ rather than an energy criterion. But mass loss was slower/faster than in the direct $N$-body models for low/high-eccentricity GC orbits. These respective deviations are at least partly due to $\Etilde>0$ not universally implying $r_{\rm apo}>r_t$, the \textit{maximum} distance to the tidal boundary, and the algorithm's inability to re-capture some escaped bodies when the boundary re-expands after each perigalacticon.

In the near term, when applying \texttt{CMC} to model streams from GCs on eccentric or inclined orbits, or in evolving, substructured $\phigal$, we will blend our collisionless integration of escape trajectories---similar to \cite{Sollima2014}---with the tidal tensor technique of \cite{Rodriguez2023}. Specifically, we intend to identify PEs like \cite{Rodriguez2023}, immediately removing them from \texttt{CMC} (neglecting collisional dynamics for PEs during their escape due to its small impact on stream properties). We can then directly integrate the PE trajectories each timestep in post-processing as in this work. This is advantageous since it will not entail building into \texttt{CMC} a separate collisionless integrator for escaping/escaped bodies. Unlike in this work, the orbit integration would occur in an inertial Galactocentric frame since the pseudo-forces (centrifugal, Coriolis, Eulerian, etc.) in the non-inertial GC frame are not generally known for an evolving, substructured $\phigal$.

In the longer term, we aspire to add a galactic dynamics integrator to \texttt{CMC} to allow collisional dynamics (and delayed escape) for PEs and re-capture of escaped bodies. Due to the lack of a general definition for $\Etilde$ in an evolving, substructured $\phigal$, and to avoid missing any bodies capable of escape, we would select PEs via a more liberal version of the apocenter criterion than used by \cite{Rodriguez2023}---e.g., $r_{\rm apo}/r_t>2/3$. This corresponds roughly to the minimum distance to the tidal boundary, and the radius beyond which the GC is significantly non-spherical. Rather than being removed immediately from \texttt{CMC}, PEs flagged in this manner would then be randomly projected into full six-dimensional phase space coordinates (as in this work) and their trajectories integrated in the full combined $\phiclus+\phigal$ with the integration time set to \texttt{CMC}'s current timestep. At the end of the integration, \texttt{CMC} would remove the PE from its collisional dynamics if $r/r_t>2/3$, but if $r/r_t\leq2/3$ the body would stay in the GC to take part in dynamics in the next timestep. (Repeating this procedure every timestep is appropriate since \texttt{CMC} operates on the relaxation timescale, over which bodies' orbits are randomized.) Collisionless integration of the escapers would continue from there, separately but still internal to \texttt{CMC}. At the end of every timestep, escapers that return within $r/r_t<2/3$ could then be transferred back to the main collisional dynamics routine. This would allow re-capture, unlike \cite{Sollima2014} or \cite{Rodriguez2023}, but implementation would not be trivial as it requires expanding \texttt{CMC}'s parallelization scheme to escapers.

\section{Summary and Future Work} \label{S:summary}
We have applied for the first time orbit-averaged star cluster models to study the formation of tidal tails and stellar streams from GCs. Specifically, we use \texttt{CMC}---a state-of-the-art, publicly-available implementation of the \cite{Henon1971a, Henon1971b} Monte Carlo method. Though this method assumes spatial spherical symmetry in its \textit{collisional} dynamics, we show that treating energetically unbound bodies (potential escapers; PEs) as collisionless enables formation in post-processing of asymmetric tidal phenomena such as tidal tails with exquisite detail. The benefits of the H\'{e}non method are that it is much faster than direct summation $N$-body methods for GCs of typical $N$ and density, making it far better suited to large-parameter-space modeling of tidal tails and stellar streams from GCs---an essential feature due to the vast array of considerations internal and external to the GC relevant to stream properties. While even faster but more approximate stream simulation is possible via common particle-spray techniques, \texttt{CMC} can generate streams with many of the advantages of a full GC-modeling code. These include highly accurate detail on time of escape, escaper kinematics, and details on streams' internal populations (e.g., masses, star types, binary properties, etc.). \texttt{CMC} may be especially ideal for rapid but reliable exploration of tidal tails from large grids of non-standard GCs, such as those with atypical stellar IMFs, extreme density, unusual binary fractions, or intermediate-mass BHs (work in progress). Key limitations are that the H\'{e}non method's assumptions of evolution on the relaxation timescale, virial equilibirum, and high $N\gtrsim10^5$ (for realistically broad stellar IMFs), make it unsuited for modeling tidal shocking or the final stage of GC dissolution. Yet as discussed in Section~\ref{S:discussion}, further refinements can allow the H\'{e}non method to simulate streams from GCs on non-circular orbits, even in an evolving, highly substructured external potential.

The main findings of this work, the second in our series on escape from GCs, are as follows:
\begin{enumerate}
\item We demonstrate that the H\'{e}non method can accurately reproduce known features of tidal tails and stellar streams from GCs on circular orbits within an unevolving, spherical Galactic potential. This is achieved by collisionlessly integrating the trajectories of potential escapers in the full tidal field.
\item We examine for the first time the in-cluster survival timescale (escape timescale $\Delta t_{\rm esc}$) of PEs in a realistically evolving GC potential, and also account roughly for the impact of ongoing collisional dynamics on PEs via the \cite{Giersz2008} $\alpha$ escape criterion. Escape in this case occurs on a timescale of ${\approx}100\,\Myr$ instead of ${\approx}10\,\Gyr$ in the static, collisionless case.
\item GC mass loss leads to a new scaling of $\Delta t_{\rm esc}$ with excess relative energy $\Etilde$---namely, $\Delta t_{\rm esc}\propto (\Etilde^{-0.1},\Etilde^{-0.4})$ for $\Etilde(\lesssim0.03,\gtrsim0.56)$, respectively---much shallower than the scaling $\Delta t_{\rm esc}\propto\Etilde^{-2}$ for a static GC potential.
\item We highlight discreteness in the $\Delta t_{\rm esc}$ distribution arising from chaotic scattering. Different characteristic $\Delta t_{\rm esc}$ within distinct locally smooth regions of the phase space of PEs introduce successive plateaus in the $\Delta t_{\rm esc}$ distribution for $0.03 \lesssim\Etilde\lesssim 0.56$. This hinders a clean power-law scaling of $\Delta t_{\rm esc}$ on $\Etilde$ in this interval.
\item We analyze for the first time the impact of return trajectories circumnavigating the Galactic center, finding they produce robust returning tidal tails on a timescale of ${\sim}10\,\Gyr$ for GCs at $\rgc=8,\kpc$ in our idealized circumstances. Though a realistically evolving Galaxy with significant substructure is likely to disperse such tails, they may eventually be observable in proper motion space and could excellently constrain the history and substructure of the Galaxy over longer timescales than typical streams. Returning tails may be more relevant for GCs in dwarf galaxies, where the timescale for low-$\Etilde$ escapers to return to the GC is much shorter.
\item In our ideal case, the returning tails increase velocity dispersion in stellar streams by several $\kms$, an effect similar in magnitude to the perturbative influence on streams by giant molecular clouds, the Galactic disk, and dark matter subhalos. Though the velocity dispersion enhancement is likely much smaller for streams dispersed by a realistically evolving and substructured Galaxy, the boost to velocity dispersion may still be comparable to the ${\lesssim 0.5\,\kms}$  boost from other observing biases such as unresolved binaries.
\end{enumerate}

In future work, we will further refine escape prescriptions in \texttt{CMC} to model streams from GCs on non-circular orbits in an evolving, substuctured Galaxy. To do so, we will combine our collisionless integration of escape trajectories with the general computation of an approximate spherical tidal boundary relying on tidal tensors \citep{Rodriguez2023}. We will also study in more detail the internal stellar populations inhabiting the tidal tails and streams from our simulated GCs. This will include an analysis of observability by continuing stellar evolution for escapers and converting their luminosities to common observing magnitudes (e.g., with \texttt{Gaia}), following the technique of \cite{Rui2021a}. Other focii will be binaries in our streams (relevant to velocity dispersion biases) and mass segregation both parallel and perpendicular to the stream axis, which respectively would bias the observed length and width of streams due to lower observing completeness at lower mass. The former case \citep{Webb2022} naturally develops since GCs tend to eject low-mass stars faster, and therefore earlier on average. Mass segregation perpendicular to the stream has not been considered before, but may arise from velocity-dependent ejection speeds of stars---e.g., due to the strength of the velocity kick scaling inversely with mass, or due to the higher rate of strong encounters for high-mass stars segregated in the GC's core. Such analysis would improve our understanding of observing biases relevant to stream morphology and kinematics, useful to reducing uncertainties when applying stream observations to, e.g., place firm constraints on the nature of dark matter. In summary, \texttt{CMC}'s speed and aforementioned capabilities/planned improvements should enable us to construct for the first time large grids of stream models using a dedicated GC-modeling code, along with detailed comparisons to stream observations.

\begin{acknowledgements}
We thank the referee for constructive feedback that led to a clearer discussion of our method and implications.
This work was supported by NSF grant AST-2108624 and NASA grant 80NSSC22K0722, as well as the computational resources and staff contributions provided for the \texttt{Quest} high-performance computing facility at Northwestern University. \texttt{Quest} is jointly supported by the Office of the Provost, the Office for Research, and Northwestern University Information Technology. F.A.R. and G.F. acknowledge support from NASA grant 80NSSC21K1722. N.C.W acknowledges support from the CIERA Riedel Family Graduate Fellowship. S.C. acknowledges support from the Department of Atomic Energy, Government of India, under project No.~12-R\&D-TFR-5.02-0200 and RTI~4002. F.K. acknowledges support from a CIERA Board of Visitors Graduate Fellowship. Support for K.K. was provided by NASA through the NASA Hubble Fellowship grant HST-HF2-51510 awarded by the Space Telescope Science Institute, which is operated by the Association of Universities for Research in Astronomy, Inc., for NASA, under contract NAS5-26555.
\end{acknowledgements}

\software{\texttt{CMC} \citep{CMC_v1.0}, \texttt{COSMIC} \citep{COSMIC}, \texttt{fewbody} \citep{Fregeau2004,Antognini2014,AmaroSeoane2016}, \texttt{Gala} \citep{Gala_v1.3},  \texttt{matplotlib} \citep{Matplotlib}, \texttt{SciPy} \citep{SciPy}, \texttt{NumPy} \citep{NumPy}, \texttt{Astropy} \citep{Astropy}, \texttt{pandas} \citep{Pandas}.}

\vspace{0pt}
\begin{figure*}
\centering
\includegraphics[width=\linewidth]{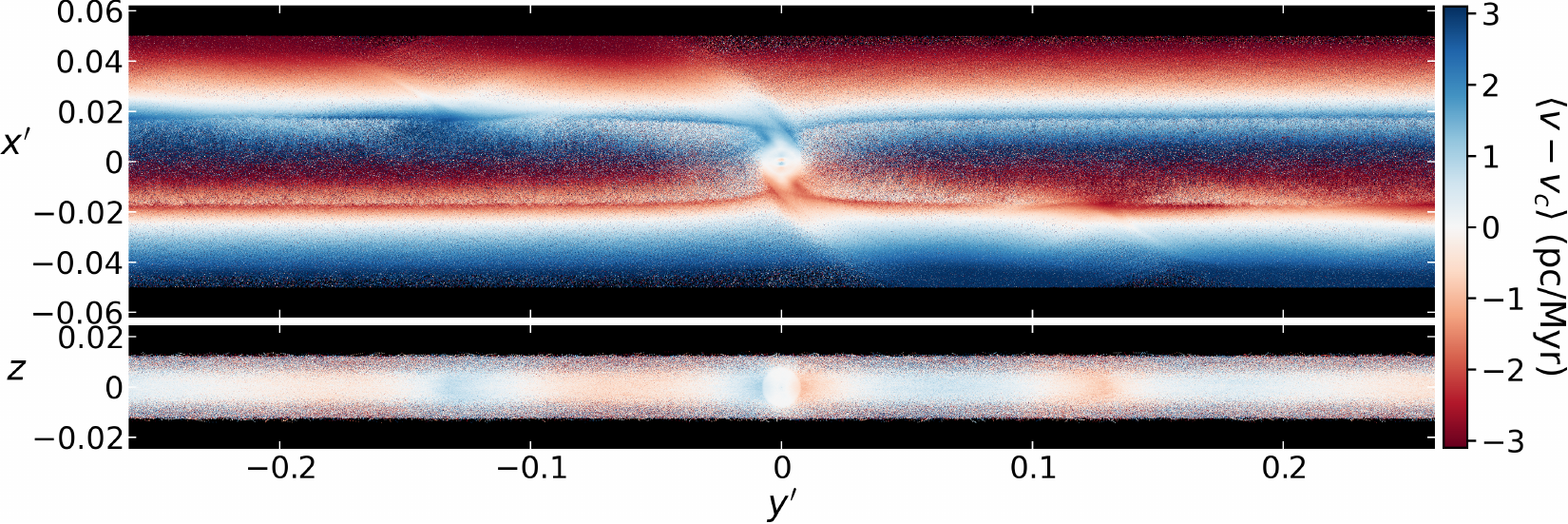}
\caption{As Figure~\ref{fig:fancy_tail}, but mapping the time-averaged local mean stream speed $\langle v-v_c \rangle$ relative to the GC circular speed instead of the surface density, and counting only PEs with $|x'(t)|<0.05$ and $z(t)<r_t(t)/\rgc$. Black regions indicate bins outside this interval or where no escapers are present in any of the 2001 snapshots across ages $t\in[11,13]\,\Gyr$.}
\label{fig:fancy_tail_speed}
\end{figure*}

\begin{figure*}
\centering
\includegraphics[width=\linewidth]{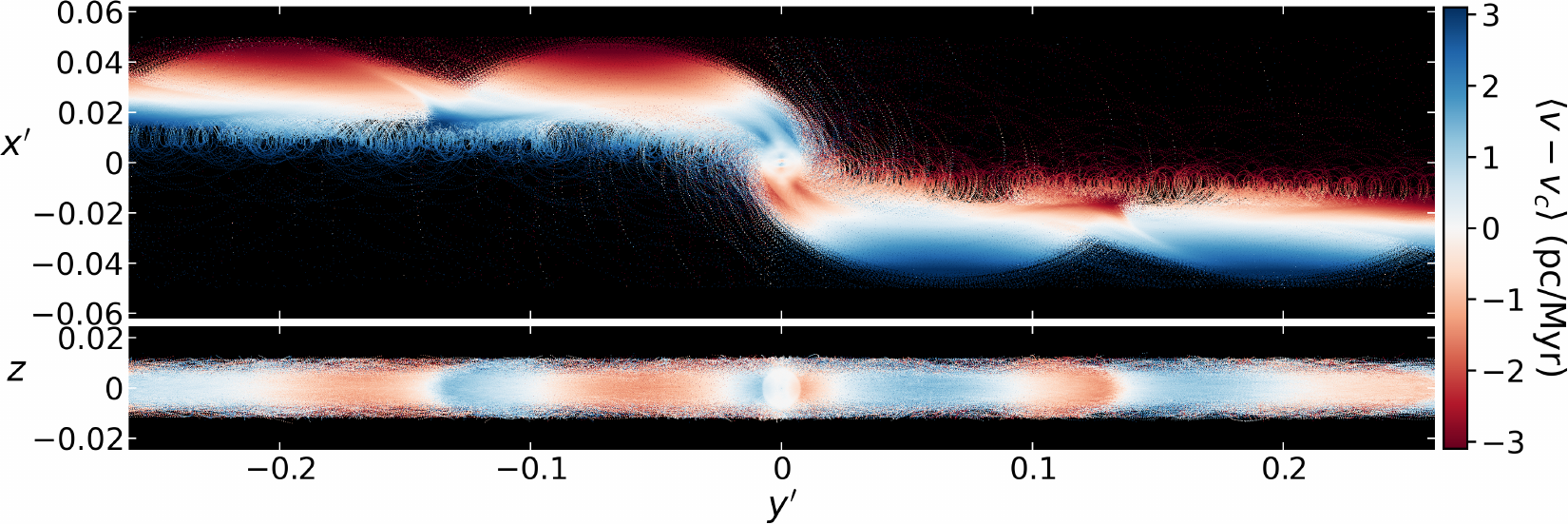}
\caption{As Figure~\ref{fig:fancy_tail_speed}, but excluding the returning tails by cutting out PEs $500\,\Myr$ after their first `escape' (crossing to $r>r_t$).}
\label{fig:fancy_tail_speed2}
\end{figure*}

\appendix
\section{Tidal Tail Speed Maps} \label{S:tail_speed_maps}
To aid visual interpretation of the Figures~\ref{fig:rhov_tail} and \ref{fig:rhov_tail2}, we include here a more detailed pair of two-dimensional maps of the speed profile $\langle v-v_c \rangle$ in Figures~\ref{fig:fancy_tail_speed} and \ref{fig:fancy_tail_speed2}. Each show the CC'd GC under the $\alpha$ escape criterion and use the same binning strategy and coordinates of Figure~\ref{fig:fancy_tail} described in Section~\ref{S:stream_morphology}, except that only escapers with $|x'(t)|<0.05$ and $z(t)<r_t(t)/\rgc$ are counted (consistent with Figures~\ref{fig:rhov_tail} and \ref{fig:rhov_tail2}). Figure~\ref{fig:fancy_tail_speed2} further excludes the returning tails by cutting out PEs $500\,\Myr$ after their first `escape' (crossing to $r>r_t$). Doing so clearly increases $\langle v-v_c \rangle$ from the perspective of the Galactic center (lower panels) and reduces the dispersion in $\langle v-v_c \rangle$ (spread in color) at any $y'$ because there is no longer a returning tail to oppose the local flow of the outgoing tail.

\bibliography{CMCejecta}

\end{document}